\documentclass[11pt,a4paper]{article}
\usepackage[utf8]{inputenc}

\usepackage{jheppub}

\usepackage{amsfonts}
\usepackage{amsmath}
\usepackage{amssymb}
\usepackage{amsthm}
\usepackage{bm}
\usepackage{subcaption}
\usepackage[scr=boondoxo, bb=pazo]{mathalpha}
\usepackage{extarrows}
\usepackage{float}
\usepackage{graphicx}
\usepackage{hyperref}
\usepackage{mathtools}
\usepackage{physics}
\usepackage{setspace}
\usepackage{verbatim}
\usepackage{wrapfig}

\def\tilde{\widetilde}
\def\t{\tilde}
\def\hat{\widehat}
\def\h{\hat}
\def\bar{\overline}
\def\b{\bar}

\def\p{\partial}
\DeclareMathOperator{\Hom}{Hom}
\newtheorem{definition}{Definition}[section]

\newcommand{\Z}{{\mathbb Z}}
\newcommand{\C}{{\mathbb C}}
\newcommand{\R}{{\mathbb R}}

\def\1{{\mathbb 1}}

\def\CC{{\mathcal C}}

\def\CH{{\mathcal H}}

\title{2d TQFTs and baby universes}

\author[]{John Gardiner}
\author[]{and Stathis Megas}

\affiliation[]{Mani L. Bhaumik Institute for Theoretical Physics \\
Department of Physics and Astronomy, University of California, Los Angeles, CA 90095, USA }

\emailAdd{johngardiner@ucla.edu}
\emailAdd{stathismegas@gmail.com}

\abstract{In this work, we extend the 2d topological gravity model of \cite{Marolf:2020xie} to have as its bulk action any open/closed TQFT obeying Atiyah's axioms.
The holographic duals of these topological gravity models are ensembles of 1d topological theories with random dimension.
Specifically, we find that the TQFT Hilbert space splits into sectors, between which correlators of boundary observables factorize,
and that the corresponding sectors of the boundary theory have dimensions independently chosen from different Poisson distributions.
As a special case, we study in detail the gravity model built from the bulk action of 2d Dijkgraaf-Witten theory, with or without end-of-the-world branes, and for arbitrary finite group $G$.
The dual of this Dijkgraaf-Witten gravity model can be interpreted as a 1d topological theory whose Hilbert space is a random representation of $G$ and whose aforementioned sectors are labeled by the irreducible representations of $G$.

These holographic interpretations of our gravity models require projecting out negative-norm states from the baby universe Hilbert space, which in \cite{Marolf:2020xie} was achieved by the (only seemingly) ad hoc solution of adding a nonlocal boundary term to the bulk action. In order to place their solution in the completely local framework of a TQFT with defects, we couple the boundaries of the gravity model to an auxiliary 2d TQFT in a non-gravitational (i.e.\ fixed topology) region. In this framework, the difficulty of negative-norm states can be remedied in a local way by the introduction of a defect line between the gravitational and non-gravitational regions.
The gravity model is then holographically dual to an ensemble of boundary conditions in an open/closed TQFT without gravity.
}

\begin{document}
\maketitle

\section{Introduction}
Recent work has re-emphasized the role of the Euclidean path integral in quantum gravity.
Specifically we have seen the revitalization of the old idea that quantum gravity should be described by a sum over spacetime topologies and an integration over all metrics within each topology.
% The past year has seen a revitalization of the original approach of doing quantum gravity, which is by summing over spacetime topologies and integrating over all the spacetime geometries admissible for a given topology.
Making sense of such a path integral is, of course, difficult or impossible in general, so this course of action has mostly focused on two dimensions.
Despite the notoriously hard problem of classifying topological manifolds in 3d, some attempts at path integrating over a subset of 3d manifolds (often Seifert manifolds) have also been made \cite{maloney2020averaging, cotler2020ads3, maxfield2020path}. 

A helpful point of view into this sum over topologies of spacetimes is afforded by the old ideas of baby universes \cite{Coleman:1988cy, Giddings:1988wv, Polchinski_1994}, where one thinks of non-trivial topologies as the emission and re-absorption of baby-universes. 
In Euclidean signature, this emission/re-absorption process takes the form of a spacetime wormhole.\footnotemark
\footnotetext{Spacetime wormholes differ from the usual Einstein-Rosen wormholes in that the latter are topological connections between otherwise distant regions of \emph{space}, whereas the former are additional topological connections between regions of spacetime.}
Specifically, \cite{Coleman:1988cy} considers a formalism in which the Hamiltonian of the universe contains couplings between the fundamental fields and ``baby universe field" operators $A_i$ which describe the creation and annihilation of different types of baby universes.
Because couplings between the $A_i$ and other fields appear in the Hamiltonian, tracing over the number and types of baby universes in existence (facts that are presumably unknowable to an observer in the parent universe) produces an effective evolution that is non-unitary.
Crucially however, the baby universe operators $A_i$ commute with each other and with the Hamiltonian.
This renders the non-unitarity relatively benign, as \emph{within} any eigenspace of $A_i$ the evolution remains unitary.
This means ignorance of the number and types of baby universes does not, in fact, lead to observable decoherence.
Rather, such ignorance is just operationally equivalent to ignorance of some number of coupling constants in the theory describing parent universe physics.
In other words, the physics of the parent universe is described by a statistical ensemble of unitary theories.
The theories in this ensemble are parametrized by the simultaneous eigenstates $\ket{\alpha}$ of the $A_i$, which are called alpha-states. The unknown coupling constants of parent universe physics are the eigenvalues of $A_i$, also called alpha-parameters.

The baby universe idea and the appearance of ensembles has a counterpart in the context of holography.
In this context, where partition functions of a boundary theory are dual to gravity in the bulk, there is a question of whether the calculation of a boundary partition function with disconnected spacetime components should include bulk geometries that connect those components \cite{Maldacena:2004rf}.
However natural a sum including connected topologies may otherwise be, it destroys the manifest factorization between spacetime components of the boundary partition function.
That is to say, a rule including contributions from connected geometries cannot a priori be expected to give boundary partition functions satisfying $Z[M\sqcup N]=Z[M]Z[N]$ for disconnected spacetimes $M$ and $N$. %\footnotemark
%\footnotetext{Lack of such factorization is of course a violation of locality. Locality in quantum field theories can be understood as conditions on consistently cutting and gluing manifolds. Factorization between disconnected components of spacetime is the ``zeroth order" such condition, which says that "cutting and gluing along co-dimension $0$ manifolds" can be done consistently. So failure of factorization is in some sense an extreme violation of locality.}
One way to proceed is to reinterpret the boundary partition functions appearing in this holographic context as ensemble averages of partition functions, where bulk gravity is then interpreted as dual to a statistical ensemble of boundary theories.\footnotemark
\footnotetext{Equally, we could turn around the logic and stipulate factorization as a \emph{criterion} for any well-behaved quantum theory of gravity, thus declaring most generic gravity path integrals to be in the so-called ``swampland" \cite{McNamara:2020uza}.}
This leads to what you could call a new entry in the holographic dictionary:
\newline
\begin{align} 
\substack{  \text{\large ensemble of non-gravitational} \\ \text{\large boundary theories on} \\  \text{\large disconnected boundaries} }  
 ~~ \Leftrightarrow \quad
\substack{\text{\large Euclidean gravity path integral} \\   \text{\large with contributions connecting} \\  \text{\large disconnected boundary components} }  \label{entry}
\;,     
\end{align}
\newline
which now has evidence from a number of lines of study, including the JT-gravity \cite{maldacena2016conformal} and SYK model \cite{Kitaev_2018} correspondence \cite{saad2019jt,witten2020matrix}, replica wormhole calculations of the Page curve \cite{Almheiri:2019qdq,penington2020replica}, and recent work suggesting a possible correspondence between 3d gravity and ensembles of 2d conformal field theories \cite{Afkhami-Jeddi:2020ezh,maloney2020averaging,cotler2020ads3,Cotler:2020hgz}.

All these different ideas merge in \cite{Marolf:2020xie}, wherein the authors demonstrate, via logic reminiscent of the older baby universe picture, how ensembles of boundary theories can naturally emerge from gravity path integrals, and connect with the replica wormhole discussions of the Page curve.
Important to the present work, the authors of \cite{Marolf:2020xie} introduce and analyze a simple topological model of a gravity path integral as an explicit example of these ideas.
Their gravity model is a sum over 2d surfaces weighted by a topological action.
Subsequently, the authors of \cite{Balasubramanian:2020jhl} extend this model to include surfaces with spin structures.
The present work picks up from where \cite{Marolf:2020xie}, and in a certain sense \cite{Balasubramanian:2020jhl} left off, by considering more general topological bulk theories.

The rest of the paper is organized as follows.
In section \ref{review}, we review the simple topological model described in \cite{Marolf:2020xie} which this work can be seen as generalizing.
In section \ref{BU}, we consider a gravity path integral constructed from Dijkgraaf-Witten theory in two dimensions.
We find (not unexpectedly \cite{Kapec:2019ecr}) that the correlators of boundary insertion operators factorize between sectors described by irreducible representations of $G$.
What's more, the boundary dual theory can be interpreted as a 1d topological theory with global symmetry group $G$ whose Hilbert space is a random representation of $G$.
Specifically, the number of times a copy of an irreducible representation of $G$ appears in the Hilbert space is given by a random integer, with integers for different irreducible representations chosen independently.
In section \ref{generalcase}, we consider a gravity path integral constructed from any 2d TQFT as defined by Atiyah's axioms \cite{Atiyah}.
We show that the Hilbert space of the boundary theory is the direct sum of sectors labeled by the eigenstates of the TQFT handle operator with the number of dimensions in each sector chosen independently from a different Poisson distribution.
In section \ref{EofW} we generalize further by considering open/closed TQFTs, that is to say TQFTs with boundary.
We describe a gravity path integral constructed from open/closed TQFTs, again using Dijkgraaf-Witten as a first example.
Similar to the closed case, for open/closed TQFTs we find that correlation functions factorize between sectors labeled by eigenstates of the handle operator.
We discuss the interpretation of the gravity path integral in terms of a boundary ensemble theory.
We encounter the same difficulty encountered by \cite{Marolf:2020xie} with defining a theory without negative-norm baby universe states, a difficulty for which they proposed the solution of adding a nonlocal boundary term to the bulk action.
We discuss this difficulty in section \ref{reflectionpos}, reframing the solution in the language of 2d TQFTs with defects.
Along the way, we introduce generalized boundary observables representing non-gravitational regions coupled to the gravity region with fluctuating topology.
The gravity region then is dual to an ensemble of boundary conditions for an open/closed TQFT.
Besides motivating a solution to the negative-norm states in terms of a defect line, this picture also motivates an understanding of the additional alpha-parameters associated with having end-of-the-world branes (or more general boundary conditions).
In section \ref{discuss} we discuss directions for future work, including the aim of studying more realistic models of gravity, as well as the possibility of exploring aspects of holography such as bulk reconstruction in the simple setting of topological theories.
In appendix \ref{DWappendix}, we review a state sum formulation of 2d Dijkgraaf-Witten theory with defects, which is sufficient to calculate all the Dijkgraaf-Witten theory results used in the paper.

\subsection{Review of a simple gravity model}\label{review}
The model of \cite{Marolf:2020xie} describes a gravity path integral built from a sum over spacetime topology.
The authors consider 2d orientable spacetimes without metric or other geometric structure and an action that is (nearly) just the Euler characteristic.
It is, in fact the Euler characteristic together with an additional term depending on the number of boundaries.
Specifically, for a 2d manifold with genus $g$ and $n$ boundaries they assign the action
\begin{align}
S[M_{g,n}]
=
S_0\,\chi(M_{g,n})+n\,S_{\p}
=
S_0(2-2g)+n(S_{\p}-S_0)
\;,\label{action}
\end{align}
where $\chi$ is the Euler characteristic $2-2g-n$ and $S_0$ and $S_{\p}$ are parameters of the theory.
The gravity path integral for spacetime with fixed $n$ number of circle boundary components then becomes
\begin{align}
    Z^{\text{QG}}[\text{$n$ boundaries}]
    =
    \sum_{
        \substack{
            \text{$M$ s.t.\ $\p M$ has}\\
            \text{$n$ components}
        }
    } \mu(M) e^{S_0 \chi(M) +S_{\p} n}
\;.\label{cf}
\end{align}
The measure $\mu(m)$ here is $\mu(M)=\frac{1}{\prod_g m_g!}$, where $m_g$ is the number of components with genus $g$ that are not connected to any boundary.
This accounts for the residual gauge symmetries permuting identical connected components of $M$.\footnotemark
\footnotetext{Permutations that involve components with boundaries are of course not gauge redundancies.}
We point out here that for the particular choice $S_\p=S_0$, the number of boundaries $n$ affects the combinatorics both in the sum over topologies and in the measure $\mu$, but it has no effect on the action.

We should now make two comments.
First, by analogy with a correlation function being computed as a sum over Feynman diagrams with fixed external legs, we notate the output of the gravity path integral (a sum over spacetimes with fixed boundaries) as a correlation function, where the fixed boundaries play the role of operator insertions.
So for example the gravity path integral over manifolds with $n$ circle boundary components will be notated
\begin{align}
\expval{\h{Z}^n}
\equiv
Z^{\text{QG}}[\text{$n$ boundaries}]
\;,
\end{align}
so that the operator $\h{Z}$ simply denotes the insertion of an additional circle boundary component.
Second, we point out that the action \eqref{action}, at least with $S_\p$ set to zero, produces the partition function of a (particularly simple) TQFT, so that the gravity path integral is simply a sum over TQFT partition functions:
\begin{align}
    Z^{\text{QG}}[\text{$n$ boundaries}]
    =
    \sum_{
        \substack{
            \text{$M$ s.t.\ $\p M$ has}\\
            \text{$n$ components}
        }
    }
    \mu(M)
    Z^{\text{TQFT}}[M]
    \;.
\end{align}
We mention this, as our aim in this paper is to allow $Z^{\text{TQFT}}$ to be a more general TQFT partition function.

The values of the correlators $\expval{\h{Z}^n}$ can be gathered in the generating function $\expval{e^{u\h{Z}}}$.
The generating function for the connected correlators is simply $\log\expval{e^{u\h{Z}}}$, which evaluates to
\begin{align}
    \log\expval{e^{u\,\h{Z}}}
    =
    \sum_{n=0}^{\infty}
    \frac{u^n}{n!}
    \expval{\h{Z}^n}_{\text{conn}}
    =
    \lambda e^{u\,e^{S_{\p}-S_0}}
    \;,
\end{align}
where $\lambda=\sum_g e^{S_0(2-2g)}=e^{2S_0}/(1-e^{-2S_0})$ is the value of the connected vacuum correlator.
From the resulting expression for $\expval{e^{u\h{Z}}}$, the correlators $\expval{\h{Z}^n}$ can be extracted
\begin{equation}
    \expval{\h{Z}^n}
    =
    e^{\lambda}
    B_n(\lambda)
    e^{(S_\p-S_0)n}
    \;,
\end{equation}
where $B_n$ here is the $n$-th Touchard (or Bell) polynomial.
This can be written equivalently as
\begin{equation}
    \expval{\h{Z}^n}
    =
    \sum_{d=0}^{\infty}
    \frac{\lambda^d}{d!}
    (d\,e^{S_{\p}-S_0})^n
    \;.
\end{equation}
The normalized correlators are thus
\begin{align}
    \expval{\h{Z}^n}/\expval{\1}
    =
    e^{-\lambda} \sum_{d=0}^{\infty} \frac{\lambda^d  }{d!}
    (d\,e^{S_{\p}-S_0})^n
    =
    \sum_{d=0}^{\infty} p_d(\lambda) (d\,e^{S_{\p} - S_0})^n
    \;,\label{zn}
\end{align}
where $ p_d(\lambda) = e^{-\lambda}\frac{\lambda^d  }{d!} $ is the Poisson distribution with mean $\lambda$.

The gravitational path integral also gives a way to define the Hilbert space of quantum gravity.
The construction of the Hilbert space begins by picking the vacuum, often called the Hartle-Hawking state $\ket{\text{HH}}$, to be the empty set thought of as a 1d manifold.
This is the state of no boundaries.
The rest of the Hilbert space is constructed by application of the $\h{Z}$ operators which insert boundaries, $\h{Z}^n\ket{\text{HH}} = \ket{\h{Z}^n}$.
The gravitational path integral then provides the means to calculate inner products of any two such states, thereby defining the Hilbert space.

The authors of \cite{Marolf:2020xie} also consider the above model with the addition of so-called end-of-the-world (EofW) branes.
These are boundaries on which spacetime ends, but unlike the $\h{Z}$ boundaries we have discussed above, they are taken to be dynamical, in that the gravity path integral includes a sum over all configurations of such branes.
By constrast the $\h{Z}$ boundaries are fixed boundaries and can be considered observables of the gravity path integral.
With the presence of EofW branes, on which spacetime can end, we now have, in addition to the $\h{Z}$ fixed boundaries, the possibility of another type of fixed boundary.
This is an interval that is bounded on both sides by EofW branes.\footnotemark
\footnotetext{In this work, to match the language of open/closed TQFTs we will sometimes call these interval boundaries ``open sector" boundaries, and we call circle boundary components like $\h{Z}$ ``closed sector" boundaries.}
These intervals bounded by branes are not dynamical (only the EofW branes are taken to be dynamical).
The gravity path integral now includes manifolds whose boundary components are of three different types, fixed circles (inserted by operators $\h{Z}$), circles with EofW brane boundary conditions running completely around the circle, and circles made from alternating fixed and EofW brane intervals.
Given some number of fixed circle and interval boundaries, the gravity path integral will only include manifolds whose fixed boundaries match those given, but will include a sum over all the different possible ways of consistently configuring the EofW branes.

Like the circle boundary components inserted by $\h{Z}$, we can associate an operator with the inclusion of an additional fixed interval boundary, which we will call $\h{S}$.
The model of \cite{Marolf:2020xie} considers the possibility of having some number $K$ of ``flavors" of EofW branes, index by $a=1,\ldots,K$.
These different types of EofW brane differ only in their label $a=1,\ldots,K$, with the rule that only EofW branes with the same label can be connected together.
The fixed intervals $\h{S}$ now have their endpoints labeled by the flavor of EofW brane on which they end, giving operators $\h{S}_{ab}$.
Correlation functions of the gravity model include insertions of both $\h{Z}$ and $\h{S}_{ab}$ operators.
For example:
\begin{align}\label{ZandScorrelator}
    \bra{\text{HH}}
    \h{Z}^n
    \h{S_{ab}}\h{S_{ba}}
    \h{S_{cd}}\h{S_{de}}\h{S_{ec}}
    \ket{\text{HH}}
    \;.
\end{align}
The operators $\h{Z}$ and $\h{S}_{ab}$ all commute within these Euclidean path integral correlators.
Given a configuration of fixed boundaries, there are many ways we can partition them into ``future" and ``past" boundaries, and reinterpret the gravity path integral correlator as an inner product of states in the baby universe Hilbert space.
For example we can write \eqref{ZandScorrelator} variously as
$\braket{\h{Z}^n\h{S_{ab}}\h{S_{ba}}}{\h{S_{cd}}\h{S_{de}}\h{S_{ec}}}$,
$\braket{\h{Z}^{n-m}\h{S_{ab}}}{\h{Z}^m\h{S_{ba}}\h{S_{cd}}\h{S_{de}}\h{S_{ec}}}$,
$\braket{\h{Z}^n\h{S_{ab}}\h{S_{ba}}\h{S_{cd}}\h{S_{de}}}{\h{S_{ec}}}$,
etc.
The baby universe Hilbert space is spanned by states of the form
\begin{equation}
\ket{\h{Z}^n\prod_{a,b}\h{S}_{ab}^{n_{ab}}}
\equiv
\h{Z}^n\prod_{a,b}\h{S}_{ab}^{n_{ab}}\ket{\text{HH}}.
\end{equation}

From AdS/CFT we recall that the non-normalizable asymptotic modes of fields in AdS specify sources in the boundary CFT,
and a bulk path integral with given boundaries $\p M$ is dual to the appropriate CFT partition function on that boundary.
Inspired by this, we might expect that the boundary conditions for the 2d gravity path integral correspond to partition functions of a 1d theory.
Or, invoking the idea of \eqref{entry}, we might lower our expectations only slightly to include the possibility that boundary conditions in this case are dual to an average of partition functions in an ensemble of 1d theories.
Indeed, the authors of \cite{Marolf:2020xie} find a dual description of this sort for the gravity path integral \eqref{cf}.
From this point of view, the correlator $\expval{\h{Z}}$ is no longer a correlator, but the average value of a partition function $Z$ in a 1d topological theory.
Each additional boundary component $\h{Z}$ is another copy of this boundary partition function, so that the correlators $\expval{\h{Z}^n}$ with multiple fixed boundaries probe higher moments, $\expval{Z^n}$, in the ensemble distribution.

The only parameter in topological quantum mechanics is the dimension $d$ of the Hilbert space, so an ensemble of 1d topological theories is a probability distribution for $d$.
We immediately run into a problem, though.
For a single theory within the ensemble
\begin{align}
    Z=\tr_{\mathcal{H}}\1=d\;,\label{normaliz}
\end{align}
but \eqref{zn} suggests that $\h{Z}$ takes the value $de^{S_\p-S_0}$ for nonnegative integer $d$.
Expecting a dual boundary interpretation of $\h{Z}$ as a partition function in a 1d topological theory, thus forces us to set $S_{\p}=S_0$ in \eqref{zn}.

Alternatively (taking a perspective more in line with that taken in the rest of this paper) we can forgo adding the term $S_\p$ to the boundary and instead take the holographic map to be rescaled by some factor $e^{S_{\p}}$, so that
\begin{equation}\label{reviewrescaling}
    e^{S_\p}\h{Z}
    \leftrightarrow
    \tr(\1)=Z^{\text{boundary}}
    \;.
\end{equation}
Then a choice of rescaling given by $e^{S_\p}=e^{S_0}$ gives a sensible boundary dual, whose theories all have integer dimensional Hilbert spaces.
This perspective has the downside, of course, of making the choice of notation $\h{Z}$ for the boundary insertion operators something of a misnomer.

The boundary interpretation just introduced motivates a conceptually useful basis for our baby universe Hilbert space: the basis of eigenvectors of $\hat{Z}$,
\begin{align}
    \hat{Z}\ket{\alpha}=\alpha\ket{\alpha}\;.
\end{align}
The eigenstates $\ket{\alpha}$, called alpha-states, are of course orthogonal: 
$\braket{\alpha'}{\alpha}\sim\delta_{\alpha',\alpha}$.
But they have one very special property.
The boundary theories in our ensemble are characterized by the values they give to the observables $\h{Z}$, so the set of alpha states is precisely the sample space of boundary theories in our ensemble.
The probability distribution over the theories in our ensemble can be extracted from the overlap between a given alpha-state and the Hartle-Hawking state:
\begin{equation}
    p(\alpha)
    =
    \frac{\braket{\text{HH}}{\alpha}\braket{\alpha}{\text{HH}}}{\braket{\alpha}{\alpha}\braket{\text{HH}}{\text{HH}}}.
\end{equation}

Whereas obtaining a sensible boundary interpretation for a theory without end-of-the-world branes necessitated only a judicious choice of rescaling \eqref{reviewrescaling} for the $\h{Z}$ operators, a potentially more serious problem manifests when we include end-of-the-world branes and their attendant $\h{S}_{ab}$ operators.
In the 1d dual theory, the interval boundary insertions $\h{S}_{ab}$ have a natural interpretation as inner products of states induced by boundary conditions $a$ and $b$ at the endpoints of the interval.
So within a particular boundary theory in the ensemble, the operators $\h{S}_{ab}$ should take as values the components of a $K$ by $K$ positive semidefinite Hermitian matrix $M$.
A boundary ensemble will be given by a joint probability distribution over the dimension $d$ and the matrix $M$.
Viewing correlators of $\h{Z}$ and $\h{S}_{ab}$ operators as averages in such an ensemble, they are moments of the ensemble probability distribution.
In particular, the generating function for normalized correlators, which \cite{Marolf:2020xie} calculate to be
\begin{align}\label{reviewgenfunc}
    \expval{e^{iu\h{Z}+\sum_{a,b}^K it_{ab}\h{S}_{ab}}}
    /
    \expval{\1}
    =
    \exp(
        \lambda
        \left(\frac{e^{iu}}{\det(\1-it)}\right)^{e^{S_\p-S_0}}
        -
        \lambda
    )
    \;,
\end{align}
should be the inverse Fourier transform of the probability density $p(d,M)$ defining the ensemble.
As we have described above, the $\h{Z}$ operators take values $de^{S_\p-S_0}$ with probability $p_\lambda(d)=e^{-\lambda}\lambda^d/d!$, giving an expansion of \eqref{reviewgenfunc} as
\begin{align}
    \expval{e^{iu\h{Z}+\sum_{a,b}^K it_{ab}\h{S}_{ab}}}
    /
    \expval{\1}
    =
    \sum_{d=0}^\infty
    p_\lambda(d)e^{iude^{S_\p-S_0}}
    \expval{
        e^{
            \sum_{a,b}^K it_{ab}\h{S}_{ab}
        }
    }_{Z=d\,e^{S_\p-S_0}}
    \;.
\end{align}
For a given $d$, the residual probability distribution $p_d(M)$ over the matrices $M$ will then be the Fourier transform of the generating function
\begin{equation}\label{reviewdistribution}
    \expval{
        e^{
            \sum_{a,b}^K it_{ab}\h{S}_{ab}
        }
    }_{Z=d\,e^{S_\p-S_0}}
    =
    \det(\1-it)^{-de^{S_\p-S_0}}
    \;.
\end{equation}
Unfortunately, only for certain values does the above generating function have an inverse Fourier transform that can be interpreted as a valid probability distribution \cite{graczyk2003}.\footnotemark
\footnotetext{In which case, it is known as the Wishart distribution.}
Specifically, the exponent $de^{S_\p-S_0}$ must lie in the set $\{0,1,2,\ldots,K-1\}\cup [K-1,\infty)$, where, as a reminder, $K$ is the number of flavors of end-of-the-world brane included in the theory.
As $d$ runs over all nonnegative integers, the factor $e^{S_\p-S_0}$ must lie in the set $\{0,1,2,\ldots,K-1\}\cup [K-1,\infty)$.
A natural choice is to take $S_\p=S_0$.

The above argument highlights an important point.
It need not be the case that a theory without factorization has a description as an ensemble.
As in the situation where $S_\p=0$, the correlation functions of a non-factorizing theory are not necessarily the moments of a probability distribution.
Failure to have an ensemble description is linked with the presence of negative-norm states in the baby universe Hilbert space.
To see this, consider a gravity theory with boundary insertion operators $\h{Z}_i$.
The theories in the ensemble will be parametrized by values these operators $\h{Z}_i$ take.
Assume for simplicity, these values are continuous, real, and independent.
Then we can formally construct the alpha-states as
\begin{equation}\label{alpha}
    \ket{\alpha}
    =
    \int\!\prod_i\left(\frac{du_i}{2\pi}\right)
    e^{i\sum_i u_i \left(\h{Z}_i-\alpha_i\right)}
    \ket{\text{HH}}
    \;,
\end{equation}
where $\alpha_i$ are the values which $Z_i$ takes in the theory described by $\ket{\alpha}$.
From this, the inner product of two alpha states is
\begin{equation}
    \begin{aligned}
        \braket{\alpha'}{\alpha}
        &=
        \int\!
        \prod_i\left(\frac{du'_i}{2\pi}\frac{du_i}{2\pi}\right)
        e^{-i\sum_iu'_i(\alpha_i-\alpha'_i)}
        \expval{e^{i\sum_iu_i(\h{Z}_i-\alpha_i)}}\\
        &=
        \delta(\alpha'-\alpha)
        \int\!
        \prod_i\left(\frac{du_i}{2\pi}\right)
        e^{-i\sum_iu_i\alpha_i}
        \expval{e^{i\sum_iu_i\h{Z}_i}}\\
        &=
        \delta(\alpha'-\alpha)\expval{\1}p(\alpha)\;.
        \label{norm}
    \end{aligned}
\end{equation}
The last equality comes from viewing the correlators of $\h{Z}_i$ as moments of a putative probability distribution $p(\alpha)$.
Viewed thus, the generating function $\expval{e^{i\sum_iu_i\h{Z}_i}}$ is simply the inverse Fourier transform of $p(\alpha)$.
This equation \eqref{norm} suggests something about theories that fail to be ensembles.
If the distribution $p(\alpha)$ takes negative values, this means both that the theory does not have an ensemble description and that the baby universe Hilbert space contains negative-norm states.

Returning again to the gravity model with end of the world branes,
one could complain that in order to cure the boundary interpretation we have ruined the locality of the bulk TQFT theory.
Indeed, the action for $S_\p\neq 0$ no longer depends only on the Euler characteristic of the manifold, so it is no longer consistent with cuttings and gluings of the spacetime.
Alternatively, if we insist on locality, the $S_\p$ term can be interpreted as the contribution of local degrees of freedom associated to the boundaries, both brane and fixed.
In that case, however, a question arises of whether or not we should consider additional $\h{S}$ operators corresponding to these additional degrees of freedom.
Doing so would be equivalent to considering the theory with $S_\p=0$ just with more flavors of end-of-the-world brane, so the problem would arise again.
The problem seems to require that the degrees of freedom propagate, unprobed, along the boundaries.
We refer the reader to the discussion of this boundary term and its meaning in \cite{Marolf:2020xie}.
We address the problem as it shows up in our case in section \ref{reflectionpos}, where we offer a somewhat different description for these degrees of freedom, and some speculation on their meaning.

The topological action of the model described in this section is, in fact, that of a 2d TQFT with a one-dimensional (closed sector) Hilbert space.
In some sense it describes the simplest possible 2d TQFT.
In what follows, we will analyze gravity models built from more complicated 2d TQFTs.
We will find that much of this analysis can be reduced to that of the simpler one-dimensional TQFT.

\section{2d Dijkgraaf-Witten theory} \label{BU}
The simple model of a gravity path integral from \cite{Marolf:2020xie} can be generalized to include any topological action in the bulk.
As a first example, we will examine 2d Dijkgraaf-Witten theory, a topological gauge theory, as our bulk theory and construct a gravity path integral from that.
We will find a dual interpretation of the gravity path integral in terms of a one-dimensional ensemble theory whose Hilbert space is a random representation of the gauge group.
We also find that the correlation functions of boundary insertion operators factorize between the irreducible representations of the gauge group, similar to the results in \cite{Kapec:2019ecr}.
In section \ref{generalcase} we will see that analogous features hold in the case of more general 2d TQFTs.
Before describing the gravity path integral, however, we will first briefly review Dijkgraaf-Witten theory and present the results of Dijkgraaf-Witten theory in two-dimensions that will be relevant to our construction.

\subsection{Review of Dijkgraaf-Witten theory}
Dijkgraaf-Witten theory is a topological gauge theory with finite symmetry group $G$ \cite{Dijkgraaf:1989pz, Freed:1991bn}.
The path integral is given as a sum over $G$-principal bundles on the spacetime manifold.
Given a connected, manifold without boundary $M$, let $\CC_M$ be the set of $G$-principal bundles on $M$.
A $G$-principal bundle on $M$ can be identified with a homomorphism from the fundamental group $\pi_1(M,x)$ to the group $G$, where $x$ is some chosen basepoint in $M$.
So we will take $\CC_M$ to be the set $\Hom(\pi_1(M,x),G)$.
Identified this way, some of the principal bundles can be related to each other via residual gauge symmetries.
Specifically, a gauge transformation $g\in G$ with support over all of $M$ will act on a bundle $\phi:\pi_1(M,x)\rightarrow G$ via conjugation, like so: $\phi(\cdot)\mapsto g\,\phi(\cdot)\,g^{-1}$.
The gauge invariant path integral must take these gauge symmetries into account, and is thus over the space of $G$-principal bundles $\CC_M$, orbifolded by this action of $G$.
We'll denote this orbifolded space by $\b{\CC_M}$.

The measure over $\b{\CC_M}$ induced by this orbifolding will weight bundles inversely to the size of their stabilizer subgroup under the action of $G$.
Without any further weighting of the bundles beyond this, the sum over $\b{\CC_M}$ gives the ``untwisted" version of Dijkgraaf-Witten theory.
In that case the partition function for $M$ is \cite{Freed:1991bn}
\begin{equation}
    Z^{\text{DW}}[M]
    =
    \sum_{\phi\in\b{\CC_M}}\frac{1}{\left|\operatorname{Stab}(\phi)\right|}
    =
    \frac{\left|\CC_M\right|}{\left|G\right|}
    \;.
\end{equation}
The numerator $\left|\CC_M\right|$ is $\left|\Hom(\pi_1(M,x),G)\right|$, the number of homomorphisms from $\pi_1(M,x)$ to $G$.
When $M_g$ is the connected, closed, oriented surface of genus $g$, this count is given by a result known as Mednykh's formula:
\begin{equation}
	\left|\CC_{M_g}\right|
	=
	\left|G\right|
	\sum_q\left(\frac{d_q}{\left|G\right|}\right)^{2-2g}
	\;,
\end{equation}
where $q$ labels the irreducible representations of $G$, and $d_q$ are the dimensions of each irreducible representation.
(See \cite{jones_1998} and references therein.)
This gives the Dijkgraaf-Witten partition function of $M_g$ as
\begin{equation}
    Z^{\text{DW}}[M_g]
    =
    \frac{\left|\CC_{M_g}\right|}{\left|G\right|}
    =
    \sum_q\left(\frac{d_q}{\left|G\right|}\right)^{2-2g}
    \;.
\end{equation}

The above partition function is for a closed surface.
If our manifold is a surface with boundaries, the path integral is a sum over $G$-principal bundles that satisfy given boundary conditions.
Specifically each boundary component is a circle with boundary condition given by a conjugacy class of $G$, representing the holonomy around that circle.
Let $\CC_{M_{g,n}}(k_1,\ldots,k_n)$ denote the set of $G$-principal bundles on the genus $g$ surface with $n$ boundary components having holonomy boundary conditions $k_1$, \ldots, $k_n$ respectively.
The path integral is again over the orbifolded space $\b{\CC_{M_{g,n}}(k_1,\ldots,k_n)}$, and comes out to $\operatorname{vol}(\b{\CC_{M_{g,n}}(k_1,\ldots,k_n)})=\left|\CC_{M_{g,n}}(k_1,\ldots,k_n)\right|/\left|G\right|$.

A $G$-principal bundle on a surface $M_{g,n}$ with boundaries is still a choice of homomorphism from $\pi_1(M_{g,n},x)$ to $G$, but now with the restriction that it is compatible with the boundary conditions on $M_{g,n}$ in the following sense: if a path $a_i\in\pi_1(M_{g,n},x)$ is homologous to the $i$-th boundary, the bundle $\phi:\pi_1(M_{g,n},x)\rightarrow G$ must map $a_i$ to an element of $k_i$.
(Note this notion of compatibility is well-defined, as all paths in $\pi_1$ homologous $a_i$ will be conjugates of each other and hence must map to the same conjugacy class in $G$.)
The partition function will be given by the count $\left|\CC_{M_{g,n}}(k_1,\ldots,k_n)\right|$ of such compatible homomorphisms $\phi$.
Mednykh's formula can be generalized to the case of a connected surface with boundaries as
\begin{equation}
    \left|\CC_{M_{g,n}}(k_1,\ldots,k_n)\right|
    =
    \left|G\right|
    \sum_q
    \left(\frac{d_q}{\left|G\right|}\right)^{2-2g-n}
    \prod_i
    \frac{\left|k_i\right|}{\left|G\right|}
    \chi_q(k_i),
\end{equation}
where $g$ is the genus of the surface, $k_1$, \ldots, $k_n$ are the respective boundary conditions of the $n$ boundaries, and $\chi_q(k)$ is the irreducible character $q$ evaluated on an element in $k$.
(See Proposition 1 in \cite{mednykh_nonequivalent_1984}.
This can also be obtained by first obtaining the result for an $(n+2g)$-holed sphere $M_{0,n+2g}$ with given holonomies on the boundaries.
This is done by counting maps from $\pi_1(M_{0,n+2g})$, the free group on $n+2g$ generators, to $G$ that satisfy the boundary constraints.
Then one can glue $2g$ of the boundaries together in pairs by summing over boundary conditions.)
The path integral on $M_{g,n}$ with boundary conditions $k_1$, \ldots, $k_n$ is thus
\begin{equation}\label{kDWpartfunc}
    Z^{DW}[M_{g,n}\,;\,k_1,\ldots,k_n]
    =
    \sum_q
    \left(\frac{d_q}{\left|G\right|}\right)^{2-2g-n}
    \prod_i
    \frac{\left|k_i\right|}{\left|G\right|}
    \chi_q(k_i).
\end{equation}
We can interpret this path integral on a manifold with boundaries as defining a multilinear map, from the Hilbert space of $n$ circles to $\C$, that takes the state $\ket{k_1}\otimes\cdots\otimes\ket{k_n}$ to the complex number $Z^{{DW}}[M_{g,n}\,;\,k_1,\ldots,k_n]$.

Note that the space of states on a circle is evidently spanned by states labeled by conjugacy classes.
We can take the path integral $Z^{\text{DW}}[M_{0,2}\,;\,k_1,k_2]$ as defining a bilinear pairing on this Hilbert space.
We get
\begin{align}
    \big(\ket{k_1},\ket{k_2}\big)
    =
    Z^{\text{DW}}[M_{0,2}\,;\,k_1,k_2]
    =
    \sum_q
    \frac{\left|k_1\right|}{\left|G\right|}
    \chi_q(k_1)
    \frac{\left|k_2\right|}{\left|G\right|}
    \chi_q(k_2)
    =
    \frac{\left|k_1\right|}{\left|G\right|}
    \delta_{k_1,k_2^{-1}}
    \;.\label{pairc}
\end{align}
Under this pairing the states $\ket{k}$ are evidently linearly independent, but they are not orthogonal.
We can switch to a diagonal basis, $\ket{q}\equiv\sum_{k}\chi_q(k^{-1})\ket{k}$, where $\chi_q(k^{-1})$ is the character for irreducible representation $q$ evaluated at an element whose inverse is in the conjugacy class $k$.
In this basis labeled by irreps of $G$, the pairing above is simply $\big(\ket{q_1},\ket{q_2}\big)=\delta_{q_1,q_2}$, so the $\ket{q}$ are orthogonal.\footnotemark
\footnotetext{If we take the $\ket{q}$ basis as a real basis, then complex conjugation induces the antiunitary map $\ket{k}\mapsto\ket{k^{-1}}$ with the interpretation of a reflection. This antiunitary composed with the pairing described above defines an inner product on the Hilbert space, under which the states $\ket{k}$ are orthogonal and the states $\ket{q}$ are orthonormal.}
The partition function for a connected, genus $g$ surface with boundaries labeled (in this irrep basis) by $q_1$, \ldots, $q_n$ is simply
\begin{equation}\label{qDWpartfunc}
    Z^{\text{DW}}[M_{g,n}\,;\,q_1,\ldots,q_n]
    =
    \sum_q
    \left(\frac{d_q}{\left|G\right|}\right)^{2-2g-n}
    \delta_{qq_1\cdots q_n}.
\end{equation}
Note that this partition function for a connected manifold with boundaries evaluates to zero, unless all boundaries are labeled by the same irreducible representation.

In addition to ``untwisted" Dijkgraaf-Witten described above, one can also define a ``twisted" generalization of Dijkgraaf-Witten by further weighting each $G$-principal bundle in the path integral by a $U(1)$-valued characteristic class $\alpha$ of the bundle.
This is equivalent to adding a term $iS_\alpha$ to the action satisfying $\alpha[\phi]=e^{iS_\alpha[\phi]}$.
So the twisted partition function for a closed, connected manifold $M$ is
\begin{equation}\label{twistedpartfunc}
    Z^{\text{DW},\alpha}[M]
    =
    \sum_{\phi\in\b{\CC_M}}\frac{1}{\left|\operatorname{Stab}(\phi)\right|}\alpha[\phi]
    =
    \sum_{\phi\in\b{\CC_M}}\frac{1}{\left|\operatorname{Stab}(\phi)\right|}e^{iS_\alpha[\phi]}.
\end{equation}
In what follows, we will consider the untwisted case before discussing in section \ref{twistedDW} how the results are altered in the twisted case.

\subsection{A Dijkgraaf-Witten gravity path integral}\label{DWGPI}
In preparation for defining a gravity path integral, we will include, in addition to the Dijkgraaf-Witten action, the topological action term $S_0\chi(M)$ of the simple theory described in \cite{Marolf:2020xie}.
This will have the effect of suppressing higher genus manifolds in an eventual sum over topology.
With this addition to the action, the bulk theory partition function is given by
\begin{equation}\label{qDWMMpartfunc}
    Z^{\text{bulk}}[M_{g,n}\,;\,q_1,\ldots,q_n]
    =
    \sum_q
    \left(e^{S_0}\frac{d_q}{\left|G\right|}\right)^{2-2g-n}
    \delta_{qq_1\cdots q_n}.
\end{equation}
In the model of \cite{Marolf:2020xie} the action includes an additional, nonlocal term $n S_\p$ proportional to the number of boundaries.
This can be regarded as either the contribution to the action of some additional degrees of freedom living on the boundary or as a rescaling of boundary insertion operators in the gravity path integral.
We won't include this term here and will discuss its meaning and inclusion in section \ref{DWboundary}.
For now, we take our partition function to be that described above, which is the partition function of a TQFT; in other words, it is local, in the sense of being compatible with cutting and gluing.

A gravity path integral defined from \eqref{qDWMMpartfunc} will take as input a boundary manifold (so, for a 2d bulk, some number of circles) with specified boundary conditions, and will output the partition function \eqref{qDWMMpartfunc} summed over all manifolds with the given boundary conditions.
Following the notation of \cite{Marolf:2020xie}, we denote the inclusion of a circle with boundary condition $k$ by the operator $\h{Z}[k]$.
The gravity path integral is
\begin{equation}
    \expval{\h{Z}[k_1]\cdots\h{Z}[k_n]}
    =
    \sum_{
        \substack{
            \text{$M$ s.t. $\p M$}\\
            \text{is $n$ circles}
        }
    }
    \mu(M)
    Z^{\text{bulk}}[M\,;\,k_1,\ldots,k_n].
\end{equation}
Where $\mu(M)$ is the appropriate measure, with a factor of $1/m!$ whenever $M$ has $m$ identical closed components.

The connected contribution to the vacuum correlator, $\lambda\equiv\expval{\1}_\text{conn.}=\log\expval{\1}$, is a sum over connected surfaces with no boundary, so in effect a sum over genus:
\begin{equation}
    \lambda
    =
    \sum_g
    \sum_q
    \left(e^{S_0}\frac{d_q}{\left|G\right|}\right)^{2-2g}
    =
    \sum_q
    \frac{
        \left(\frac{e^{S_0}d_q}{\left|G\right|}\right)^2
    }{
        1-\left(\frac{e^{S_0}d_q}{\left|G\right|}\right)^{-2}
    }
    =
    \sum_q
    \lambda_q.  \label{l}
\end{equation}
Here we've denoted $\sum_g\left(e^{S_0}d_q/\left|G\right|\right)^{2-2g}$ by $\lambda_q$.
The full vacuum correlator is correspondingly $\expval{\1}=e^\lambda=\prod_qe^{\lambda_q}$.

Calculating the correlators of boundary insertion operators will be easier in the basis labeled by irreducible representations.
To that end, define operators $\h{Z}_q\equiv \sum_k\chi_q(k^{-1})\h{Z}[k]$, corresponding to the TQFT states $\ket{q}$.
We can define a generating function for the general correlator, $F(u_q)=\expval{e^{\sum_qu_q\h{Z}_q}}$
with chemical potentials $u_q$ for the insertion of each operator $\h{Z}_q$.
The logarithm of this generating function will simply be the corresponding generating function for connected correlators
\begin{equation}
    \log F(u_q)
    =
    \sum_{\cdots, n_q,\cdots}
    \prod_q\frac{1}{n_q!}u_q^{n_q}
    \expval{\prod_q\h{Z}_q^{n_q}}_{\text{conn.}}.
\end{equation}
Connected correlators are simple to calculate as the surfaces to be summed over in the corresponding gravity path integral are parametrized by genus.
For example,
\begin{equation}
    \begin{aligned}
        \expval{\h{Z}_{q_1}\cdots\h{Z}_{q_n}}_{\text{conn.}}
        &=
        \sum_g
        Z^{\text{bulk}}[M_{g,n}\,;\,q_1,\ldots,q_n]\\
        &=
        \sum_g
        \sum_q
        \left(e^{S_0}\frac{d_q}{\left|G\right|}\right)^{2-2g-n}
        \delta_{qq_1\cdots q_n} \label{conv}\\
        &=
        \lambda_{q_1}
        \left(e^{S_0}\frac{d_{q_1}}{\left|G\right|}\right)^{-n}
        \delta_{q_1\cdots q_n}.
    \end{aligned}
\end{equation}
Note that this connected correlator evaluates to zero unless all boundaries are labeled by the same irreducible representation.
This fact simplifies the resulting expression for $\log F(u_q)$.
The correlator with $n_q$ boundaries for each type $q$ will be zero unless no more than one of the $n_q$ is nonzero.
So the sum over all possible numbers of boundary $n_q$ for each $q$ reduces to a sum over just one of the $n_q$, followed by a sum over $q$.
We obtain
\begin{equation}
    \log F(u_q)
    =
    \sum_q
    \sum_{n_q}
    \frac{1}{n_q!}u_q^{n_q}
    \lambda_{q}
    \left(e^{S_0}\frac{d_{q}}{\left|G\right|}\right)^{-n_q}
    =
    \sum_q
    \lambda_q
    \exp(
        u_q
        \frac{\left|G\right|}{e^{S_0}d_{q}}
    ).
\end{equation}
The final result of this simplification is that the full generating function $F(u_q)$ factorizes between the different labels $q$:
\begin{equation}\label{qDWgenfunc}
    F(\ldots,u_q,\ldots)
    =
    \prod_q
    e^{\lambda_q\exp(u_q\left|G\right|/e^{S_0}d_{q})}.
\end{equation}

From \eqref{qDWgenfunc} the full correlators can be extracted.
They are
\begin{equation}\label{qDWcorrelator}
    \expval{\prod_q\h{Z}_q^{n_q}}
    =
    \prod_q
    e^{\lambda_q}
    B_{n_q}\!(\lambda_q)
    \left(
    \frac{\left|G\right|}{e^{S_0}d_q}
    \right)^{n_q},
\end{equation}
where $B_m$ denotes the $m$-th Touchard, or Bell, polynomial.
Note the normalized correlation functions have the property of factorizing between boundaries labeled by different irreducible representations: 
\begin{equation}
\expval{\prod_q\h{Z}_q^{n_q}}/\expval{\1}
=
\prod_q\left(\expval{\h{Z}_q^{n_q}}/\expval{\1}\right),
\end{equation}
whereas no such factorization holds between boundaries label by the same irrep, e.g.\ $\expval{\h{Z}_q^{n+m}}\nsim\expval{\h{Z}_q^n}\expval{\h{Z}_q^m}$.
The correlators for the operators $\h{Z}[k]$ can be gotten through the change of basis back from the $\h{Z}_q$ operators to the $\h{Z}[k]$ operators,
namely $\h{Z}[k]=\sum_q\frac{\left|k\right|}{\left|G\right|}\chi_q(k)\h{Z}_q$.

\subsection{Boundary interpretation}\label{DWboundary}
The above result \eqref{qDWcorrelator} is analogous to having multiple copies of the model presented in \cite{Marolf:2020xie}. In fact, each factor labeled by $q$ is equivalent to one copy of the model of \cite{Marolf:2020xie}, where $e^{S_0} \rightarrow e^{S_0} \frac{d_q}{|G|}$.
We will see that our gravity path integral with a Dijkgraaf-Witten bulk likewise has a dual interpretation as a random theory living on the boundary.
This will in fact be equally true of any 2d TQFT satisfying Atiyah's axioms, as we will demonstrate in section \ref{generalcase}.
We present Dijkgraaf-Witten theory here as a representative example.

In a standard holographic boundary interpretation the $\h{Z}$ operators would get reinterpreted as the partition functions of a one-dimensional theory.
In our case, however, this is impossible because the correlators of the $\h{Z}$ operators do not factorize.
Instead we will look for an ensemble of one-dimensional theories and interpret the correlator $\expval{\h{Z}[k_1]\cdots\h{Z}[k_n] }/\expval{\1}$ as the average of the quantity $Z[k_1]\cdots Z[k_n]$.
The boundary theories in our ensemble are characterized by the value they assign to each operator $\h{Z}[k]$,
so the ensemble will be a probability distribution over the space $\C^r$, where $r$ is the number of irreducible representations of $G$ and where each copy of $\C$ represents the values that one of the $\h{Z}[k]$ can take.
This makes the correlators $\expval{\h{Z}[k_1]\cdots\h{Z}[k_n]}/\expval{\1}$ interpretable as moments of the ensemble probability distribution.
The problem of finding the ensemble probability distribution from the correlators \eqref{qDWcorrelator} is thus an instance of the so-called moment problem, wherein one attempts to find a probability distribution from its moments.

If we let $\vec{\alpha}\in\C^r$ index the theories in our ensemble, the probability distribution $p_\alpha$ over the theories should satisfy $\expval{\h{Z}_{q_1}\cdots\h{Z}_{q_n}}/\expval{\1}=\int\!d^r\!\alpha\,p_{\alpha}\alpha_1\cdots\alpha_r$.
In terms of the generating function $F(u_1,\ldots,u_r)$ for the correlators,
\begin{equation}
    F(u_1,\ldots,u_r)/\expval{\1}
    =
    e^{\sum_q\lambda_q\left(\exp(u_q\left|G\right|/e^{S_0}d_{q})-1\right)}
    =
    \int\!d^r\!\alpha\,p_{\alpha}
    e^{\sum_qu_q\alpha_q}.
\end{equation}
We can extract the function $p_\alpha$ by performing a Fourier transform with respect to the variables $iu_q$.
The result is
\begin{equation}
    p_\alpha=
    \prod_q
    \sum_{N_q=0}^\infty
    \frac{\lambda_q^{N_q}}{{N_q}!e^{\lambda_q}}
    \delta\left(\alpha_q-\frac{\left|G\right|}{e^{S_0}d_q}{N_q}\right)
    \;.\label{distr}
\end{equation}
In other words, each $\alpha_q$ takes the values $\frac{\left|G\right|}{e^{S_0}d_q}N_q$ where the $N_q$ are random integers drawn independently from Poisson distributions with respective means $\lambda_q$.
Recall that the $\alpha_q$ are the values of $Z_q$ in the different theories in our ensemble, so
\begin{equation}
    Z_q
    =
    \frac{\left|G\right|}{e^{S_0}d_q}
    N_q
    \;.
\end{equation}
Switching from the $Z_q$ to the $Z[k]$ basis gives
\begin{equation}\label{kbasispartfunc}
    Z[k]
    =
    \sum_q
    \frac{\left|k\right|}{e^{S_0}d_q}
    \chi_q(k)
    N_q
    \;.
\end{equation}

Our bulk theory is topological, so it's boundary dual will likewise be topological.
As the $\h{Z}[k]$ operators represent the insertion of a boundary with holonomy $k$, we are tempted to interpret $Z[k]$ as a partition function in a one-dimensional topological quantum mechanics theory, with an insertion of a $G$-symmetry operator with conjugacy class $k$.
That is to say, $Z[k]=\tr(U(g))$, where $U(g)$ is the representation on the Hilbert space of a group element $g\in k$.
In fact, this is only nearly so.
Looking at \eqref{kbasispartfunc} we see that $Z[k]$ has the form of a trace of an element of $k$ in a representation that has $\frac{\left|k\right|}{e^{S_0}d_q}N_q$ copies of the representation $q$, for each $q$.
Unfortunately for this interpretation, $\frac{\left|k\right|}{e^{S_0}d_q}N_q$ is not necessarily an integer, which it would have to be to avoid the absurdity of a theory with a fractional number of copies of a representation.
One immediate fix would appear to be picking a specific value for $S_0$ such that $\frac{\left|G\right|}{e^{S_0}d_q}\in\Z$.
This is not possible though, as it would ruin the convergence of eq.\ \eqref{conv} and, what's worse, would render $\lambda_q$ negative, giving negative probabilities in our ensemble distribution.

On the other hand, going back to the $Z_q$ operators, in light of their definition $Z_q=\sum_k\chi_q(k^{-1})Z[k]$ a natural interpretation would be for $Z_q$ to be the contribution to the partition function of states with charge $q$.
That is to say, $Z_q=\tr(P_q)$, where $P_q$ is a projection onto states living in copies of irreducible representation $q$.
Again, this is nearly so, but unfortunately $Z_q=\frac{\left|G\right|}{e^{S_0}d_q}N_q$ is not an integer for all $N_q$.
One possible solution is to identify the size of the $q$-sector, $\tr(P_q)$, with a rescaled operator $e^{S_q}Z_q$ rather than with $Z_q$, where we choose $e^{S_q}$ so that $e^{S_q}Z_q$ is an integer in every $\alpha$-state.
(We will discuss a possible motivation for this rescaling in section \ref{reflectionpos}.)
For now, we can interpret this rescaling as a modification of the expected holographic map $Z_q\leftrightarrow \tr{P_q}$ to be
$e^{S_q}Z_q=\tr(P_q)$.
Similarly, we can to identify $\tr(U(g))$, with an appropriately rescaled operator $e^{S_k}Z[k]$, rather than with $Z[k]$ as is, in order to avoid the situation of having a fractional number of copies of a representation.
A choice of rescalings for the $\h{Z}_q$ and the $\h{Z}[k]$ that avoids noninteger dimensions, that avoids noninteger copies of irreducible representations, and that respects the identity $\tr(U(g))=\sum_q\tr(P_q)\chi_q(g)/d_q$ is
\begin{align}
     S_k &= S_0 + \log(\left|G\right|/\left|k\right|)  \\
     S_q&=S_0 + \log d_q\;.  \label{Sq}
\end{align}
These result in a consistent interpretation of the gravity model as a boundary theory with $\frac{|G|}{d_q}N_q\in\Z$ copies of the irreducible representation $q$.\footnotemark
\footnotetext{
This is an integer by the basic result from representation theory that $|G|/d_q\in\Z$ for any irreducible representation.}

\subsection{Twisted Dijkgraaf-Witten}\label{twistedDW}
We now turn our attention to twisted Dijkgraaf-Witten theories. After a review of the essential background, we use twisted Dijkgraaf-Witten in our gravity models, and describe their boundary interpretation.

\subsubsection{Background}  \label{sec_backgr}
The purpose of this subsection is to describe a practical way to compute the partition function of ``twisted" 2d Dijkgraaf-Witten.
Here we follow the exposition in \cite{Kapec:2019ecr}, and refer the reader there for details.
See also \cite{Turaev_2007}.
We will see that the partition functions of 2d twisted Dijkgraaf-Witten are given as a sum similar to \eqref{kDWpartfunc} but over projective representations of $G$.

As was briefly mentioned above, the different twisted theories are labeled by characteristic classes of $G$-principal bundles.
The different characteristic classes that describe the possible twisted action terms are classified by elements of $H^2(BG,U(1))$, the second $U(1)$-valued cohomology classes of the classifying space $BG$ \cite{Dijkgraaf:1989pz, Freed:1991bn}.
Specifically, given a class $\alpha$ in $H^2(BG,U(1))$, and viewing a $G$-principal bundle on a manifold $M$ as a map $\phi$ from $M$ to $BG$, consider the pullback of $\phi^*\alpha$ along this map.
Evaluating $\phi^*\alpha$ on $M$ gives a phase $\phi^*\alpha(M)=\alpha(\phi(M))\in U(1)$ for the bundle $\phi$.
The partition function on $M$ is a sum over inequivalent $G$-principal bundles $\phi$ weighted by these phases.
See equation \eqref{twistedpartfunc}.
For simplicity, in what follows we will only consider the particular case where $H^2(BG,U(1))=\Z_N$.

One simplification afforded by restricting to the case $H^2(BG,U(1))=\Z_N$ is that all the twisted Dijkgraaf-Witten actions can be described in terms of any characteristic class $\alpha_0\in H^2(BG,U(1))$ which generates the others, so $H^2(BG,U(1))=\Z_N = \{\alpha_0^k : ~ k=0,...,N-1\}$.
The first part of this section is devoted to the construction of such a $\alpha_0$. Then, after eq.\ \eqref{om} we consider $\alpha_0^k$ with general $k$ and show how the partition functions for general $k$ become sums over projective representations of $G$.

We will relate $\alpha_0$ to the failure of the bundle $\phi$ to lift to a $\t{G}$-principal bundle when $\t{G}$ is a central extension of $G$ by $Z_N$.
% Specifically, the failure of $\phi:M\rightarrow BG$ to uplift to a $\t{G}$-principal bundle will be described by an element of $Z_N$ and we take $\alpha_0$ to be the map that assign this element of $Z_N$ to the 2-chain $\phi(M)\subset BG$.
% 
% Now let's explain how the Dijkgraaf-Witten action weighs differently the isomorphism classes of $G$-bundles.
% Having chosen $G$ such that $H^2(BG,U(1))=\Z_N$, we can weigh each $G$-bundle by its obstruction to being lifted to a $\t{G}_S$-bundle.
Consider a $G$-bundle on a surface $M$ described in terms of transition functions $g_{ij}$ which satisfy the triple overlap condition
\begin{align}
    g_{ij}g_{jk}g_{ki}=1\;.
\end{align}
Picking a lift of each transition function $g_{ij} \mapsto \t{g}_{ij}$, we get
\begin{align}
    \t{g}_{ij}\t{g}_{jk}\t{g}_{ki}
    =
    \t{c}_{ijk}
    \;\in\;
    \Z_N
    \;,
\end{align}
which for nontrivial $c_{ijk}$ is a violation of the $\t{G}_S$-bundle cocycle condition.
Gauge transformations and changes of lift can change $\t{c}_{ijk}$ locally but in general there is a global obstruction to removing all such violations.
The assignment of an element of $\Z_N$ to each triple intersection, modulo gauge transformations and changes of lift, defines a 2-cocycle
\begin{align}\label{om}
    [\omega_N]\in H^2(M,\Z_N)\;,
\end{align}
which can be paired with the 2-cycle $[M]$ to give an element $\omega=\expval{[\omega_N],[M]}\in\Z_N$.
Note that the element $\omega\in\Z_N$ depends on the bundle $\phi$.
The characteristic class $\alpha_0\in H^2(BG,U(1))$ is then defined via $\alpha_0(\phi(M)) = e^{2 \pi i \omega(\phi) / N}$.
In fact, these are the phases by which the Dijkgraaf-Witten theory described by $\alpha_0$ weights each bundle.

% From $\alpha_0$ we can get additional twisted Dijkgraaf-Witten theories $\alpha_0^k$ that weight each bundle $\phi$ by $e^{2 \pi i k \omega(\phi) / N}$.
% % Thus a general Dijkgraaf-Witten action $\alpha_0^k$ is one that weights each bundle $\phi$ by $e^{2 \pi i k \omega(\phi)}$.
% This last expression shows how a Dijkgraaf-Witten action assigns an $N$-th root of unity to each isomorphism class of bundles and how this assignment can be picked in $N$ different ways (indicated by $k$).

The characteristic class $\alpha_0$ obtained this way depends on the choice of the central extension $\t{G}$.
We would like to choose $\t{G}$ so that $\alpha_0$ is a generating element of $H^2(BG,U(1))=\Z_N$ so that any Dijkgraaf-Witten action can be obtained as $\alpha_0^k$ for some $k$.
We are also interested in $\t{G}$ with the property that all irreducible representations of $G$, projective or linear, can be lifted to linear representations of $\t{G}$.
There is always a central extension of $G$ that satisfies these criteria, namely the Schur covering group of $G$.
The Schur covering group $\t{G}_S$ is a central extension
\begin{align}
    1
    \rightarrow
    H_2(G,\Z)
    \stackrel{i}{\rightarrow}
    \t{G}_S
    \stackrel{\pi}{\rightarrow}
    G
    \rightarrow
    1
    \;.
\end{align}
For finite groups $G$,
\begin{align}
    H^2(BG,U(1))=H^2(G,U(1))=H_2(G,\Z)\;.
\end{align}
Because we are restricting to the case $H^2(BG,U(1)) = Z_N$, this means the kernel of the quotient map $\pi: \t{G}_S\rightarrow G$ is $H_2(G,\Z) = Z_N$.
So $\t{G}_S$ is indeed a central extension of $G$ by $\Z_N$.

For what follows it is convenient to classify the irreducible representations of $\t{G}_S$ by how they represent the subgroup $\ker \pi = Z_N$.
Notate the generator of $\ker \pi = Z_N$ by $e^{2\pi i/N}$.
For any irreducible representation $\t{q}$ of $\t{G}_S$, the element $e^{2\pi i/N}$ must get represented as
\begin{align}
    U^{(\t{q})}(e^{2\pi i/N}) = e^{2\pi i k'(\t{q})/N}\1\;,
\end{align}
for some $k'(\t{q})\in\{0,1,\ldots,N-1\}$.
Refer to the value $k'(\t{q})$ for a given $\t{q}$ as the $N$-ality of $\t{q}$.
% which implies that a choice of Dijkgraaf-Witten action amounts to a choice of how to represent the generator $z$ of $\ker(\pi) = \Z_N=\{1,z,z^2,\ldots,z^{N-1}\}$,
% \begin{align}
%     \t{q}(z) = e^{2\pi ik'(\t{q})/N}\;,
% \end{align}
% where $\t{q}$ is a given representation of $\t{G}_S$ and where we have now restricted to the cases when $H^2(BG,U(1))=\Z_N$.
% We will refer to $k'$ as the $N$-ality of the projective representation $\t{q}$, and like we mentioned, $k'$ also fixes a choice of DW action.
Notice that, of course, a choice of how to represent $e^{2\pi i / N}$ doesn't completely specify $\t{q}$, so that there are many $\t{q}$ of the same $N$-ality.
We will see that the partition function of Dijkgraaf-Witten theory with ``twist" $k'$ can be written as a sum over representations with $N$-ality $k'$.

Bringing everything together, let's calculate the partition function of twisted Dijkgraaf-Witten on a manifold with boundaries.
First recall that in the untwisted case (section \ref{DWGPI}), we were simply counting isomorphism classes of $G$-bundles (without weighting them with phases, i.e.\ $k=0$) consistent with conjugacy classes specified on closed boundaries.
A gauge transformation keeps us in the same isomorphism class and changes the boundary holonomy only up to conjugacy, so that the counting is well-defined.
However, in the twisted case, although the counting is still a well-defined problem, the weighting is not well-defined, when there are boundaries of fixed conjugacy class.

On a closed manifold, a gauge transformation will not change the cohomology class $\omega_N$ defined in \eqref{om}.
This is not neccesarily the case, however on a manifold with boundaries. A gauge transformation that is nonzero at the boundary can nontrivially transform $\omega_N$:
\begin{align}
\omega_N \rightarrow \omega_N + db
\end{align}
where $b$ is a 1-cocycle.
Then on a manifold with boundaries
\begin{align}
\delta \omega = \expval{db,[M]} = \expval{b, \p [M]}
\end{align}
is potentially nonzero, causing phase ambiguities in the partition function
\begin{align}
    e^{2\pi ik\omega(\phi)/N}
    \rightarrow
    e^{2\pi ik\omega(\phi)/N}
    e^{2\pi ik\delta\omega(\phi)/N}
    \;.
\end{align}
% To see this, remember that \eqref{om} changes trivially due to $G$ gauge transformations
% \begin{align}
% \omega_N \rightarrow \omega_N + \delta b
% \end{align}
% but this change doesn't integrate to zero on a surface $M$ when the gauge transformation is non zero at the boundaries
% \begin{align}
% \delta \omega = \expval{\delta b,[M]} = \expval{b, \p [M]} \neq 0.
% \end{align}
% This creates phase ambiguities in the partition function:
% \begin{align}
%     e^{2\pi ik\omega(\phi)/N}
%     \rightarrow
%     e^{2\pi ik\omega(\phi)/N}
%     e^{2\pi ik\delta\omega(\phi)/N}
%     \;.
% \end{align}
The presence of these phase ambiguities implies that twisted Dijkgraaf-Witten is not gauge invariant on manifolds with boundary, so on the boundary there has to be a theory with an 't Hooft anomaly.
To have an unambiguously defined bulk partition function, we must make an ad hoc choice for those phases.
We can do this by picking a lift $U_i \rightarrow \t{U}_i$ for the each specified boundary holonomy $U_i$.

We are now equipped to calculate the partition function of twisted Dijkgraaf-Witten theory on a manifold with boundaries.
Remember that for a Riemann surface with genus $g$ and $n$ boundary components, the holonomies have to be such that 
\begin{align}
    V_1W_1V_1^{-1}W_1^{-1}
    \cdots
    V_gW_gV_g^{-1}W_g^{-1}U_1
    \cdots
    U_n
    =
    1_G
    \;,
\end{align}
where $g$ is the genus and $U_i$ are the holonomies on the $n$ boundaries of $M$.
Trying to uplift in a straightforward way would give 
\begin{align}
    u
    =
    \t{V}_1\t{W_1}\t{V_1}^{-1}\t{W_1}^{-1}
    \cdots
    \t{V}_g\t{W}_g\t{V}_g^{-1}\t{W}_g^{-1}\t{U}_1
    \cdots
    \t{U}_n
    \in
    \Z_N
    \;,
\end{align}
which means that this $G$-bundle can't be uplifted to a $\t{G}_S$-bundle.
In such a case $u$ is in fact $u=e^{2\pi i\omega(\phi)/N}$ where $\omega$ is that defined earlier.
Although this $G$-bundle doesn't uplift, we could excise a disk and impose the holonomy $u^{-1}$ on this new boundary to create a $\t{G}_S$-bundle, albeit over a manifold with an extra boundary.
This gives us a way to count the number of $G$-bundles that are weighted by the same value $\omega$.
Specifically, we get the number of $G$-bundles on $M_{g;U_1,\ldots,U_n}$ with weight $\omega$ by (correctly) counting the number of $\t{G}_S$-bundles on $M_{g;\t{U}_1,\ldots,\t{U}_n,u^{-1}}$.
Counting correctly here means accounting for the fact that when a $G$-bundle can be uplifted there are $N^{2g+n}$ possible upliftings.
We have
\begin{equation}\label{twistedpartfunccount}
    Z_G^{(k)}\big[g\,;\,\t{U}_1,\ldots,\t{U}_n\big]
    = 
    \sum_{\phi}
    \frac{e^{2\pi ik\omega(\phi)/N}}{|\operatorname{Stab}(\phi)|}
    =
    \sum_{\omega=0}^{N-1}
    e^{2\pi ik\omega/N}
    \frac{
        \big|\t{G}\big|\,
        Z^{(0)}_{\t{G}_S}\big[
            g\,;\,\t{U}_1,\ldots,\t{U}_n,e^{-2\pi i\omega}
        \big] 
    }{
        N^{2g+n-1}
    }
\end{equation}
where $Z^{(0)}_{\t{G}_S}\big[g\,;\,\t{U}_1, \ldots, \t{U}_n, e^{-2\pi i\omega}\big]$ is the partition function of untwisted Dijkgraaf-Witten theory for $\t{G}_S$ on $M_{g;\t{U}_1,\ldots,\t{U}_n, u^{-1}}$.
Using our known expression for the partition function of untwisted Dijkgraaf-Witten theory
\begin{equation}
    Z^{(0)}_{\t{G}_S}\big[g\,;\,\t{U}_1,\ldots,\t{U}_n,u^{-1}\big]
    =
    \sum_{\t{q}}
    \left(
        \frac{d_{\t{q}}}{\big|\t{G}\big|}
    \right)^{2-2g-n}
        \prod_{i=1}^n
    \left(
        \frac{\big|\t{U}_i\big|}{\big|\t{G}\big|}  \chi_{\t{q}}(\t{U}_i)
    \right)
    \frac{|u^{-1}|}{\big|\t{G}\big|}
    \chi_{\t{q}}(u^{-1})
\end{equation}
and substituting this in to \eqref{twistedpartfunccount} gives the result
\begin{equation}
    Z_G^{(k)}\big[g\,;\,\t{U}_1,\ldots,\t{U}_n\big]
    =
    \sum_{\t{q}}
    \delta_{k,k'({\t{q}})}
    \left(\frac{d_{\t{q}}}{|G|}\right)^{2-2g-n}
    \prod_{i=1}^n
    \left(
        \frac{\big|\t{U}_i\big|}{\big|\t{G}\big|}\,
        \chi_{\t{q}}\big(\t{U}_i\big)
    \right)
    \;\label{twist}
\end{equation}
after expanding $\chi_{\t{q}}(u^{-1})=d_{\t{q}}\,e^{-2\pi ik'({\t{q}})\omega/N}$ then simplifying via the identity $\sum_{\omega}e^{2\pi i\omega(k-k')/N}=N\delta_{k,k'}$.

\subsubsection{The gravity path integral}
Since the form of eq.\ \eqref{twist} is very similar to the untwisted case, the calculations proceed in the same fashion.
We define boundary conditions labeled by projective representation $\t{q}$ by 
$
    \h{Z}^{(k)}_{\t{q}}
    =
    \sum_{\t{k}}
    \chi_{\t{q}}
    \big(\t{k}^{-1}\big)
    \h{Z}^{(k)}[\t{k}]
$,
where the sum goes over all conjugacy classes $\t{k}$ of $\t{G}_S$.
This definition makes sense for any irreducible representation $\t{q}$ of $\t{G}_S$, but the representations of the wrong $N$-ality will give zero in correlators because
\begin{align}
Z^{(k)}_G\big[M_{g,n}\,;\,\t{q}_1,\ldots,\t{q}_n\big]
=
\sum_{\t{q}}
\delta_{k,k'(\t{q})}
\left(
    e^{S_0}
    \frac{d_{\t{q}}}{\left|G\right|}
\right)^{2-2g-n}
\delta_{\t{q}\t{q}_1\cdots\t{q}_n}
\;,
\end{align}
as can be readily worked out from \eqref{twist}.
Thus the most general correlator, one with $n_{\t{q}}$ insertions of each boundary with conditions $\t{q}$, is
\begin{equation}\label{twistedqDWcorrelator}
    \expval{
        \prod_{
            \substack{
                \text{$\t{q}$ with}
                \\
                \text{$N$-ality $k$}
            }
        }
        \h{Z}_{\t{q}}^{n_{\t{q}}}
    }
    =
    \prod_{
        \substack{
            \text{$\t{q}$ with}
            \\
            \text{$N$-ality $k$}
        }
    }
    e^{\lambda_{\t{q}}}
    B_{n_{\t{q}}}\!(\lambda_{\t{q}})
    \left(
    \frac{\left|G\right|}{e^{S_0}d_{\t{q}}}
    \right)^{n_{\t{q}}}
    \;,
\end{equation}
with $\lambda_{\t{q}}$ defined analogously to $\lambda_q$ in the untwisted case.
From the similarity of these correlators to those of the untwisted case, we immediately recognize the boundary dual of our twisted Dijkgraaf-Witten bulk gravity model.
Namely, (after making the rescalings described in section \ref{DWboundary}) it is a 1d topological theory whose Hilbert space consists of random numbers of \emph{projective} representations of $G$ that have the correct $N$-ality.
In other words, the boundary dual is an ensemble of 1d topological theories with anomalous global symmetry $G$.

\section{General 2d TQFTs}\label{generalcase}
In the previous section we constructed a simple gravity path integral with a bulk action of Dijkgraaf-Witten theory and found a dual interpretation as an ensemble of 1d theories on its boundary.
We will now show that a general 2d TQFT bulk theory likewise leads to a model gravity path integral with similar features.
To be precise, we will consider here TQFTs as defined by Atiyah's axioms \cite{Atiyah} and over the field $\C$.
Such TQFTs are fairly simple.
They have finite dimensional Hilbert spaces, and, as we will discuss, can be viewed as a direct sum of theories all with Hilbert space dimension 1.
Though they are simple, we speculate that the important features of our analysis will extend appropriately to 2d TQFTs more broadly defined, and perhaps even, in some form, to TQFTs in higher dimensions.

\begin{figure}[t]
    \centering
    \includegraphics[width=0.5\linewidth]{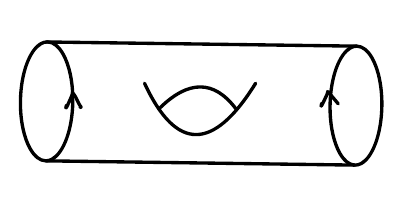}
    \caption{The handle creation operator: $\CH_{S^1} \rightarrow \CH_{S^1}$.}
    \label{handle}
\end{figure}

There are many excellent expositions of TQFTs;\footnotemark
\footnotetext{See, for example \cite{Carqueville:2017fmn}; or for a more abstract and more general exposition see \cite{lurie}.}
we present here only the most basic sketch for those not familiar.
In the language of category theory, a 2d TQFT can be defined as a functor (with certain requirements) from the category of 2d cobordisms $\mathbf{Cob}(2)$ to the category of complex vector spaces $\mathbf{Vect}(\C)$.
This definition, again, for those not familiar with this language, is a concise codification of the cutting and gluing properties that would naturally be expected of a path integral.
Like a path integral, a 2d TQFT assigns to each closed 2d manifold a number, the partition function on that manifold.
To a closed 1d manifold it assigns a Hilbert space.
To a 2d manifold $M$ with boundary $\p M$ it assigns a state in the Hilbert space associated to the boundary $\p M$.\footnotemark
\footnotetext{Or, alternatively, an object that takes as input a state and outputs a complex number.}
These assignments are compatible in the way expected of the output of a path integral.
For example, gluing two boundary components of a manifold together corresponds to a sum over matching states on the two components.
As another example, if a TQFT assigns to a circle the Hilbert space $\CH_{S^1}$, then it assigns to the ``handle creation operator" (the manifold as seen in fig. \ref{handle}) a unitary map $\CH_{S^1} \rightarrow \CH_{S^1}$.
From this map and the state in $\CH_{S^1}$ assigned to a hemisphere, we can construct any closed manifold by gluing.
This handle creation map always has positive, real eigenvalues \cite{Durhuus_1994}.
We will denote these eigenvalues by $\mu_I^{-2}$ where $I=1,\ldots,\dim(\CH_{S^1})$.
In the eigenbasis, $\ket{I}$, of the handle creation operator, the calculation of partition functions becomes a simple matter.
In particular, the partition function of a connected manifold $M_{g,n}$ with genus $g$ and $n$ boundaries, and states $\ket{I_1}$,\ldots, $\ket{I_n}$ input on the boundaries respectively, is
\begin{equation}\label{generalpartfunc}
    Z^{\text{TQFT}}[M_{g,n}\,;\,I_1,I_2,\ldots,I_n]
    =
    \sum_I\mu_I^{2-2g-n}\delta_{II_1I_2\cdots I_n}\;.
\end{equation}
Note this evaluates to zero if the boundary labels $I_i$ are not all the same.

From the result \eqref{generalpartfunc}, the generating function for connected correlation functions of the gravity path integral is
\begin{equation}
    \begin{aligned}
        \log\expval{e^{\sum_Iu_I\h{Z}_I}}
        &=
        \sum_I
        \sum_n\frac{u_I^n}{n!}
        \sum_g
        \mu_I^{2-2g-n}\\
        &=
        \sum_I
        \lambda_I
        e^{u_I/\mu_I}\;,
    \end{aligned}
\end{equation}
where $\lambda_I\equiv\mu_I^2/(1-\mu_I^{-2})$.
The full generating function of correlators is then $\expval{\exp(\sum_Iu_I\h{Z}_I)}=\prod_I\exp(\lambda_Ie^{u_I/\mu_I})$, from which we can extract the result
\begin{equation}
    \expval{\prod_I\h{Z}_I^{n_I}}
    =
    \prod_I
    e^{\lambda_I}
    B_{n_I}(\lambda_I)
    \mu_I^{-n_I}\;.
\end{equation}
This can be written
\begin{equation}
    \expval{\prod_I\h{Z}_I^{n_I}}/\expval{\1}
    =
    \prod_I
    \sum_{N_I}
    p_{\lambda_I}(N_I)
    \left(\frac{N_I}{\mu_I}\right)^{n_I}
\end{equation}
where $p_{\lambda_I}$ here is the Poisson distribution with mean $\lambda_I$.
The normalized correlators then have an interpretation as an average, where $Z_I$ takes the value $N_I/\mu_I$, and the $N_I$ are independently chosen Poisson random integers.
As we saw for the case of Dijkgraaf-Witten, the $Z_I$ don't quite have the interpretation as partition functions of a one-dimensional topological theory.
If, however, we rescale each of them by a factor of, in this case, $\mu_I$, we do find a nice interpretation for them.
Namely we have as our boundary theory a topological quantum mechanics with sectors labeled by $I$ and the number of dimensions in each sector $I$ given by the Poisson distribution with mean $\lambda_I$.
Then $\mu_IZ_I=\tr(P_I)$ where $P_I$ is a projection onto the sector $I$.

\section{2d TQFTs with boundaries} \label{EofW}
In their simple model of a gravity path integral, \cite{Marolf:2020xie} consider the addition of end-of-the-world branes (EofW branes).
These are boundaries on which bulk spacetime ends.
Generalizing this construction, we will consider general boundaries for general 2d TQFTs and the describe the model gravity path integrals built out of them.
As a first example we will again consider Dijkgraaf-Witten theory, but this time with the addition of EofW branes.
As we will explain, EofW branes are in some sense the simplest boundary conditions possible in the theory, but we will nonetheless see features that will hold in the more general case.
These include the factorization between different sectors of the theory, as before, as well as a new difficulty with interpreting the gravity path integral as a boundary ensemble theory without an additional modification.
We will discuss this difficulty and a solution to it in section \ref{reflectionpos}.

\subsection{End-of-the-world branes for Dijkgraaf-Witten}
As a first example of an open/closed TQFT we consider Dijkgraaf-Witten theory with the addition of end-of-the-world (EofW) brane boundaries.
Following the constructions of \cite{Marolf:2020xie, Balasubramanian:2020jhl}, we allow for some number of ``flavors" of otherwise identical EofW branes, which we will label by $a=1,2,\ldots,K$.
Though these $K$ differently labeled EofW branes have identical dynamics, the number $K$ of such branes will play an important role later.
The path integral for Dijkgraaf-Witten theory with EofW branes is still defined as a sum over gauge backgrounds, but now that our spacetime manifolds have boundaries, the specification of a gauge background must include two kinds of data: the holonomies around loops (as before), as well as the parallel transports along paths that begin and end on EofW branes.
Cutting a manifold with brane boundaries will result in a new, non-brane boundary.
These resulting boundaries, which we will call variously ``gluing" boundaries, ``state" boundaries, or ``Hilbert space" boundaries, 
are not EofW brane boundaries, but rather, correspond to a Hilbert space of states that represent the input of field configuration data on that boundary.
The ``gluing" boundaries, in our case, will have fixed parallel transports along them which serve as boundary conditions for the possible gauge backgrounds.
More specifically, a circle gluing boundary component will be labeled by the holonomy about that circle, and an interval gluing boundary component will be labeled by an element of $G$ representing the parallel transport across that interval.
Note, that the intervals, unlike the circles, are labeled by \emph{elements} of $G$, rather than conjugacy classes.
This is similar to the construction of \cite{Balasubramanian:2020jhl}.
Though they consider boundary conditions for spin structures, these are analogous to our case with the choice of $G=\Z_2$.\footnotemark
\footnotetext{Where $\Z_2=\{\mathrm{id},a\}$, NS boundary conditions on a circle are analogous to a holonomy of $\mathrm{id}$ about the circle, and R boundary conditions are analogous to a holonomy of $a$ around the circle.
Likewise, an interval labeled by the identity (by the non-identity $(-1)^F$) is analogous to an interval with parallel transport $\mathrm{id}$ (parallel transport $a$).
Substituting $\Z_2$ for $G$ in what follows will largely mirror much of their discussion, where one keeps in mind that the irreps of an abelian group like $\Z_2$ are all one-dimensional.}

EofW branes, unlike gluing boundaries, do not get labeled by group elements or conjugacy classes of $G$.
We take the EofW branes to be ``decoupled" from the bulk, in the sense that inserted $G$-symmetry operators are not permitted to end on EofW branes, which is to say, there is no parallel transport as one moves along an EofW brane boundary.

When boundary conditions for gauge backgrounds are specified along circle and interval gluing boundaries, the count of gauge backgrounds will only include those that match the specified holonomies and parallel transports along those boundaries.
Counting gauge backgrounds can be done using a lattice description, as explained in appendix \ref{DWappendix}.\footnotemark
\footnotetext{Alternatively, given a connected manifold $M$ with end-of-the-world-brane boundary $\p_{\text{brane}}M$, a choice of gauge background can be identified with a homomorphism from $\pi_1\left(M/\p_{\text{brane}}M,m\right)$ to $G$, where $M/\p_{\text{brane}}M$ is the quotient manifold obtained by identifying all end-of-the-world brane boundaries to a single point $m$, and where we take that point $m$ as our basepoint.}
We list some results, from which all partition functions can be calculated.
First, consider a strip with EofW branes on both sides and gluing boundaries at the ends, with boundary conditions $g_1,g_2\in G$ respectively.
There is one gauge background compatible with the boundary conditions if $g_1=g_2^{-1}$ and zero otherwise, giving the partition function
\begin{equation}\label{DW_pairing_k_basis}
\vcenter{\hbox{\includegraphics[width=0.25\linewidth]{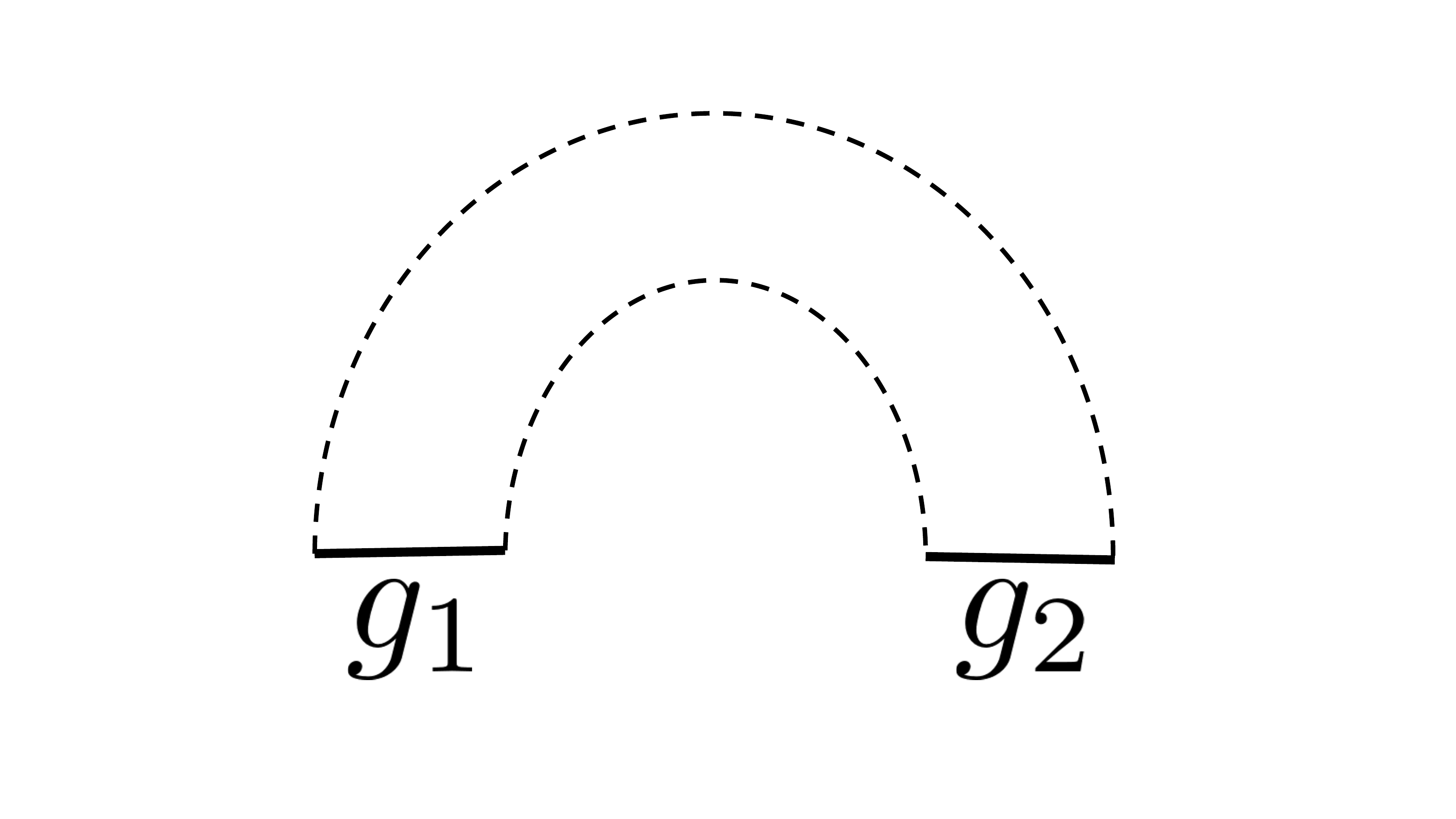}}}
=
\quad\quad
\delta_{g_1,g_2^{-1}}\;.
\end{equation}
Taking the different boundary conditions $g\in G$ as states in the Hilbert space $\CH_{\text{open}}$ associated to the interval, the above diagram gives a (nondegenerate) pairing on that Hilbert space,  $\left(\ket{g_1},\ket{g_2}\right)=\delta_{g_1,g_2^{-1}}$.
Second, consider a disk with three gluing boundary intervals labeled by $g_1,g_2,g_3\in G$, alternating with three EofW brane intervals.
This path integral gives $\delta(g_3g_2g_1,1_G)$.
We can likewise interpret this as a map from the Hilbert spaces of the intervals to $\C$, namely $\ket{g_1}\ket{g_2}\ket{g_3}\mapsto\delta(g_3g_2g_1,1_G)$.
This along with the pairing above induces a map $\CH_{\text{open}}\otimes\CH_{\text{open}}\rightarrow\CH_{\text{open}}$, represented in diagram form by
\begin{equation}\label{DW_multiplication_k_basis}
\vcenter{\hbox{\includegraphics[width=0.25\linewidth]{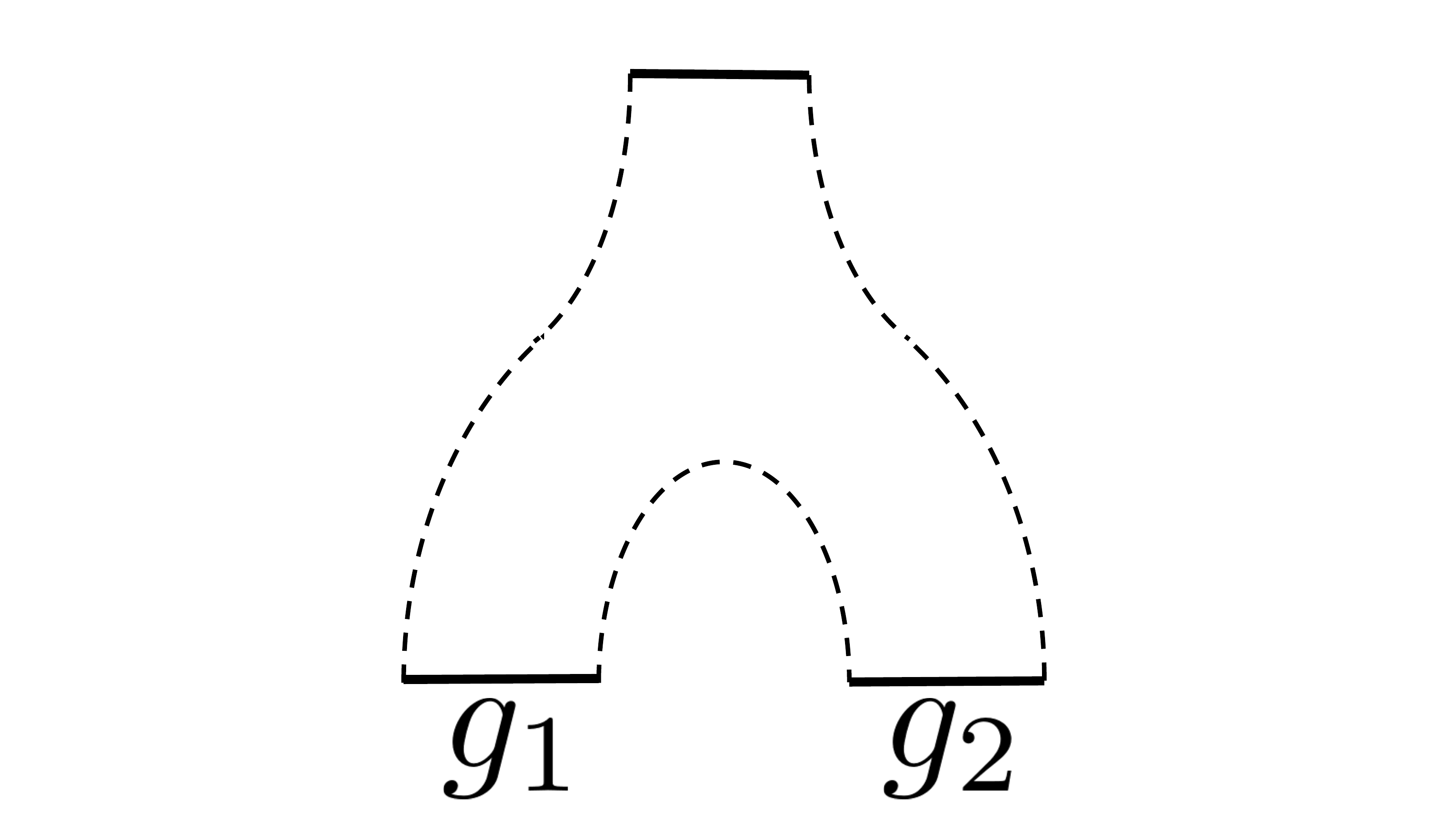}}}
=
\quad\quad
\ket{g_2g_1}\;.
\end{equation}
\begin{figure}[t]
    \centering
    \includegraphics[width=0.5\linewidth]{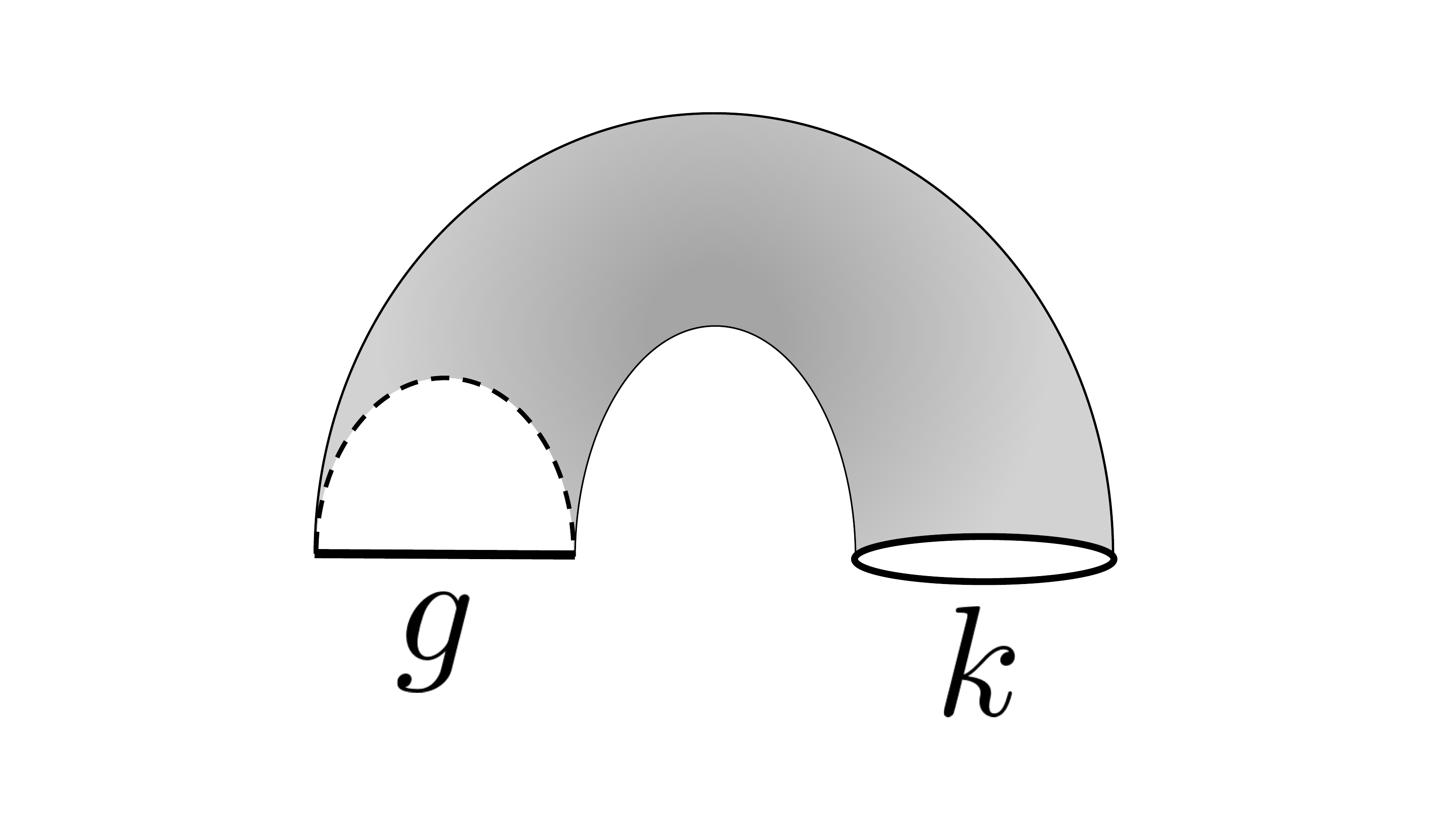}
    \caption{A cylinder with a mixture of end-of-the-world branes and gluing boundaries with boundary conditions. With the above boundary conditions the path integral evaluates to $\delta(k,[g^{-1}])$, where $[g^{-1}]$ is the conjugacy class of the element $g^{-1}\in G$.}
    \label{fig:DW_trace}
\end{figure}
Finally, consider a cylinder where one boundary circle is a gluing boundary with fixed holonomy $k$ and where the other boundary circle is made up of an EofW brane interval together with a gluing interval with fixed parallel transport $g$.
(See figure \ref{fig:DW_trace}.)
The count of gauge bundles on this manifold is $\delta(k,[g^{-1}])$, where $[g^{-1}]$ is the conjugacy class of the element $g^{-1}\in G$.
Via the pairing \eqref{pairc}, this provides a map from the open sector Hilbert space $\CH_\text{open}$ to the circle  Hilbert space $\CH_{S^1}$ given by the diagram
\begin{equation}\label{DW_trace_k_basis}
\vcenter{\hbox{\includegraphics[width=0.25\linewidth]{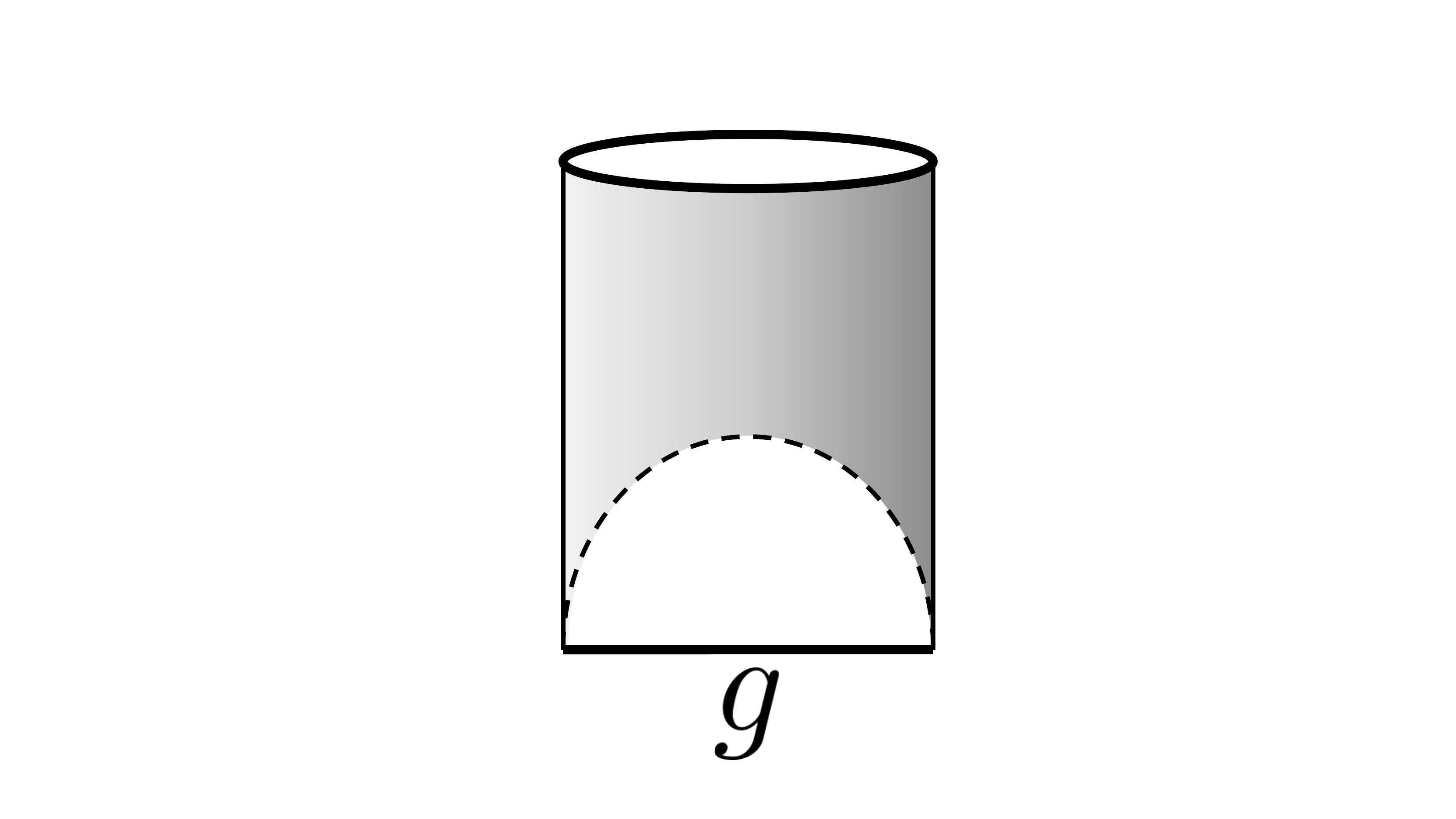}}}
=
\quad\quad
\frac{\left|G\right|}{\left|[g]\right|}
\ket{[g]}
\;,
\end{equation}
where, again, $[g]$ is the conjugacy class of $g\in G$.

These are all the diagrams necessary to compute any partition function.
The computations proceed more easily in bases for the open and closed sectors that are labeled by irreducible representation.
We define bases
\begin{align}
    \ket{q}&\equiv \sum_k\chi_q(k^{-1})\ket{k}\\
    \ket{q;i,j}&\equiv \frac{d_q}{\left|G\right|}\sum_{g\in G} U^{(q)}_{ij}(g^{-1})\ket{g}
    \label{opensectorchangeofbasis}
\end{align}
for the closed and open Hilbert spaces, respectively, where $i,j=1,\ldots,d_q$.
In these bases the maps \eqref{DW_pairing_k_basis}, \eqref{DW_multiplication_k_basis}, and \eqref{DW_trace_k_basis} are
\begin{align}
    \vcenter{\hbox{\includegraphics[width=0.35\linewidth]{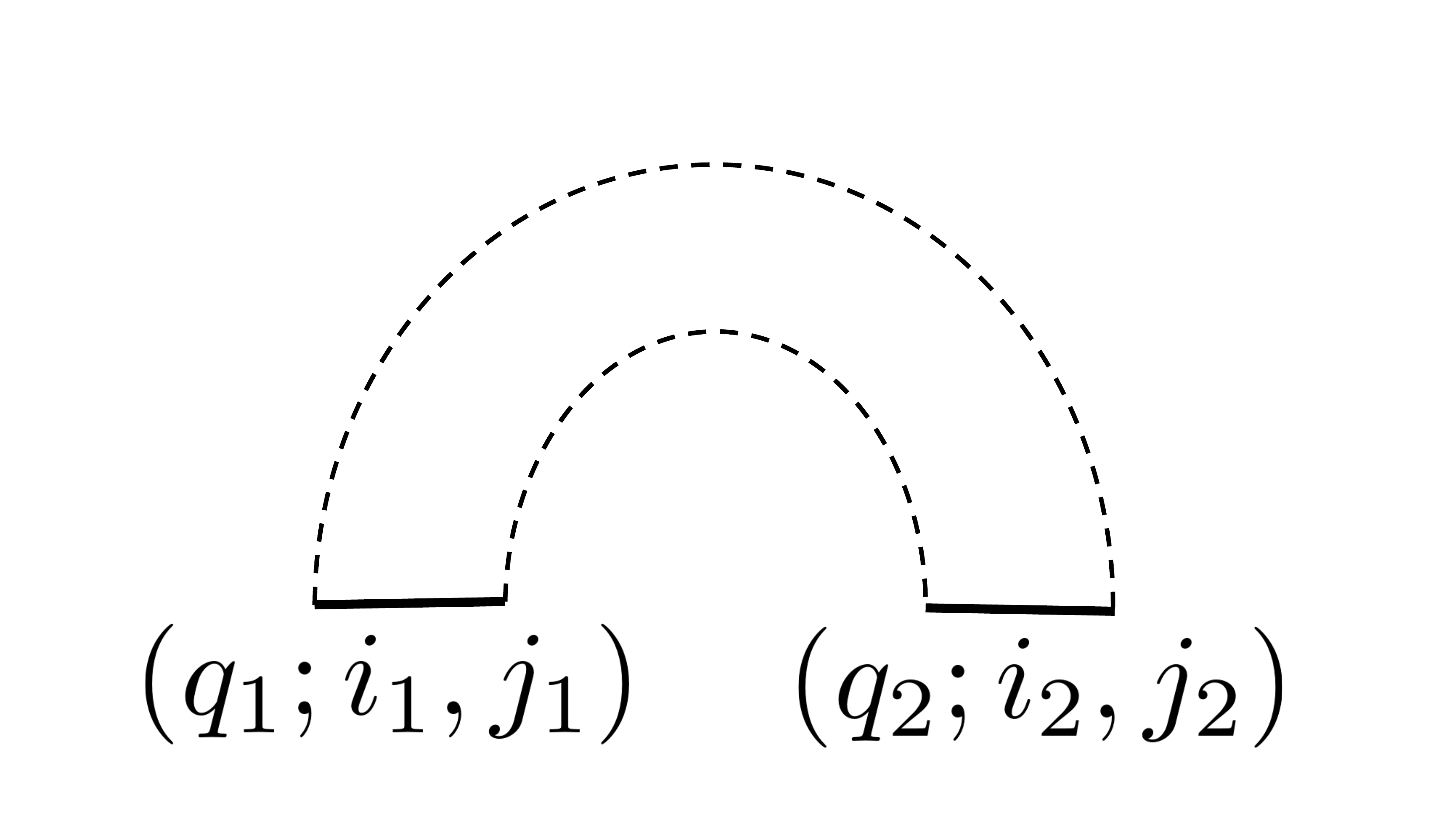}}}
    &=
    \quad\quad
    \frac{d_{q_1}}{\left|G\right|}\delta_{q_1q_2}\delta_{j_1i_2}\delta_{j_2i_1}
    \label{r1}\\
    \vcenter{\hbox{\includegraphics[width=0.35\linewidth]{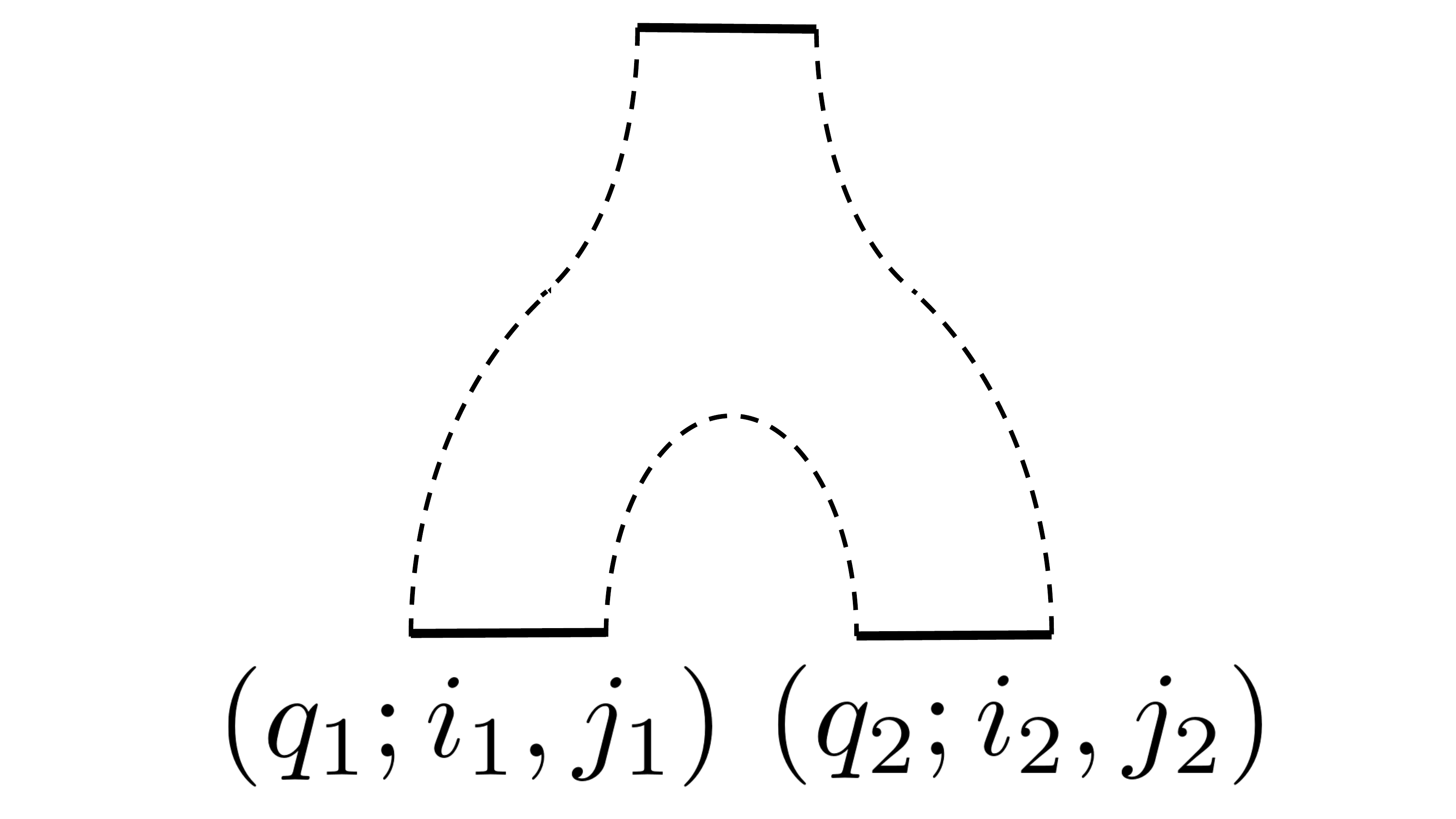}}}
    &=
    \quad\quad
    \delta_{q_1q_2}\delta_{j_1i_2}\ket{q_1;i_1,j_2}
    \label{DWmultiplication}\\
    \vcenter{\hbox{\includegraphics[width=0.35\linewidth]{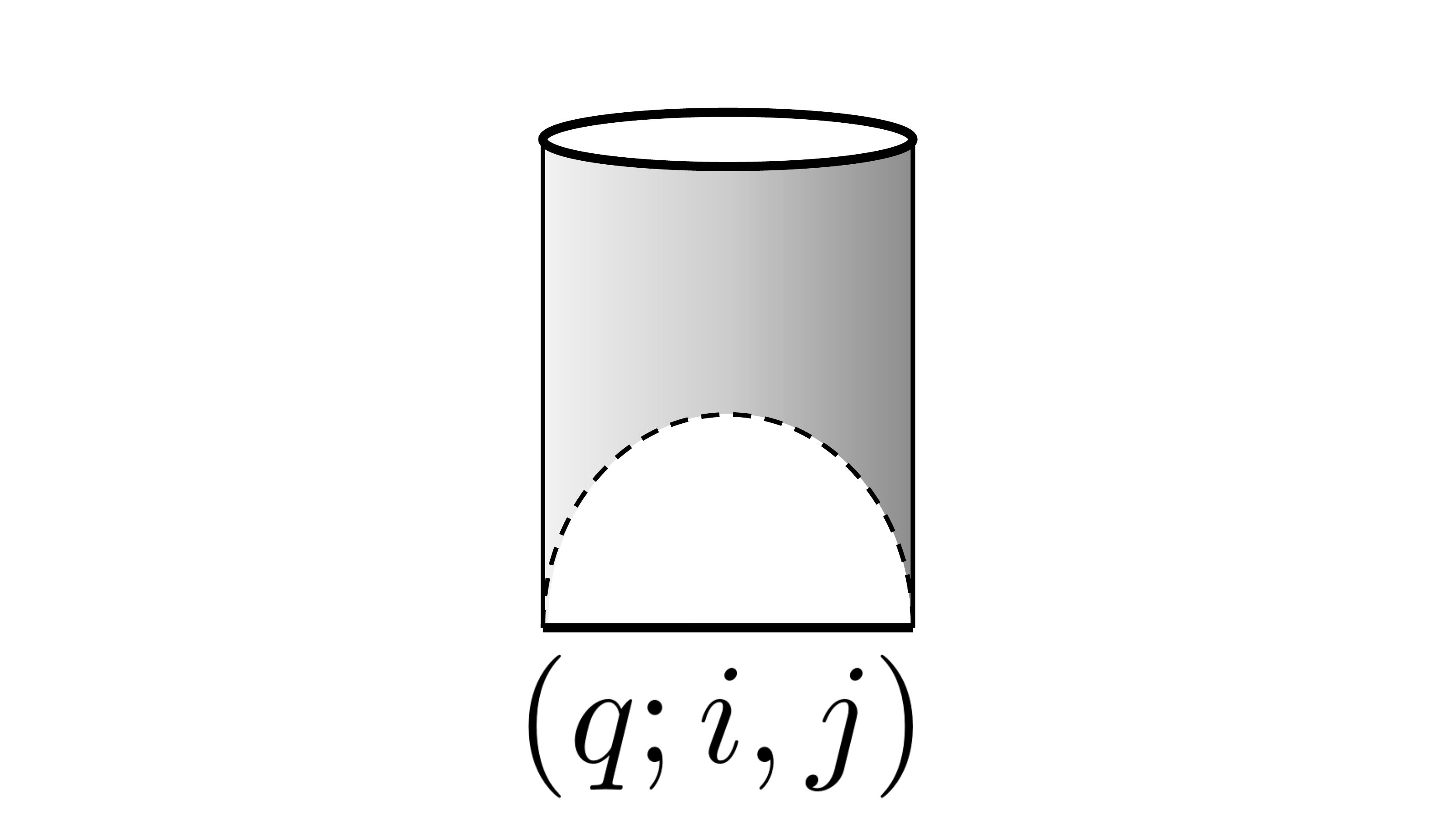}}}
    &=
    \quad\quad
    \delta_{ij}\ket{q}
\end{align}
In particular, these imply the following two diagrams:
\begin{align}
    \vcenter{\hbox{\includegraphics[width=0.4\linewidth]{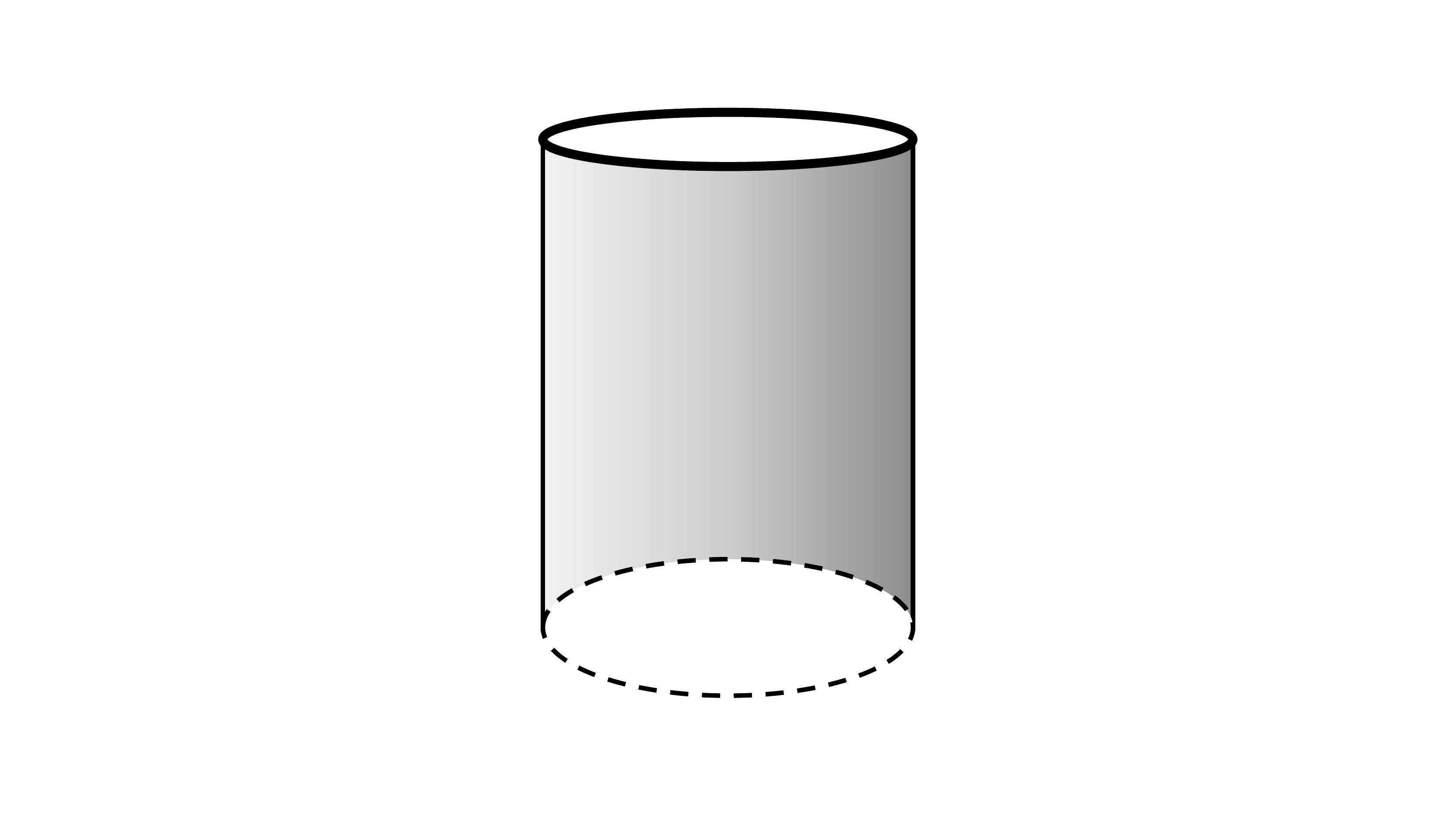}}}
    &=
    \quad\quad
    \sum_q
    d_q
    \ket{q}
    \\
    \vcenter{\hbox{\includegraphics[width=0.4\linewidth]{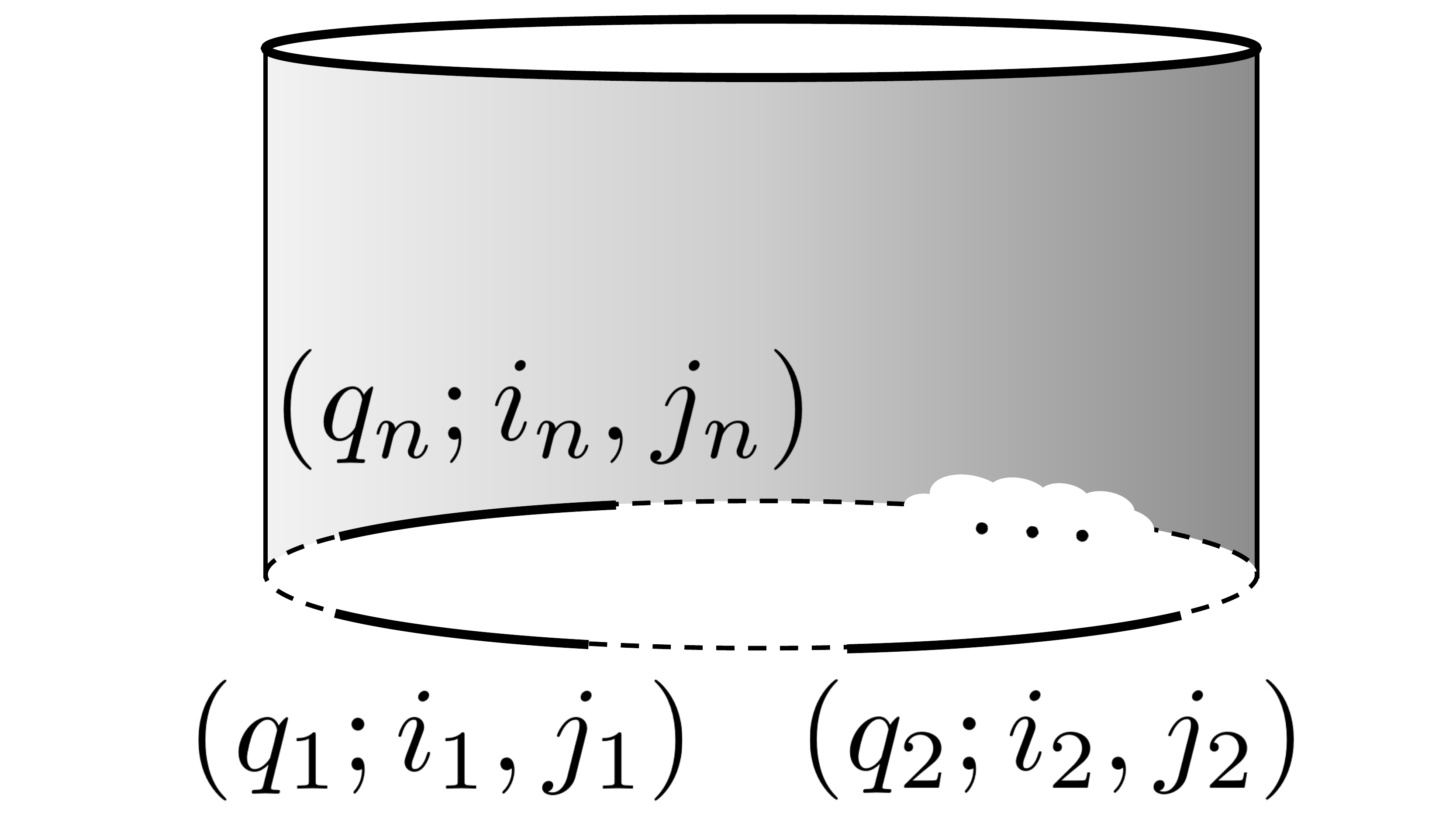}}}
    &=
    \quad\quad
    \sum_q
    \delta_{qq_1\cdots q_n}
    \delta_{j_1i_2}\delta_{j_2i_3}\cdots\delta_{j_ni_1}
    \ket{q}
    \label{r5}
\end{align}
with which, we can calculate any partition function.

A manifold can have boundary components of three different types: closed sector circles, circles made up of a single EofW brane, and circles made up of alternating open sector intervals and EofW brane intervals.
The implications of the last two diagrams above, for the partition functions of connected manifolds, include the following four facts:
\begin{itemize}
	\item Any connected manifold that has boundaries, of any type, labeled by different irreps, will evaluate to zero.
	\item Any connected manifold with a boundary made up of alternating open sector and EofW interval boundaries, where the i,j indices on the open sector labels do not match up appropriately, will also evaluate to zero.
	Matching appropriately means that the second index of one open sector interval equals the first index of the next open sector interval, and so on around the circle.
	In the case where we have more than one type of EofW brane, we also require that any indices for the species of EofW brane match similarly.
	\item A boundary made up of alternating open sector and EofW brane intervals, with all open sector intervals labeled by q and all the i,j indices matching appropriately, will simply contribute the same as a closed sector boundary labeled by q.
	\item A circular boundary made up of a single EofW brane contributes a factor of $|G|$ relative to just filling in that boundary with a disk.
\end{itemize}

Let's use these facts to obtain the partition function of a general manifold.
We will consider two cases: (1) manifolds with at least one ``gluing" boundary and (2) manifolds with no boundaries other than EofW brane boundaries.
For the first case, let $M$ be a surface of genus $g$, with $n$ closed sector boundaries, $m$ circle EofW brane boundaries, and $\ell$ boundaries made up of alternating open sector and EofW brane intervals, where all closed and open sector boundaries are labeled by $q$ and all $i,j$ indices are matched appropriately.
Note that in this first case at least one of $n$ or $\ell$ must be greater than $0$.
The partition function for such a manifold will be
\begin{equation}
    Z^{\text{DW}}[M]
    =
    \left(\frac{d_q}{\left|G\right|}\right)^{2-2g-n-m-\ell}
    d_q^m.
\end{equation}
The additional $m$ factors of $d_q$ come from the last fact above: filling in a boundary with a disk will contribute a factor of $d_q/\left|G\right|$, and a circular EofW brane boundary will contribute $\left|G\right|$ times that, so it will contribute an overall factor of $d_q$.
Now consider the second case: let $M$ be a manifold with $m$ circular EofW brane boundaries, but no closed or open sector boundaries.
Then the partition function of $M$ is
\begin{equation}
    Z^{\text{DW}}[M]
    =
    \sum_q
    \left(\frac{d_q}{\left|G\right|}\right)^{2-2g-m}
    d_q^m.
\end{equation}

\subsection{The gravity path integral}\label{openDWcalc}
We are now equipped to build a gravity path integral out of Dijkgraaf-Witten with EofW branes.
The gravity path integral will be a sum over all manifolds with compatible EofW brane boundaries.
Specifically, we take the EofW branes to be dynamical while we take the open and closed sector ``gluing" boundaries to be fixed.
The open sector boundaries are intervals with a fixed label at each end designating the type of EofW brane found there, and in the gravity path integral we only allow configurations where an EofW brane of a given type attaches only to interval endpoints labeled by that type.

Operators inserting fixed circle boundaries will be denoted by $\h{Z}[k]$ or $\h{Z}_q$, as before.
The operators inserting fixed interval boundaries will be denoted by $\h{S}_{gab}$ where $g$ specifies the fixed gauge background boundary conditions (i.e.\ the parallel transport $g\in G$ along the interval) and where $a$ and $b$ are the just mentioned labels designating the type of EofW brane allowed at the endpoints.
Note that the boundary insertion operators are in one-to-one correspondence with the states of the TQFT: the $\h{Z}$ operators correspond to states in the closed sector, and the $\h{S}$ operators labeled with $a$ and $b$ correspond to states in the open sector with boundary conditions $a$ and $b$.
Using the change of basis \eqref{opensectorchangeofbasis} we can define the more convenient interval operators $\h{S}_{qijab}$ representing the insertion of an interval with the state $\ket{q;i,j}$ fixed on it.

We will add a term $S_0\chi$ proportional to the Euler characteristic to the Dijkgraaf-Witten action, as we did before.
This will have the effect of suppressing higher genus manifolds, allowing the sum over topologies to converge.
The gravity path integral is the sum of the TQFT partition functions over every manifold that is compatible with the inserted boundaries with a measure that takes into account possible residual diffeomorphisms.

We now describe the calculation of correlators of the boundary insertion operators.
Let $m_a$ denote the number of circular EofW brane boundaries of type $a$ on a connected manifold, so that $m=\sum_a m_a$.
Then the connected vacuum correlator for Dijkgraaf-Witten with EofW branes is
\begin{equation}
\begin{aligned}
    \lambda'
    &\equiv
    \log\expval{\1}
    =
    \expval{\1}_{\text{connected}}\\
    &=
    \sum_g\sum_{m_1,\ldots,m_K}
    \frac{1}{m_1!\cdots m_K!}
    \sum_q
    e^{S_0(2-2g-\sum_a m_a)}
    \left(\frac{d_q}{\left|G\right|}\right)^{2-2g-\sum_am_a}
    d_q^{\sum_am_a}
    \\
    &=
    \sum_q
    \left(e^{S_0}\frac{d_q}{\left|G\right|}\right)^2
    \frac{1}{1-\left(e^{S_0}\frac{d_q}{\left|G\right|}\right)^{-2}}
    \prod_{a=1}^K
    \exp(
        \left(\frac{e^{S_0}}{\left|G\right|}\right)^{-1}
    )\\
    &=
    \sum_q
    \lambda_q
    e^{K\left|G\right|e^{-S_0}}\\
    &=
    e^{K\left|G\right|e^{-S_0}}\lambda,
\end{aligned}
\end{equation}
where we have used our previous definitions \eqref{l} of $\lambda_q$ and $\lambda$.
The connected manifolds without fixed boundary are entirely specified by their genus $g$ and then by the number $m_a$ of dynamical EofW brane circle boundaries they have for each type $a$, hence the sums over $g$ and each $m_a$ in the above expression.
Also note the factor of $1/m_a!$ for each brane type.
These are factors in the measure of the path integral and are due to the $m_a!$ large diffeomorphisms that permute the EofW brane circles (which are dynamical and hence taken to be indistinguishable).
Comparing this expression for the connected vacuum correlator to that of the theory without EofW branes \eqref{l}, we see that the effect of the circle EofW brane boundaries is simply to multiply $\lambda$ by a factor.

Let's now consider the generating function for all correlators
\begin{equation}
    F(\vec{u},T)
    =
    \expval{
        \exp(
            \sum_q u_q\h{Z}_q
            +
            \sum_{a,b=1}^K\sum_q\sum_{i,j=1}^{d_q}
            t_{qijab}\h{S}_{qijab}
        )
    }
    \;.
\end{equation}
For convenience, we will collect the chemical potentials $t_{qijab}$ into an object $T$ and the chemical potentials $u_q$ into the object $\vec{u}$.
Also, note that the $i,j$ indices within each $q$ sector play the same role as the EofW brane labels $a,b$, so to ease notation for the calculation of the above generating function we will combine $i$ and $a$ into a single index $A$ and similarly $j$ and $b$ into $B$.

Following \cite{Marolf:2020xie}, and analogously to the procedure in section \ref{DWGPI} for the theory without EofW branes, we calculate the logarithm of the generating function,
\begin{equation}\label{exponentialexpansion2}
\begin{aligned}
&\log F(\vec{u},T)
=
\sum_{n}
\sum_{k}
\frac{1}{n!}
\frac{1}{k!}
\expval{
\left(\sum_qu_q\h{Z}_q\right)^n
\left(\sum_{qAB}t_{qAB}\h{S}_{qAB}\right)^{k}
}_{\text{conn}}\\
&=
\sum_q
\sum_{n,k}
\frac{1}{n!}
u_q^n
\frac{1}{k!}
\sum_{A_1B_1\cdots A_kB_k}
t_{qA_1B_1}\cdots t_{qA_kB_k}
\expval{
\h{Z}_q^n
\h{S}_{qA_1B_1}\h{S}_{qA_2B_2}\cdots \h{S}_{qA_kB_k}
}_{q,\text{conn}} \; 
\end{aligned}
\end{equation}
where we have used the fact that a connected correlator with boundaries labeled by different irreps is always zero.

Each correlator in the above will be a sum over genus, a sum over numbers of circle EofW brane boundaries, and finally a sum over all the ways of connecting the $\h{S}$ operators by EofW branes.
We can express this final sum as a sum over all permutations of $k$ elements, where $k$ is the number of open sector interval boundaries.
Say $\h{S}_{qA_1B_1}$, $\h{S}_{qA_2B_2}$, \ldots, $\h{S}_{qA_kB_k}$ are the intervals in our correlator.
For a given permutation $\pi\in S(k)$ we attach the outgoing end of $\h{S}_{qA_iB_i}$ to the ingoing end of $\h{S}_{qA_{\pi(i)}B_{\pi(i)}}$ for every $i$.
A particular way of connecting the branes will only result in a nonzero partition function if all pairs of connected endpoints have the same index.
In other words, for any $\pi$ the partition function will be proportional to $\delta_{B_1,A_{\pi(1)}}\delta_{B_2,A_{\pi(2)}}\cdots \delta_{B_k,A_{\pi(k)}}$.
For a permutation $\pi$, denote the number of 1-cycles in $\pi$ by $a_1(\pi)$, the number of 2-cycles by $a_2(\pi)$, and so on.
The total number of cycles we will denote by $a(\pi)=\sum_ia_i(\pi)$, and therefore the number of alternating-type boundaries will be $\ell=a(\pi)$ for any $\pi$.

We are now ready to calculate the general correlator of $n$ closed sector boundaries and $k$ open sector interval boundaries.
We get
\begin{align}
    \langle\h{Z}^n_q\h{S}_{qA_1B_1}
    &
    \h{S}_{qA_2B_2}
    \cdots\h{S}_{qA_kB_k}\rangle_{\text{conn}}
    \nonumber\\
    &=
    \sum_{m_1,\ldots,m_K}
    \frac{1}{m_1!\cdots m_K!}
    \sum_g
    \sum_{\pi\in S(k)}
    \delta_{B_1,A_{\pi(1)}}\cdots \delta_{B_k,A_{\pi(k)}}
    Z^{\text{bulk}}\left[M_{g,n+m+\ell}\,;q,m\right]
    \nonumber\\
    &=
    \lambda'_q
    \left(e^{S_0}\frac{d_q}{\left|G\right|}\right)^{-n}
    \sum_{\pi\in S(k)}
    \delta_{B_1,A_{\pi(1)}}\cdots \delta_{B_k,A_{\pi(k)}}
    \left(e^{S_0}\frac{d_q}{\left|G\right|}\right)^{-a(\pi)}
    \;,
\end{align}
where we have used $Z^{\text{bulk}}\left[M_{g,n+m+\ell}\,;q,m\right]$ to denote the partition function of a connected manifold of genus $g$ with $n+m+\ell$ circular boundaries, $n$ of which are closed sector boundaries labeled by $q$, $m=\sum_am_a$ of which are EofW brane boundaries, and $\ell=a(\pi)$ of which are alternating interval boundaries labeled by $q$.

We can now plug this in to our expansion \eqref{exponentialexpansion2} above.
We get
\begin{equation}
    \log F(\vec{u},T)
    =
    \sum_q
    \lambda'_q
    \exp(
        u_q
        \frac{\left|G\right|}{d_qe^{S_0}}   
    )
    \sum_{k}
    \frac{1}{k!}
    \sum_{\pi\in S(k)}
    \left(e^{S_0}\frac{d_q}{\left|G\right|}\right)^{-\sum_ja_j(\pi)}
    \prod_j
    \tr(T_{(q)}^j)^{a_j(\pi)}
    \;,
\end{equation}
Note that the functions $a_j$ on the permutation group only depend on the cycle structure of the permutation.
So we can replace the sum over permutations with a sum over cycle structures, using the fact that there are $\frac{k!}{\prod_ja_j!j^{a_j}}$ permutations with $a_j$ cycles of length $j$ for each $j$.
This results in
\begin{equation}\label{EofWDWgenfunccircles}
    \log F(\vec{u},T)
    =
    \sum_q\lambda'_q\exp(u_q\frac{\left|G\right|}{d_qe^{S_0}})
    \prod_{j=0}^\infty\sum_{a_j}\frac{1}{a_j!}
    \left(
        \frac{\left|G\right|}{d_qe^{S_0}}
        \frac{1}{j}
        \tr(T_{(q)}^j)
    \right)^{a_j}
    \;.
\end{equation}
Recognizing the rightmost sum as an exponential and then using the identity $\exp(\sum_j\frac{1}{j}\tr(T^j))=\det(\1-T)^{-1}$, we get a final expression for the generating function:
\begin{equation}\label{EofWDWgenfunc}
    F(\vec{u},T)
    =
    \prod_q
    \exp(
        \lambda'_q
        \left(
            \frac{e^{u}}{\det(\1-T_{(q)})}
        \right)^{\frac{\left|G\right|}{d_qe^{S_0}}}
    )
\end{equation}
Compare to \eqref{reviewgenfunc}, the analogous result for the simple model of \cite{Marolf:2020xie}.

\subsection{Boundary Interpretation}
As explained in section \ref{DWboundary}, finding an interpretation of the gravity path integral in terms of an ensemble of boundary theories reduces to an instance of the moment problem, the problem of finding a distribution given its moments.
In the case of Dijkgraaf-Witten with EofW branes, the different boundary theories that could appear in our ensemble are characterized by the values that the boundary partition functions $Z_q$ and $S_{qijab}$ take in them.
These are the alpha-parameters and the (normalized) correlators of the $\h{Z}$ and $\h{S}$ operators are moments probing the probability distribution over the alpha-parameters.
From this point of view we obtain the following relation between the generating function \eqref{EofWDWgenfunc} obtained above and the probability distribution $p(\alpha)$ over the alpha-parameters:
\begin{equation}\label{momentgenfunc}
    \prod_q
    \exp(
        \lambda'_q
        \left(
            e^{u}/\det(\1-T_{(q)})
        \right)^{\left|G\right|/d_qe^{S_0}}
    -\lambda'_q
    )
    =
    \int\!\!d\alpha\,
    p(\alpha)
    e^{\sum_qu_q\alpha_q+\sum_q\tr(T_{(q)}M_{(q)})}.
\end{equation}
Here the integral is over all $\alpha\in\C^{r+\left|G\right|K^2}$ with $\alpha=(\alpha_1,\ldots,\alpha_r;M_{(1)},\ldots,M_{(r)})$ where the $\alpha_q$ are the values the $Z_q$ take in a particular theory and each $M_{(q)}$ is a $Kd_q$ by $Kd_q$ matrix representing the value of the matrix $S_{(q)}$ in a particular theory.
In principle we can extract the probability distribution $p(\alpha)$ by taking the appropriate Fourier transform of both sides of the above.
In fact, we already know the probability distribution for the $\alpha_q$.
These come out nearly the same as in section \ref{DWboundary}.
The result is that $Z_q$ takes the value $Z_q=\frac{\left|G\right|}{e^{S_0}d_q}{N_q}$ where $N_q$ is an integer drawn from the Poisson distribution with mean $\lambda'_q$.

So we are left with the task of determining the probability distribution for the matrices $M_{(q)}$.
To see what that will be, we can consider the generating function for the $\h{S}$ operators but conditioned on chosen values $\alpha_q$ for the $\h{Z}$ operators.
This can read off from the result of taking the Fourier transform of both sides of \ref{momentgenfunc} with respect to the variables $iu_q$.
We get
\begin{equation}
    \expval{
        \exp(
            \sum_{a,b=1}^K\sum_q\sum_{i,j=1}^{d_q}t_{qijab}\h{S}_{qijab}
        )
    }_{Z_q=\alpha_q}
    =
    \det(\1-T_{(q)})^{-\alpha_q}
    \;,
\end{equation}
so we are left with
\begin{equation}\label{conditionalgenfunc}
    \det(\1-T_{(q)})^{-\alpha_q}
    =
    \int\!\!dM\,
    p(\vec{\alpha};M)
    e^{\sum_q\tr(T_{(q)}M_{(q)})},
\end{equation}
where $dM$ represents an integral over all values for the matrices $M_{(1)},\ldots,M_{(r)}$.
The situation is simply that of \eqref{reviewdistribution}, encountered for the simple model of \cite{Marolf:2020xie}.
Namely, $\det(\1-T_{(q)})^{-\alpha_q}$ is a generating function of moments of a probability distribution only when the exponent $\alpha_q$ takes certain values \cite{graczyk2003}, in this case values in the set $\{0,1,2,\ldots,Kd_q-1\}\cup[Kd_q-1,\infty)$.
% \begin{align}
%     p_{\alpha,K}(S)&=\prod_q p_{\alpha_q,K}(S^{(q)}) \;, \\
%     p_{\alpha_q,K}(S^{(q)}) &= \mathcal{N} \det(S^{(q)})^{\alpha_q-K} e^{-TrS^{(q)}} \; ,\\
%     \mathcal{N} &= \pi^{\frac{K(K-1)}{2}} \Gamma(d_q) \Gamma(d_q-1) \cdots \Gamma(d_q-(K-1)) \; .
% \end{align}
% \footnote{We should note that more refined arguments in \cite{Marolf:2020xie} show that reflection positivity actually implies the stronger bound $\alpha_q>K-1$.}
Unfortunately, $\alpha_q$ takes values $\frac{\left|G\right|}{e^{S_0}d_q}N_q$ where $N_q$ is a nonnegative integer.
For $\alpha_q$ to always be in the set $\{0,1,2,\ldots,Kd_q-1\}\cup[Kd_q-1,\infty)$ would require that $\frac{\left|G\right|}{e^{S_0}d_q}$ be an integer.
This is not possible as it would render $\lambda'_q$ negative, which is nonsensical as $\lambda'_q$ is the mean of the Poisson distribution for $N_q$.
So we are left with the problem that the gravitational path integral correlators are not the moments of some probability distribution, meaning it cannot be dual to an ensemble of boundary theories.
This problem is worse than the one encountered earlier which necessitated identifying the boundary partition functions with rescaled $\h{Z}$ operators.
Rescaling the $\h{Z}$ or $\h{S}$ operators does not change the values that $\alpha_q$ takes.

One solution is to modify the theory by adding terms to the TQFT action proportional to the number of boundary components.
This would be for all three sorts of boundary components: circular EofW branes, closed sector fixed boundaries, and alternating EofW brane and open sector intervals.
Specifically we can add a term $S_q$ in the action for each boundary component labeled by $q$.
Retracing through the calculations of section \ref{openDWcalc} we see how an additional factor of $e^{S_q}$ for each boundary circle modifies the generating function of correlators.
First, considering the boundaries made up entirely of a single EofW brane component, we see that the connected vacuum correlator gets modified: $\lambda'_q=\lambda_qe^{K\left|G\right|e^{-S_0}}\rightarrow\lambda_qe^{K\left|G\right|e^{S_q-S_0}}$.
Second, a factor of $e^{S_q}$ for circle fixed boundaries gives $u_q\rightarrow e^{S_q}u_q$.
And finally, a factor of $e^{S_q}$ for circles made of alternating brane and fixed boundaries changes the factors $\tr\left(T_{(q)}^j\right)$ like $\tr\left(T_{(q)}^j\right)\rightarrow e^{S_q}\tr\left(T_{(q)}^j\right)$.
Making these substitutions in \ref{EofWDWgenfunccircles} leads to a modified generating function of
\begin{equation}
    F(\vec{u},T)
    =
    \prod_q
    \exp(
        \lambda'_q
        \left(
            \frac{e^{u}}{\det(\1-T_{(q)})}
        \right)^{\frac{e^{S_q}\left|G\right|}{d_qe^{S_0}}}
    )
    \;.
\end{equation}
Then with a the choice of $S_q=S_0+\ln d_q$ as in \eqref{Sq}, for example, the values of $\alpha_q$ are integers, ensuring the generating functional \eqref{conditionalgenfunc} is indeed the generating function for moments of a probability distribution.

The additional $S_q$ terms in the action can be viewed as the contributions of degrees of freedom that propagate along the boundaries.
This picture is odd, however, in that these degrees of freedom must propagate along both EofW brane \emph{and} fixed boundaries.
In this solution the ``fixed" boundary components are no longer fixed, but must include these dynamical degrees of freedom that propagate across them.
In some sense the solution requires us to not allow the full set of $\h{S}$ operators, which would otherwise allow us to completely fix the interval boundary conditions in the gravity path integral.
We discuss the problem in these terms in section \ref{reflectionpos}.

\subsection{General open/closed TQFTs}\label{generalopenclosed}
In this section we will describe the gravity path integral obtained by choosing our bulk theory to be a general open/closed TQFT (aka a general 2d TQFT with boundaries).
We'll see that this has the same features that we found above for the specific case of Dijkgraaf-Witten theory with end-of-the-world brane boundaries.

When discussing Dijkgraaf-Witten theory above, we considered only end-of-the-world brane boundaries.
These are, in the context of Dijkgraaf-Witten theory, the simplest sort of boundary compatible with gauge symmetry.
In general, we could consider more complicated boundaries, and in particular for theories without gauge symmetry, there is no condition of gauge symmetry that boundaries must be compatible with.
The only conditions that we need to require of our boundaries are that they be compatible with the possible ways of cutting and gluing spacetimes with boundaries.
For a review of open/closed TQFTs and the axioms that define them we refer interested readers to \cite{Carqueville_2018}.
Here we will simply state some of the results that apply to our case.

We'll use the index $a$ to label the different types of boundaries present in the theory.
For any two types $a$ and $b$ we have a corresponding open sector Hilbert space $\CH_{ab}$, which is the Hilbert space associated with an interval with $a$ boundary conditions on one end and $b$ boundary conditions on the other.
In addition to these open sectors we have the closed sector Hilbert space $\CH_{S^1}$, which is the Hilbert space of states on a circle.
The closed sector makes, in its own right, a 2d TQFT.
So, as described in section \ref{generalcase}, the partition function for the closed, connected manifold of genus $g$ has the form $\sum_I\mu_I^{2-2g}$,
for some positive, real numbers $\mu_I$, and accordingly, the closed sector Hilbert space $\CH_{S^1}$ has an orthonormal basis $\ket{I}$ labeled by the index $I$.

\begin{figure}[t]	
	\centering
	\begin{subfigure}[t]{0.3\linewidth}
		\centering
		\includegraphics[width=\linewidth]{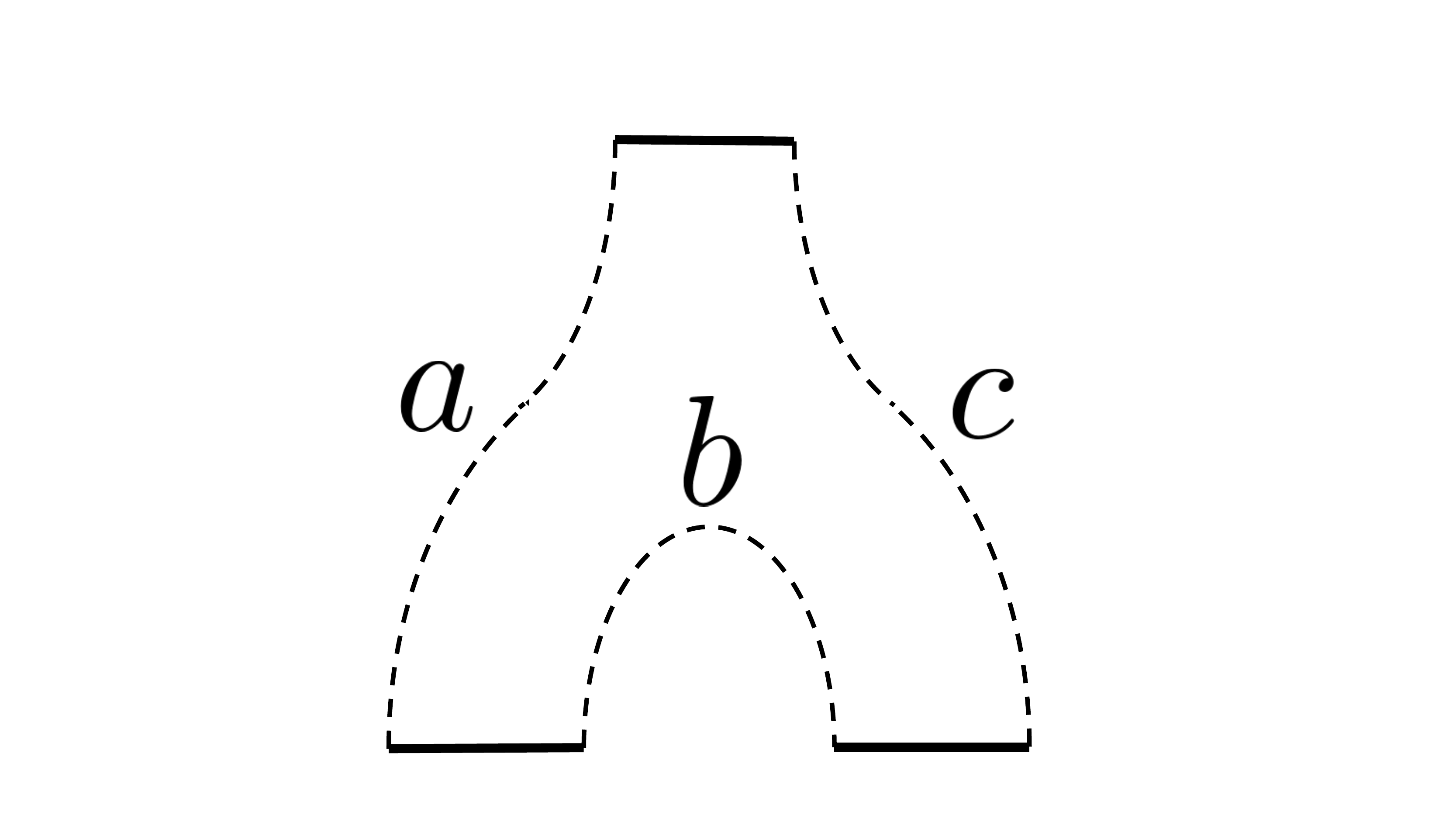}
		\caption{multiplication map}\label{fig:multiplication}		
	\end{subfigure}
	\quad
	\begin{subfigure}[t]{0.3\linewidth}
		\centering
		\includegraphics[width=\linewidth]{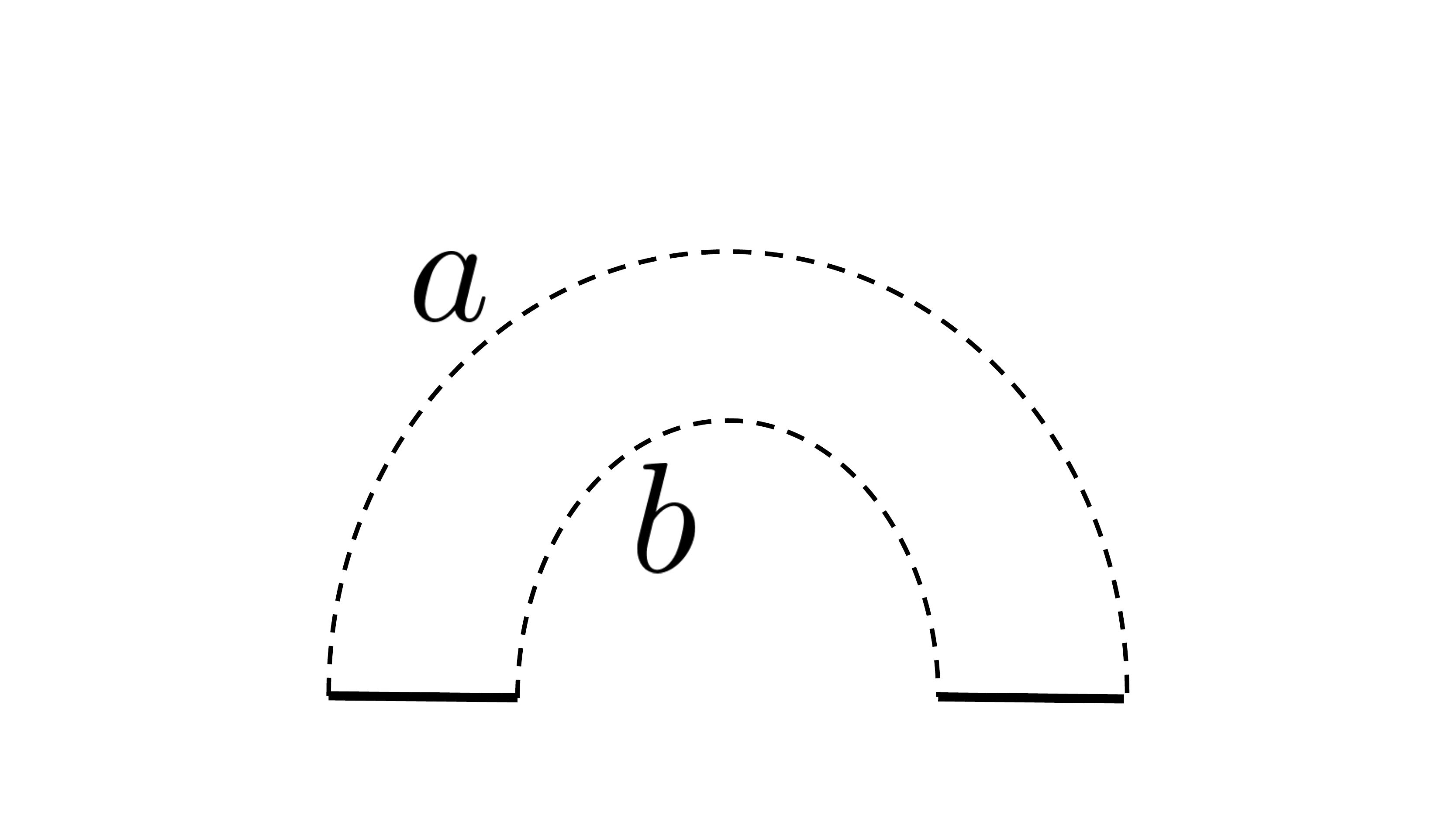}
		\caption{pairing}\label{fig:pairing}
	\end{subfigure}
	\begin{subfigure}[t]{0.3\linewidth}
		\centering
		\includegraphics[width=\linewidth]{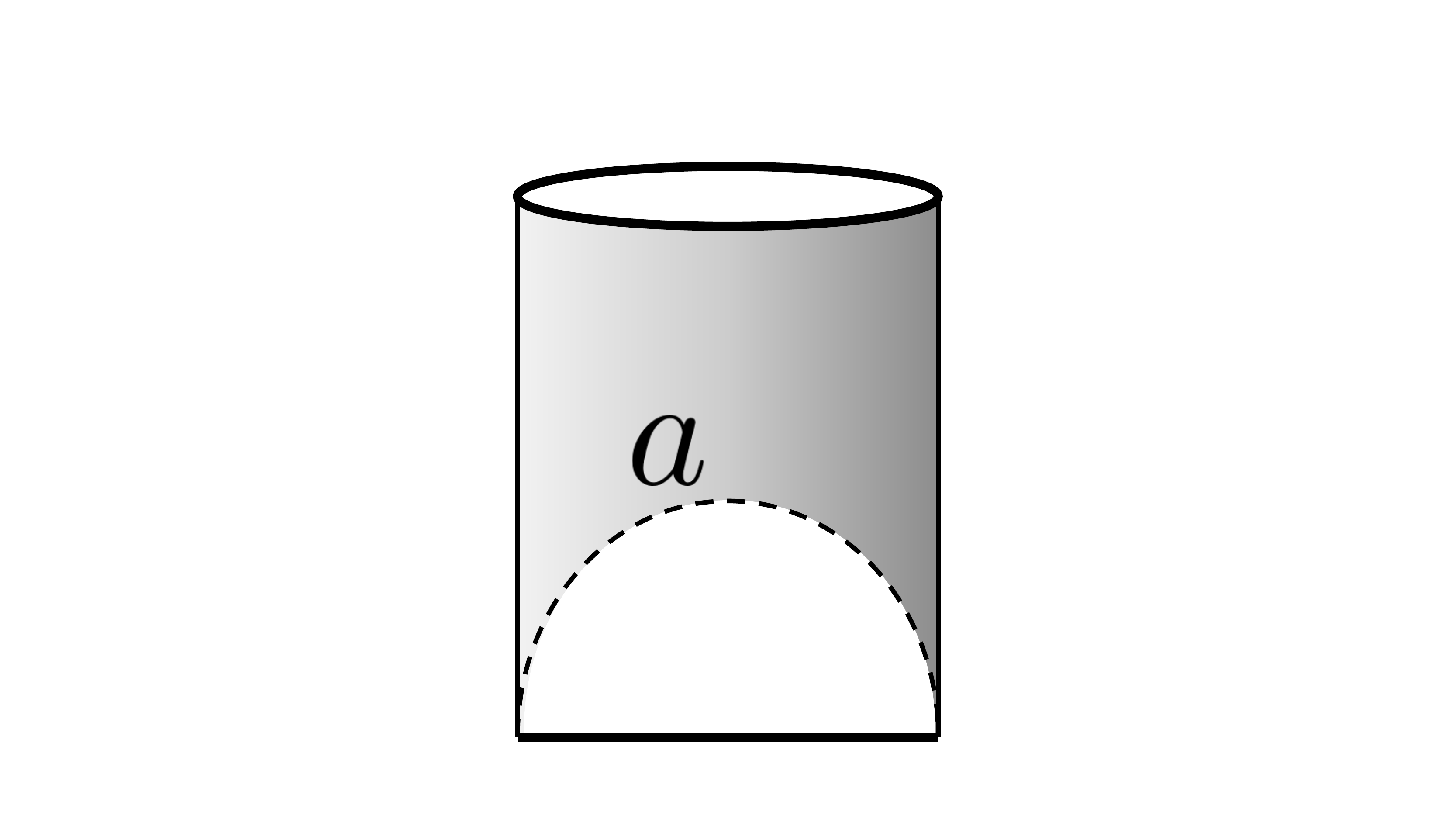}
		\caption{trace map}\label{fig:trace}
	\end{subfigure}
	\caption{
	    Three open/closed TQFT diagrams:
	    (a) A map from $\CH_{ab}\otimes\CH_{bc}$ to $\CH_{ac}$.
	    The open sector states can be represented as matrices such that this map implements matrix multiplication.
	    (b) The pairing between Hilbert spaces $\CH_{ab}$ and $\CH_{ba}$.
	    (c) The trace map, which maps $\CH_{aa}$ to $\CH_{S^1}$.
	    With the open sector states represented as matrices, this map implements the trace of these matrices.
	    }\label{fig:opencloseddiagrams}
\end{figure}

The interval Hilbert space $\CH_{ab}$ will have the form $\CH_{ab}\cong\bigoplus_I\left(\C^{d_{aI}}\otimes\C^{d_{bI}}\right)$, where $d_{aI}$ are integers.
In fact, we can think of the states $\ket{\psi}$ in $\CH_{ab}$ as being direct sums of matrices $\ket{\psi}=\bigoplus_I\Psi_I$ where the ``multiplication" diagrams as in figure \ref{fig:multiplication} are computed via matrix multiplication of the matrices $\Psi_I$.
(For example, the multiplication
$\vcenter{\hbox{\includegraphics[width=0.06\linewidth]{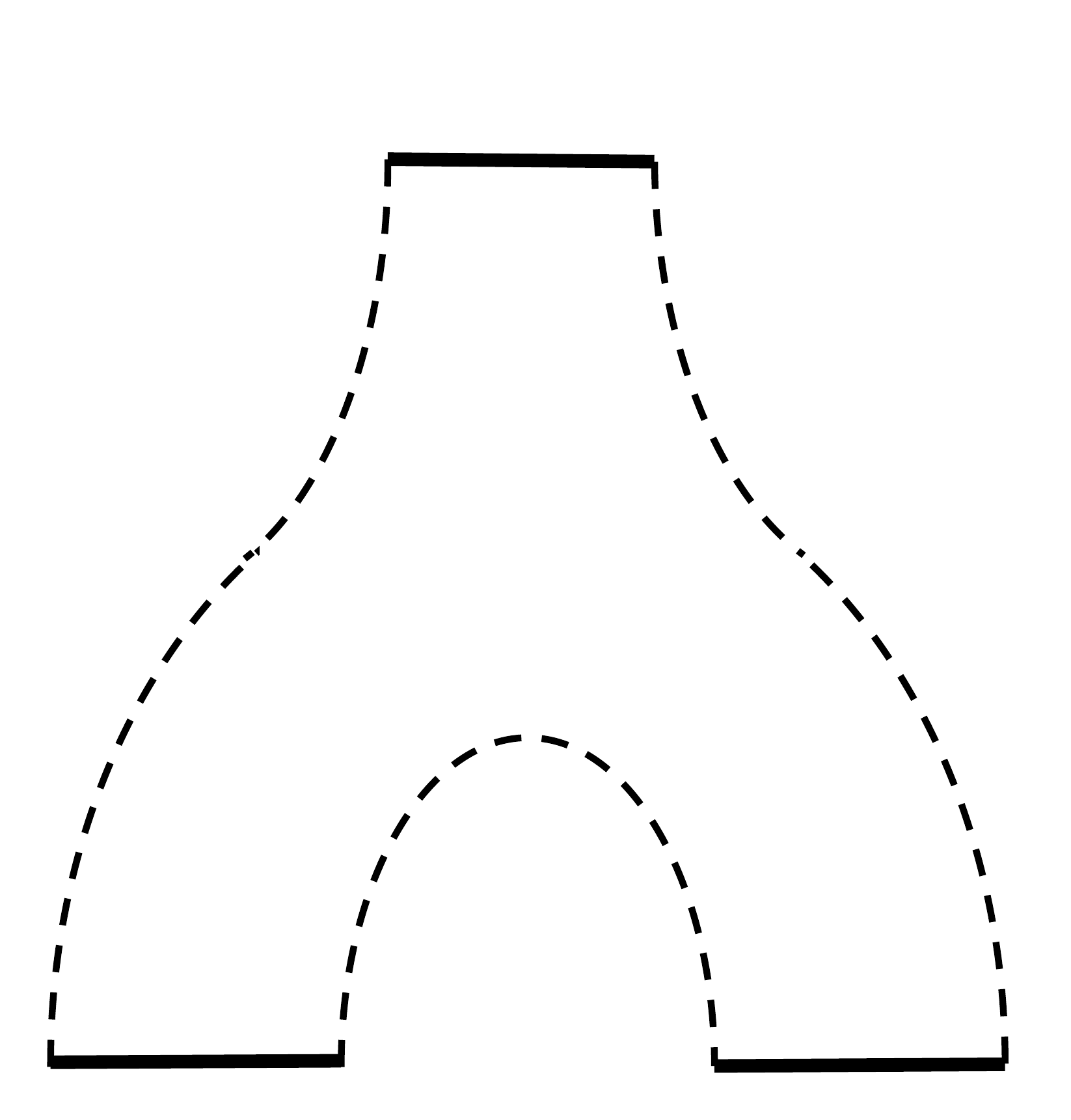}}}$
acting on states $\ket{\psi}\in\CH_{ab}$ and $\ket{\phi}\in\CH_{bc}$ produces a state in $\CH_{ac}$ described by $\bigoplus_I\Psi_I\Phi_I$.)
We can take as a basis for $\CH_{ab}$ the matrices with one entry $1$ and the rest zero.
Denote these basis states by $\ket{I;i,j}$ where $I$ labels the block and where indices $i=1,\ldots,d_{aI}$ and $j=1,\ldots,d_{bI}$ label the position of the nonzero entry.
(To make connection to Dijkgraaf-Witten theory with end-of-the-world branes, the label $I$ is in that case the label $q$ running over irreducible representations, the values $\mu_I$ are $e^{S_0}d_q/\left|G\right|$, and the dimensions $d_{aI}$ are simply $d_q$ for all boundaries.
The diagram \ref{DWmultiplication} can be seen to be describing matrix multiplication.)

There are two additional facts we make use of.
First, the partition function calculated by a strip with boundary type $a$ on one side and boundary type $b$ on the other (see figure \ref{fig:pairing}) induces a pairing between the Hilbert spaces $\CH_{ab}$ and $\CH_{ba}$.
To be entirely consistent with the cutting and gluing axioms defining an open/closed TQFT, this pairing must take the form
\begin{equation}
    \left(\ket{\psi},\ket{\phi}\right)
    =
    \sum_I
    \mu_I
    \tr(\Psi_I\Phi_I)
\end{equation}
where $\ket{\psi}\in\CH_{ab}$ and $\ket{\phi}\in\CH_{ba}$.
Second, we have the map
$\vcenter{\hbox{\includegraphics[width=0.04\linewidth]{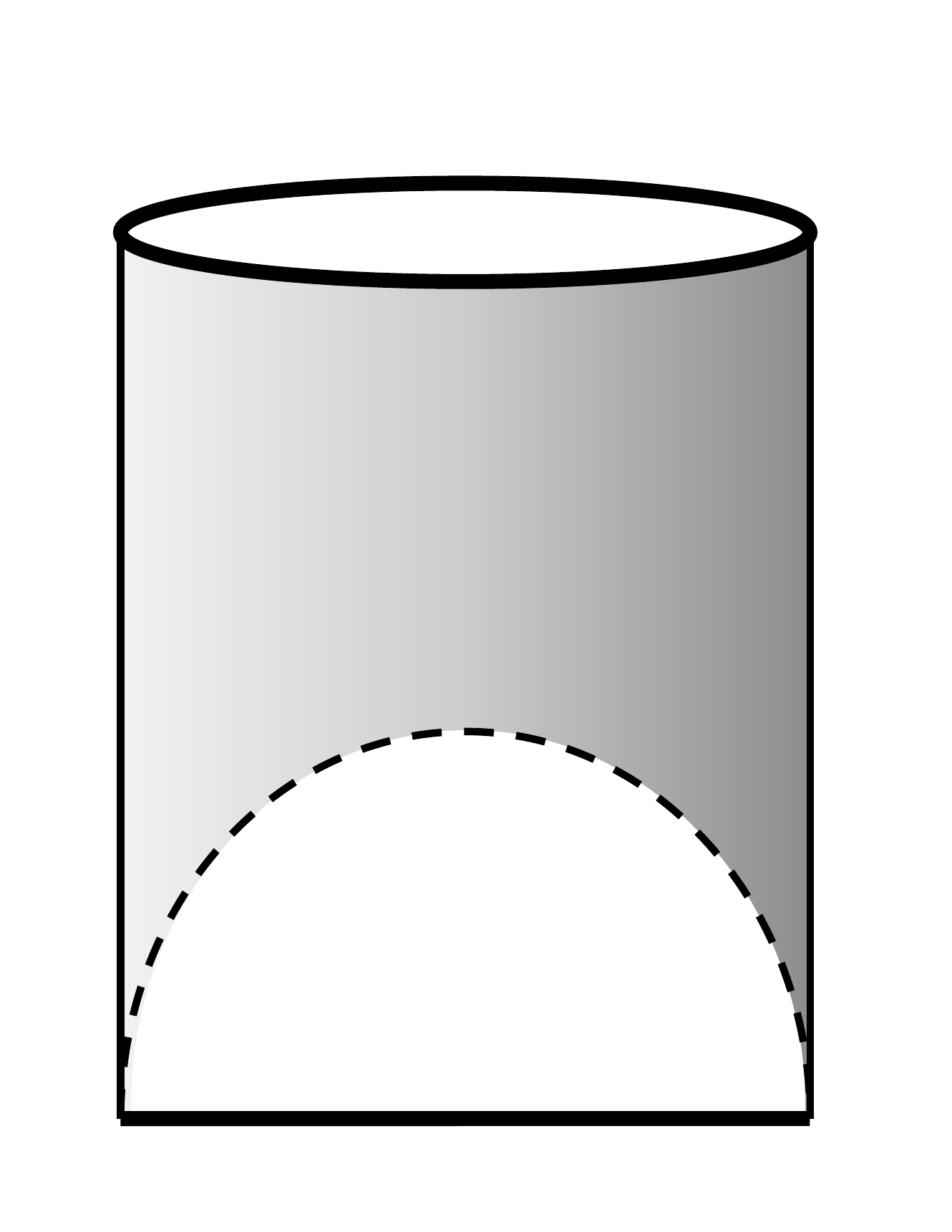}}}\!\!:\CH_{aa}\rightarrow\CH_{S^1}$
from any diagonal open sector $\CH_{aa}$ to the closed sector $\CH_{S^1}$.
This map simply implements the trace $\ket{\psi}\mapsto\sum_I\tr(\Psi_I)\ket{I}$.

With the multiplication, pairing, and trace diagrams in hand we can calculate the partition function of any manifold with boundaries, and from there define a gravity path integral as a subsequent sum over all compatible manifolds with boundaries.
The observables in this gravity theory are operators $\h{Z}_I$ that insert a circle with state $\ket{I}\in\CH_{S^1}$, and $\h{S}_{abIij}$ which insert the interval with state $\ket{I;i,j}\in\CH_{ab}$ on it.
The correlators are given by a sum over manifolds where the boundary configurations are compatible with the boundary labels of all the insertions $\h{S}_{abIij}$.
In our sum over manifolds $M$ we include the appropriate measure $\mu(M)$.
In addition to the factors for having multiple identical components without fixed boundaries, the measure $\mu(M)$ also includes a factor of $1/m_a!$ whenever a component of $M$ has $m_a$ circle boundaries with boundary conditions $a$ all around the circle.
These are due to residual diffeomorphisms that permute the identical dynamical boundaries.

The calculation of correlators is analogous to the Dijkgraaf-Witten case in section \ref{openDWcalc}.
We simply state the results.
The connected vacuum correlator is
\begin{equation}\label{openlambda}
    \lambda'
    \equiv
    \log\expval{\1}
    =
    \expval{\1}_{\text{connected}}
    =
    \sum_I
    \lambda_I
    e^{K_I/\mu_I}
    =
    \sum_I\lambda_I'\;,
\end{equation}
where $K_I=\sum_ad_{aI}$ count the boundary degrees of freedom, and the $\lambda_I=\mu_I^2/(1-\mu_I^{-2})$ are as defined in section \ref{generalcase} for the corresponding closed TQFT.
The generating function
\begin{equation}
F(u_I,t_{abIij})
=
\expval{
    e^{
        \sum_Iu_I\h{Z}_I+\sum_{ab}\sum_{I,i,j}t_{abIij}\h{S}_{abIij}
    }
}
\end{equation}
is given by
\begin{align}\label{openclosedgenfunc}
    F(u,t)
    =
    \prod_I
    e^{
        \lambda_I'
        \exp(
            \mu_I^{-1}u_I
            +
            \mu_I^{-1}
            \sum_{j=1}^{\infty}
            \frac{1}{j}\tr(T_{I}^j)
        )
    }
    =
    \prod_I
    \exp(
        \lambda_I'
        e^{u_I/\mu_I}
        \det(\1-T_I)^{-1/\mu_I}
    )
    \;,
\end{align}
where we have collected the chemical potentials $t_{abIij}$ into $K_I$ by $K_I$ matrices $T_I$.
This is of course the same form as \eqref{EofWDWgenfunc}.
Similar to the special case of Dijkgraaf-Witten with end-of-the-world branes, correlators will factorize between the $I$ sectors, and will also fail to correspond to the moments of an ensemble distribution without negative probabilities.
We discuss this failure and possible solutions in the next section.

\section{Boundaries and the ensemble problem}\label{reflectionpos}
In order to get a sensible holographic interpretation from a gravity path integral built from a closed 2d TQFT, we were forced to introduce in a seemingly ad hoc fashion a rescaling of the $\h{Z}_I$ operators.
What's more, for an open/closed theory, we find that rescalings of the operators $\h{Z}$ and $\h{S}$ are no longer enough to land on a sensible boundary interpretation.
The solution discussed in \cite{Marolf:2020xie} and reviewed here in section \ref{review} is to add a nonlocal boundary term $S_\p=S_0$ to the action.
The authors attempt to justify this in terms of a large number of additional degrees of freedom that are allowed to propogate on the boundary.
We discuss the analogous solution in our case and attempt to paint a clearer picture of how these degrees of freedom fit into a local framework.

The observables in our gravity theory are in one-to-one correspondence with states in a TQFT.
By definition, a TQFT is guaranteed to be compatible with cutting and gluing spacetime manifolds.
In particular, upon cutting a spacetime manifold the newly created boundary is what we will call a ``gluing boundary."
By this we mean this boundary corresponds to a Hilbert space of states and hence can be glued to a similar gluing boundary by appropriately summing over states on the two boundaries.
So far, the operators describing observables in the theory (the $\h{Z}$ and $\h{S}$ operators) are operators that insert such gluing boundaries in our gravitational spacetime.
This is suggestive.
Specifically, it suggests the possibility of gluing the boundaries of our gravity spacetime to nongravitational TQFT path integrals.
That is to say, we can consider spacetimes with designated nongravitational regions where topology is fixed, while the rest of the manifold has unspecified (i.e.\ summed over) topology.

\begin{figure}[t]
    \centering
    \[
        \vcenter{\hbox{\includegraphics[width=0.45\linewidth]{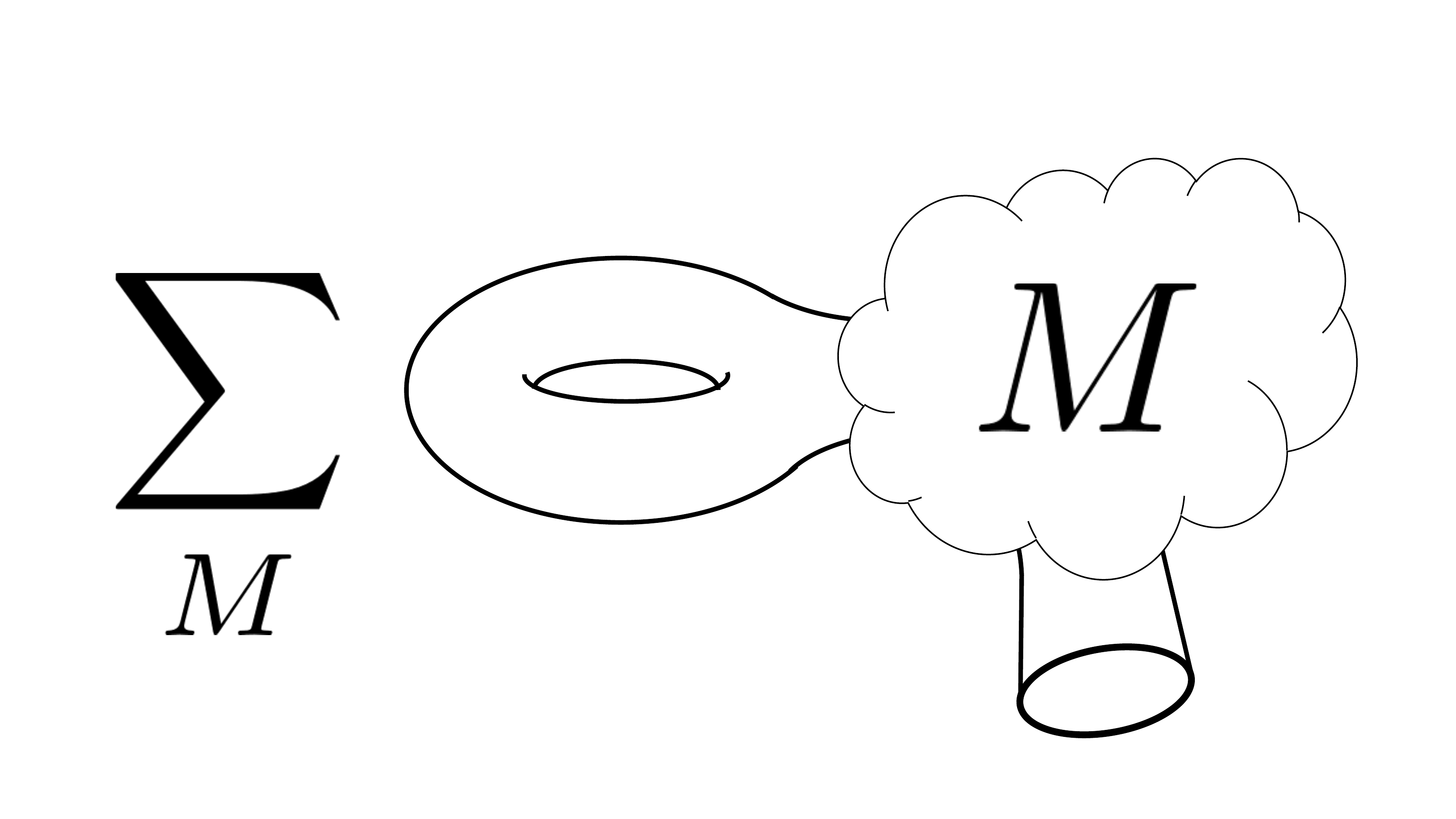}}}
        =
        \vcenter{\hbox{\includegraphics[width=0.45\linewidth]{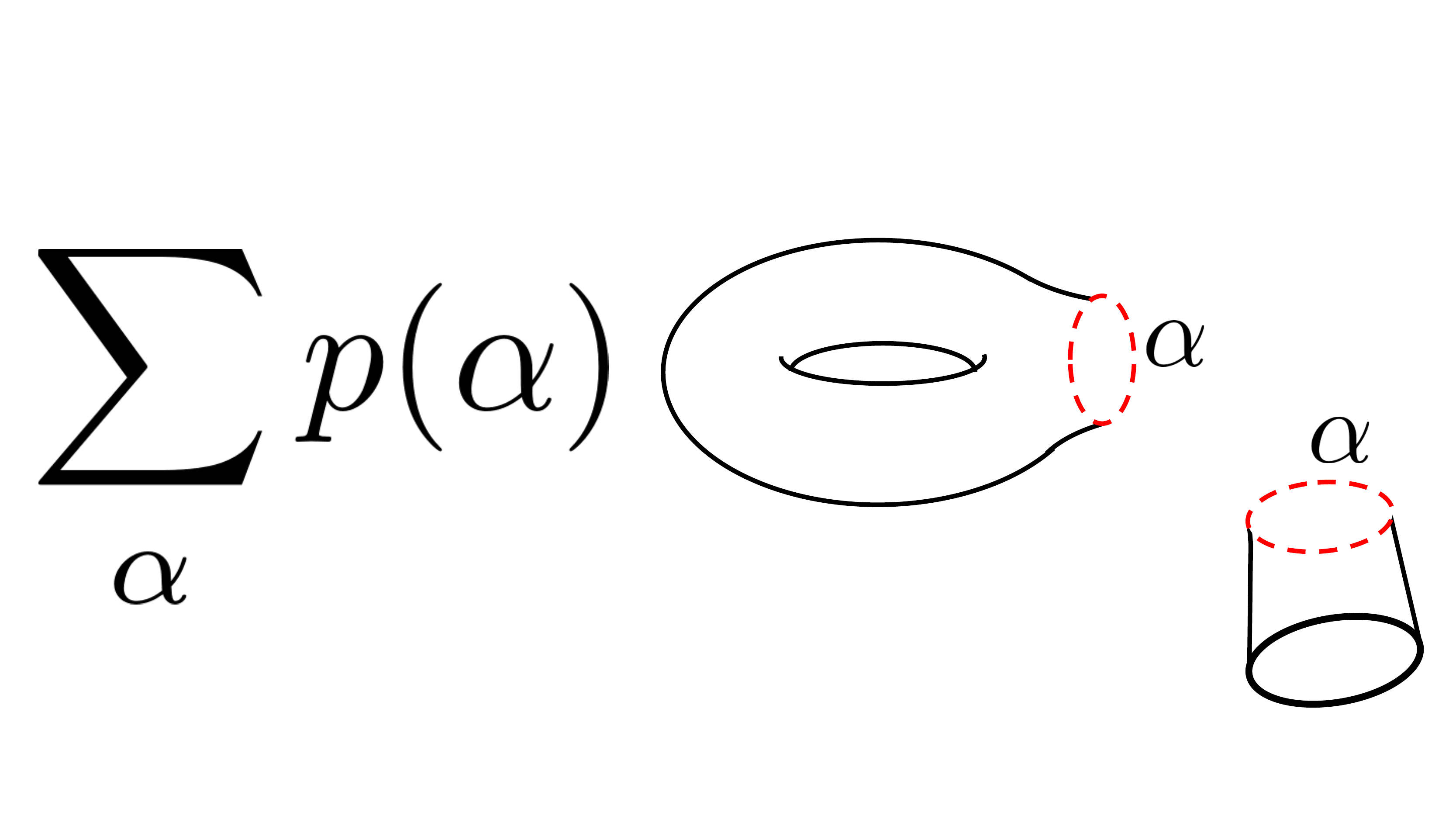}}}
    \]
    \caption{A gravitational region whose topology is summed over is holographically dual to an ensemble of boundary conditions.
    The above figure is a schematic illustration of a gravity correlator of nongravitational regions.
    The correlator is a sum over all manifolds that are compatible with the boundaries of the nongravitational regions.
    Equivalently, it is an average over different boundary conditions $\alpha$ (represented by dashed lines) that are placed on the boundaries of the nongravitational regions.}
    \label{fig:holography}
\end{figure}

Holography on such hybrid gravitational/nongravitational spacetimes will reduce the gravitational bulk region to a boundary condition on the nongravitational region.
Or, more generally, as in our case, the gravitational region will be dual to an \emph{ensemble} of boundary conditions.
Specifically, each alpha-state of the gravitational theory will correspond to a distinct boundary condition.
For each type of boundary $\alpha$, there are corresponding open sector Hilbert spaces for intervals that end on that boundary.
For example, in a theory that already has boundaries (for example, these could be end-of-the-world branes) labeled by the index $a$, we will get additional open sector Hilbert spaces like $\CH_{a\alpha}$ and $\CH_{\alpha a}$.
Even in a theory without additional boundaries, such as those considered in sections \ref{BU} and \ref{generalcase}, upon going to the holographic dual we will have an open/closed TQFT with boundary $\alpha$ and the open sector $\CH_{\alpha\alpha}$.

Taking gravity as a boundary condition provides a new perspective on the interpretational problems that necessitated rescaling the $\h{Z}$ operators and that led to negative probabilities when $\h{S}$ operators are added.
We can see how such problems arise in this framework.
Consider the correlator
$\expval{\h{\vcenter{\hbox{\includegraphics[width=0.075\linewidth]{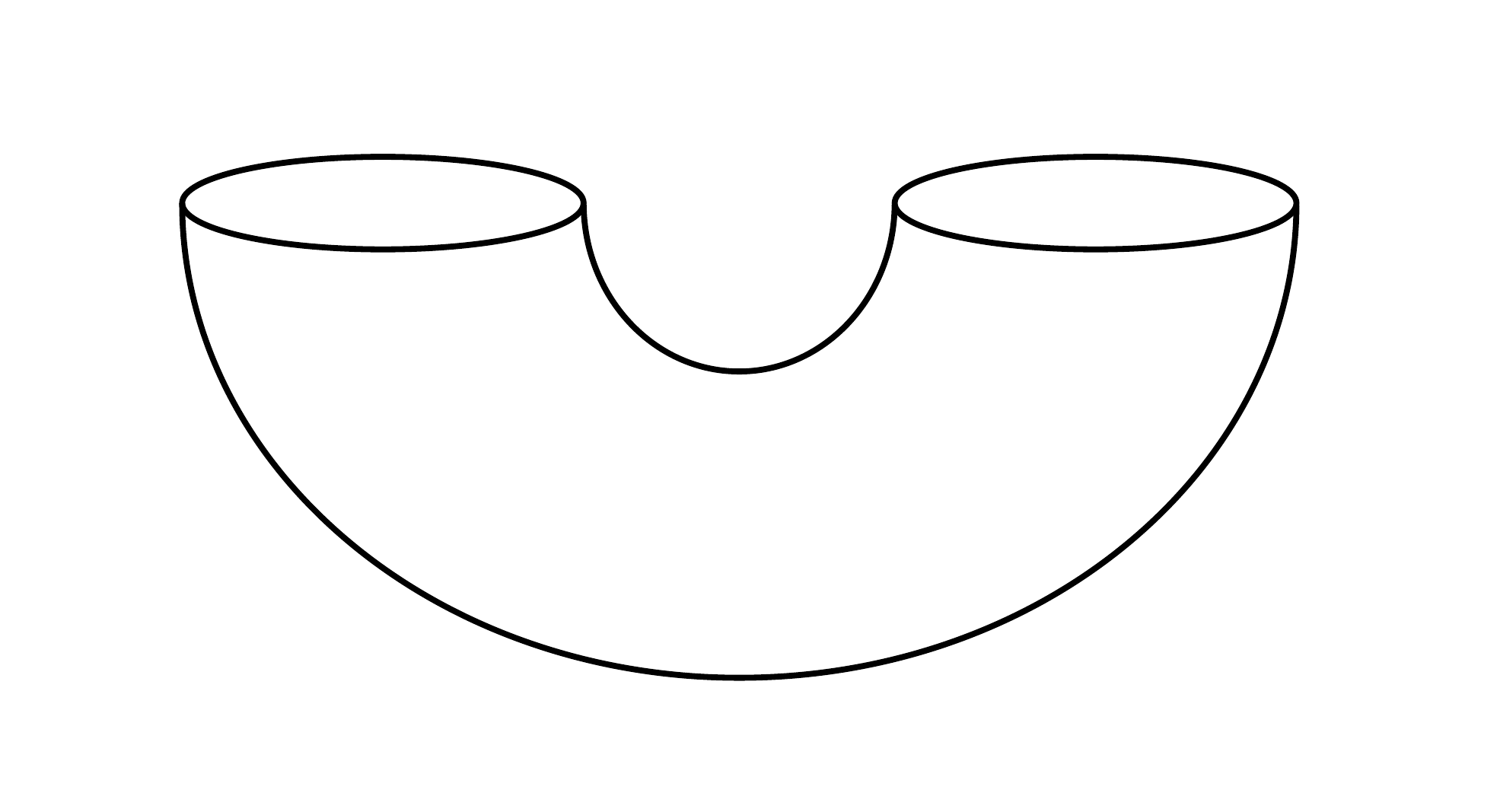}}}}}=\sum_I\expval{\h{Z}_I\h{Z}_I}$
, where $I$ labels the closed sector states of the TQFT.
This is dual to the annulus where both boundaries are the gravity boundary.
This diagram is a trace that computes the dimension of the open sector Hilbert space $\CH_{\alpha\alpha}$.
Consequently, it ought be an integer.
However, from the results of section \ref{generalcase},
\begin{equation}
    \sum_I\expval{\h{Z}_I\h{Z}_I}
    =
    \sum_{\ldots,N_I,\ldots}
    \prod_{I}\frac{\lambda_I^{N_I}}{N_I!e^{N_I}}
    \sum_I\frac{N_I}{\mu_I}\frac{N_I}{\mu_I}\;,
\end{equation}
meaning the annulus with gravity boundaries evaluates to $\sum_I\frac{N_I^2}{\mu_I^2}$ where $N_I$ is a Poisson random integer.
This is not in general an integer.
In fact, it cannot be an integer for all $N_I$ if we restrict $\mu_I$ to the values that are compatible with a convergent gravity path integral or with an ensemble without negative probabilities.
Likewise, if we start with a TQFT where boundary conditions are already present, similar problems arise for the ``mixed" open sectors $\CH_{\alpha a}$ and $\CH_{a\alpha}$, where $a$ here labels the original, nongravity, boundary conditions in the theory.
An annulus with one boundary of type $a$ and the other with gravity theory $\alpha$ on it, will evaluate to $\sum_I d_{aI}\frac{N_I}{\mu_I}$, where $d_{aI}$ are integers.
This, again, is not in general an integer.
This implies the absurbity that the Hilbert spaces $\CH_{\alpha\alpha}$ and $\CH_{a\alpha}$ have noninteger dimension.
One can equally view this as a breakdown of locality, in that cutting a gravity boundary $\alpha$ cannot be done consistently.

The perspective of gravity as dual to a boundary condition on a nongravitational TQFT also motivates an understanding of what the alpha-states for a theory with boundaries, like end-of-the-world branes, are.
Consider a state $\ket{I;i,j}$ in an open sector $\CH_{ab}$ that propogates from a nongravitational region into the gravitational region.
\begin{figure}[t]
    \centering
    \[
        \vcenter{\hbox{\includegraphics[width=0.475\linewidth]{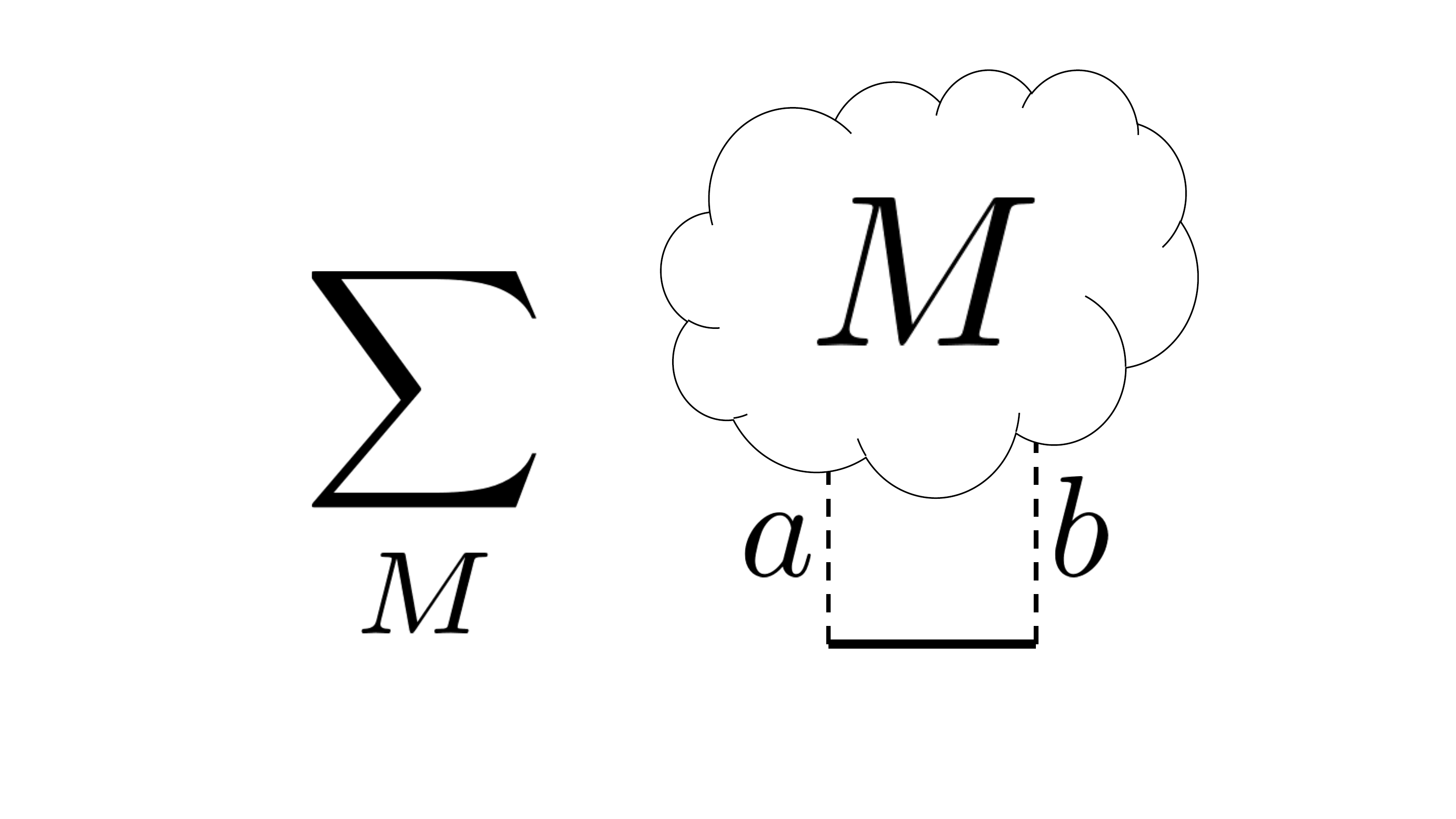}}}
        =
        \vcenter{\hbox{\includegraphics[width=0.475\linewidth]{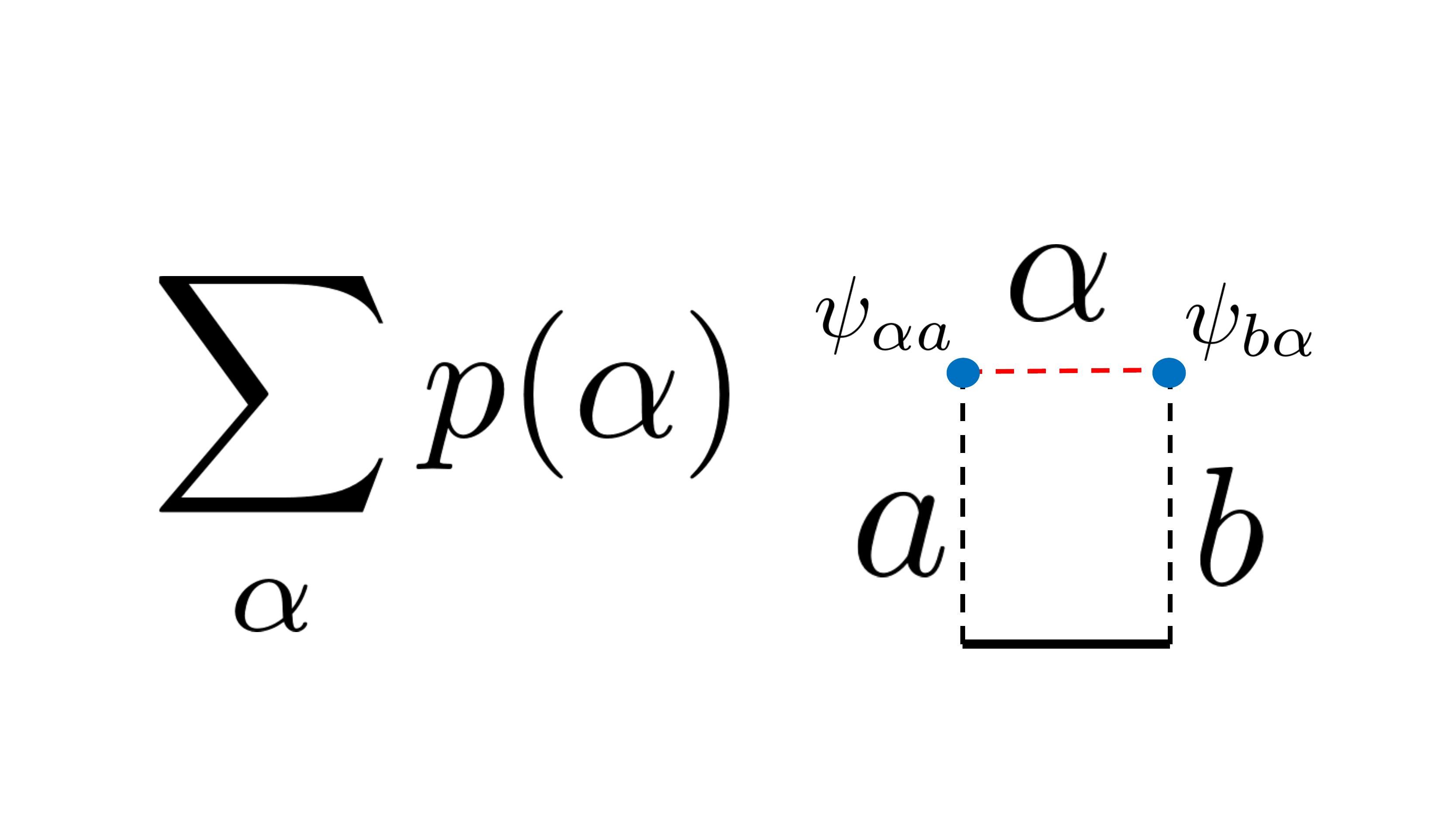}}}
    \]
    \caption{A strip with $a$ and $b$ boundary conditions that enters a gravitational region is dual to a strip that ends with gravity boundary conditions $\alpha$ interpolating between $a$ and $b$.
    At the junction between $\alpha$ and $a$ boundary conditions there must be a boundary condition changing operator, which we call $\psi_{\alpha a}$.
    Likewise for the junction between $b$ and $\alpha$ boundary conditions.}
    \label{fig:GPI_open_sector}
\end{figure}
This is holographically dual to a half-disk bounded by three segments: a segment with boundary conditions $a$, a segment with gravity boundary conditions $\alpha$, and a segment with boundary conditions $b$.
(See figure \ref{fig:GPI_open_sector}.)
At the two points on the edge of the half-disk where the boundary conditions switch, there are so-called boundary condition changing operators \cite{Cardy:2004hm}.
A boundary condition changing operator between boundaries of two types, say from $a$ to $b$, is a state in the open sector Hilbert space $\CH_{ab}$.
(This can be seen from considering the slice surrounding the point where the boundary conditions change.
As this slice ends on the boundaries with conditions $a$ and $b$, the Hilbert space associated to it is $\CH_{ab}$.)
So in this case of a half-disk with three different boundary types, there must be specified two states, in $\CH_{b\alpha}$ and $\CH_{\alpha a}$ respectively, which represent the boundary condition changing operators.
We take the specification of these states, call them $\ket{\psi_{b\alpha}}\in\CH_{b\alpha}$ and $\ket{\psi_{\alpha a}}\in\CH_{\alpha a}$, to be additional alpha parameters.
That is to say, fully specifying the alpha state, and hence the boundary theory, includes not just specifying the dimensions $d_{\alpha I}$, but also the boundary condition changing states $\ket{\psi_{a\alpha}}$ that must appear when switching from the gravity path integral to its holographic dual.\footnotemark
\footnotetext{Reflection positivity ensures that the state $\ket{\psi_{\alpha a}}$ switching $\alpha$ boundaries to $a$ boundaries and the state $\ket{\psi_{a\alpha}}$  in the other direction, are determined by each other.
Represented as matrices they are each other's Hermitian conjugates.}
Recalling from section \ref{generalopenclosed} that the states of an open sector $\CH_{ab}$ can be regarded as direct sums of $d_{aI}$ by $d_{bI}$ matrices, we can represent the states $\ket{\psi_{\alpha a}}$ as $\bigoplus_I\Psi_{(\alpha aI)}$.
We see that the probability distribution $p(\alpha)$ defining our ensemble should be over the nonnegative integers $d_{\alpha I}$, for each $I$, and the $d_{\alpha I}$ by $d_{aI}$ complex matrices $\Psi_{(\alpha aI)}$, for each choice of $a$ and $I$.

We see an immediate problem, however.
For the naive gravity path integral the dimensions $d_{\alpha I}$ are $d_{\alpha I}=N_I/\mu_I$, and hence, as explained above, not nonnegative integers for every $N_I\in\{0,1,2,\ldots\}$.
Defining the boundary theories in our ensemble then seems to entail choosing the elements of a matrix with noninteger dimension.
We saw that, for a TQFT without additional boundaries, the holographic dual already has the problem that the Hilbert space $\CH_{\alpha\alpha}$ does not exist. (It would have noninteger dimension.)
This can be viewed as a violation of locality, insofar as it forbids us from making cuts in our manifolds that intersect $\alpha$ boundaries.
With that restriction on making cuts in place, perhaps it might otherwise makes sense.
In the presence of additional boundaries $a$, however, the problems with the holographic dual theory become worse.
The inability to make cuts intersecting $\alpha$ boundaries prevents us from relating boundary condition changing operators to states $\ket{\psi_{\alpha a}}\in \CH_{\alpha a}$, as the Hilbert space $\CH_{\alpha a}$ does not exist.
Instead, while such states $\ket{\psi_{b\alpha}}$ and $\ket{\psi_{\alpha a}}$ don't exist, we \emph{do} have access to the state in $\CH_{ba}$ that would be their product under open sector multiplication described in section \ref{generalopenclosed} (see figure \ref{fig:multiplication}).
When represented as a matrix, the entries of this state in $\CH_{ba}$ are simply the values that the operators $\h{S}_{abIij}$ take.
As we have explained, the generating function \eqref{openclosedgenfunc}, by analogy with the arguments of \cite{Marolf:2020xie}, does not lead to a nonnegative probability distribution for the values $S_{abIij}$, for all values of $d_{\alpha I}=N_I/\mu_I$.
Thus, even with the locality-violating restriction of disallowing cuts that intersect gravity boundaries, we are left without a consistent boundary ensemble theory.

So should we give up hope of consistently viewing the gravity path integral as dual to an ensemble of boundary conditions?
In the case of the simple model with end-of-the-world branes considered in \cite{Marolf:2020xie}, they discuss, as a solution to the negative probabilities, adding $e^{S_0}$ degrees of freedom that propagate along the end-of-the-world branes. These degrees of freedom contribute to the existing bulk action a boundary term $S_{\p}=S_0$ for every boundary in the theory.
This effectively rescales the boundary insertion operators $\h{Z}$ and likewise ensures that $d$ in eq.\ \eqref{reviewgenfunc} is an integer.
Then the Fourier transform of eq.\ \eqref{reviewdistribution} does, indeed, give a valid probability distribution, namely the Wishart distribution.

This solution unavoidably has one of two problems, however.
On the one hand, if the degrees of freedom are taken to propagate only along the end-of-the-world branes, then the open sectors $\CH_{ab}$ consequently have $e^{2S_0}$ times as many states as they did before, and we have a corresponding $e^{2S_0}$ times as many interval insertion operators $\h{S}$.
As the problem of negative probabilities is, roughly speaking, the problem of having too many boundary flavors, the additional interval insertion operators $\h{S}$ reintroduce the problem that adding the $e^{S_0}$ degrees of freedom might have solved.
On the other hand, if we allow the additional boundary degrees of freedom to propagate along not just the branes, but also the ``gluing" boundaries, these boundaries lose their interpretation as gluing boundaries.
In other words, the bulk action is now nonlocal.

The perspective of a gravitational region being dual to boundary conditions on a nongravitational region clarifies what the problem is.
Adding, in our case, $\mu_I$ degrees of freedom to each boundary type $a$, implements the change $d_{aI}\rightarrow d'_{aI}=\mu_Id_{aI}$.
On its face, this seems to solve the problem of $\dim\CH_{a\alpha}$ being a noninteger, as $\sum_Id'_{aI}\frac{N_I}{\mu_I}=\sum_Id_{aI}N_I\in\Z$, but the new size of the open sector Hilbert space $\CH_{ba}$ is then inconsistent with a multiplication rule $\CH_{b\alpha}\otimes\CH_{\alpha a}\rightarrow\CH_{ba}$ that involves matrices with $d_{bI}N_I$ and $d_{aI}N_I$ number of entries.\footnotemark
\footnotetext{It also does not solve the problem of noninteger $\dim\CH_{\alpha\alpha}$.}
That is to say, there are no sizes that two matrices could have such that
\begin{itemize}
    \item the number of entries in each are $d_{bI}N_I$ and $d_{aI}N_I$ respectively, and
    \item their matrix product has $\mu_Id_{bI}\mu_Id_{aI}$ entries,
\end{itemize}
other than their inner dimensions being $N_I/\mu_I$, which is not an integer for all $N_I$.
(This makes precise the statement we made above that the problem is in some sense having too many flavors of boundary.)

The difficulties with the boundary interpretation all ultimately stem from the fact that the $d_{\alpha I}=N_I/\mu_I$ are not integers for all $N_I$.
This suggests that a solution should be something that adds $\mu_I$ additional degrees of freedom to the gravity boundary $\alpha$, rather than merely adding degrees of freedom to the original boundaries $a$.
If we substitute $d_{\alpha I}\rightarrow d'_{\alpha I}=\mu_Id_{\alpha I}$, the Hilbert spaces $\CH_{\alpha\alpha}$, $\CH_{a\alpha}$, etc.\ are all defined as well as the boundary condition changing operators.
With $\CH_{a\alpha}$ well-defined, locality would then imply that the $S_{abI}$ matrices factorize as $S_{abI}=\Psi_{\alpha aI}^\dag\Psi_{\alpha bI}$.
Such a solution would presumably cure the problem of negative probabilities for the values of $S_{abIij}$.

\begin{figure}[t]
    \centering
    \includegraphics[width=0.75\linewidth]{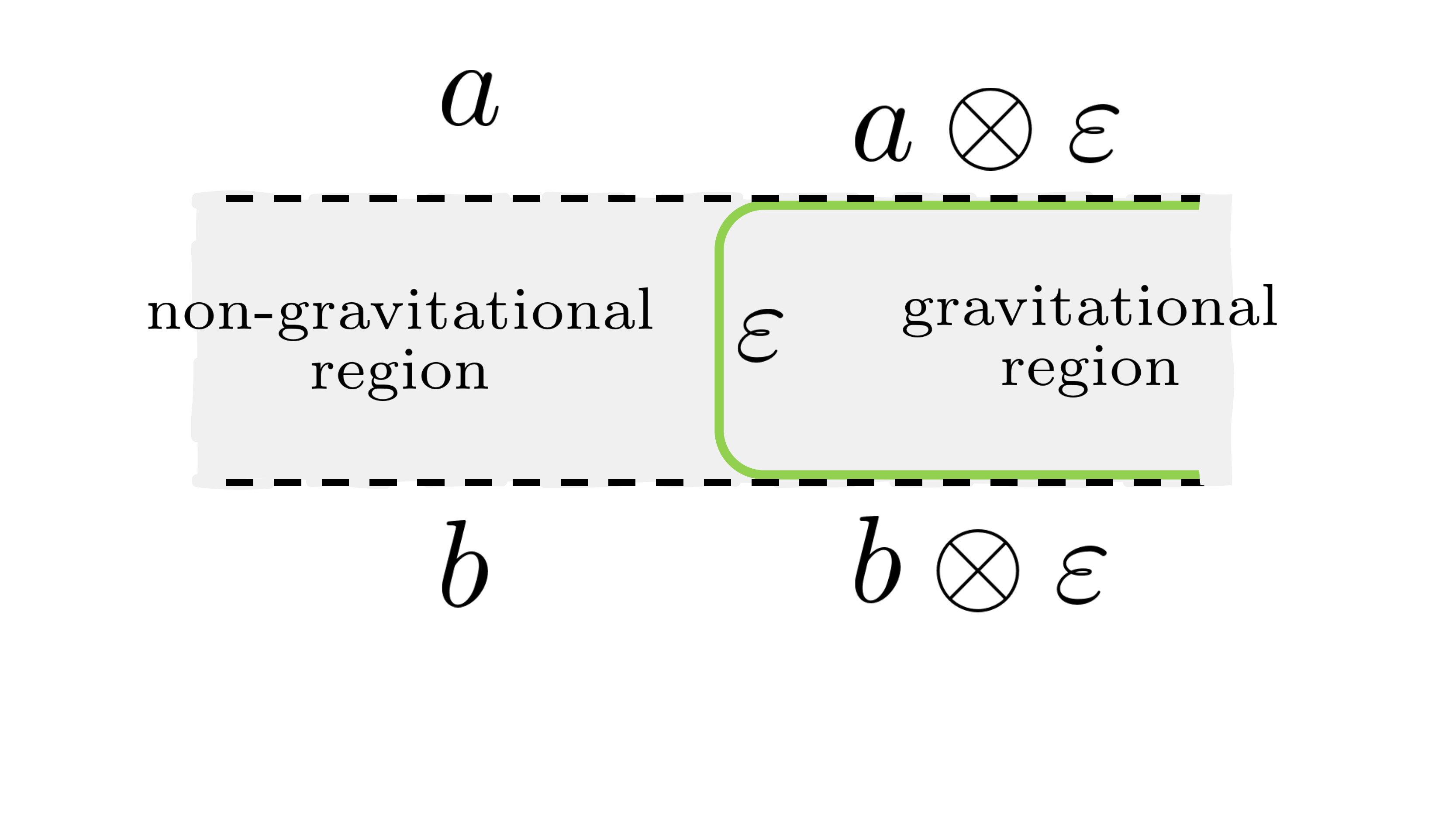}
    \caption{The gravitational and non-gravitational regions are separated by the defect.
    At the junctions where, for example, $a$, $\varepsilon$, and $a\otimes\varepsilon$ meet, we have the degrees of freedom of $a$ and the degrees of freedom of $\varepsilon$ propagate in the obvious way.}
    \label{fig:open_sector_defect}
\end{figure}

\begin{figure}[t]
    \centering
    \[
        \vcenter{\hbox{\includegraphics[width=0.45\linewidth]{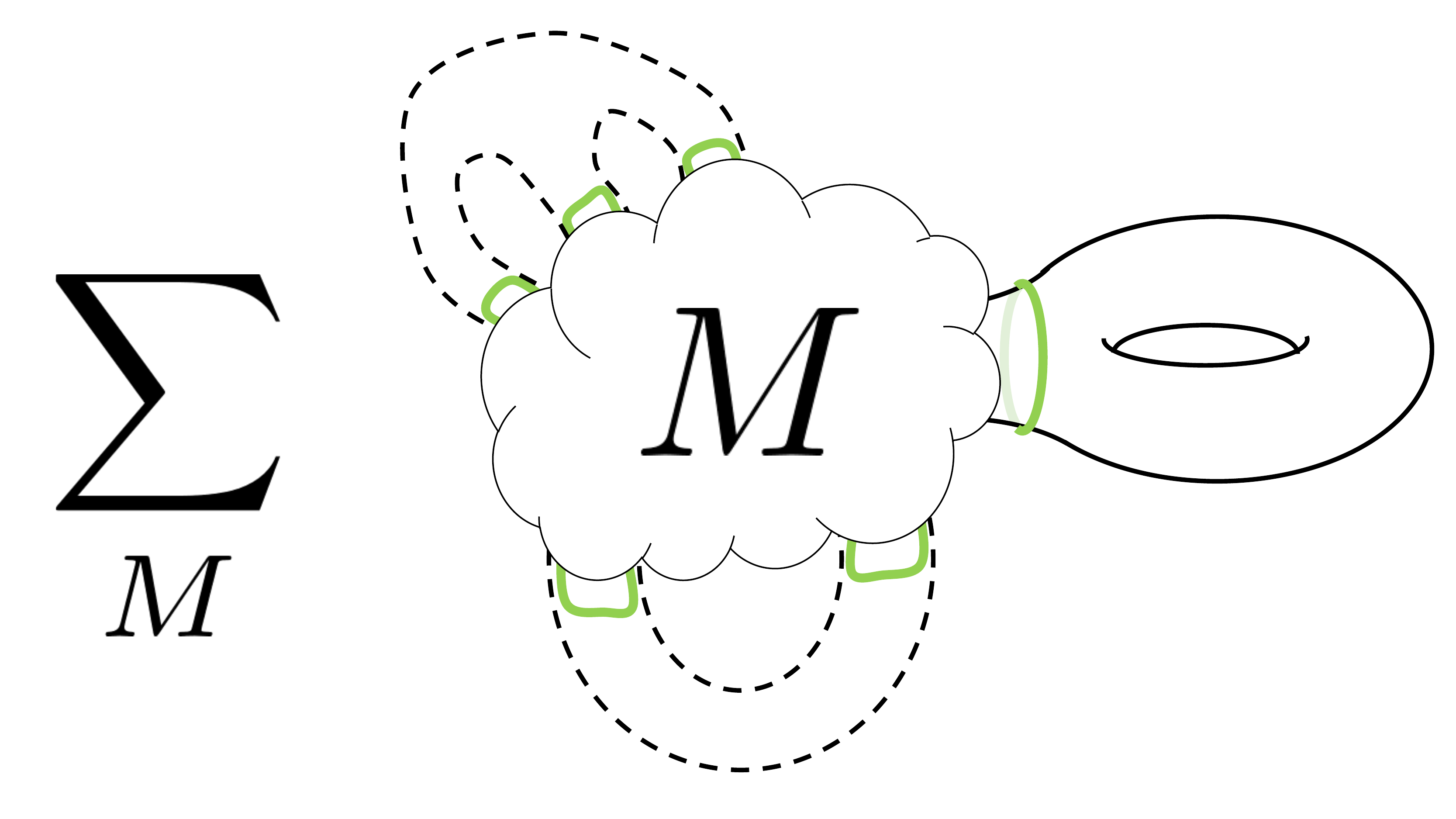}}}
        =
        \vcenter{\hbox{\includegraphics[width=0.5\linewidth]{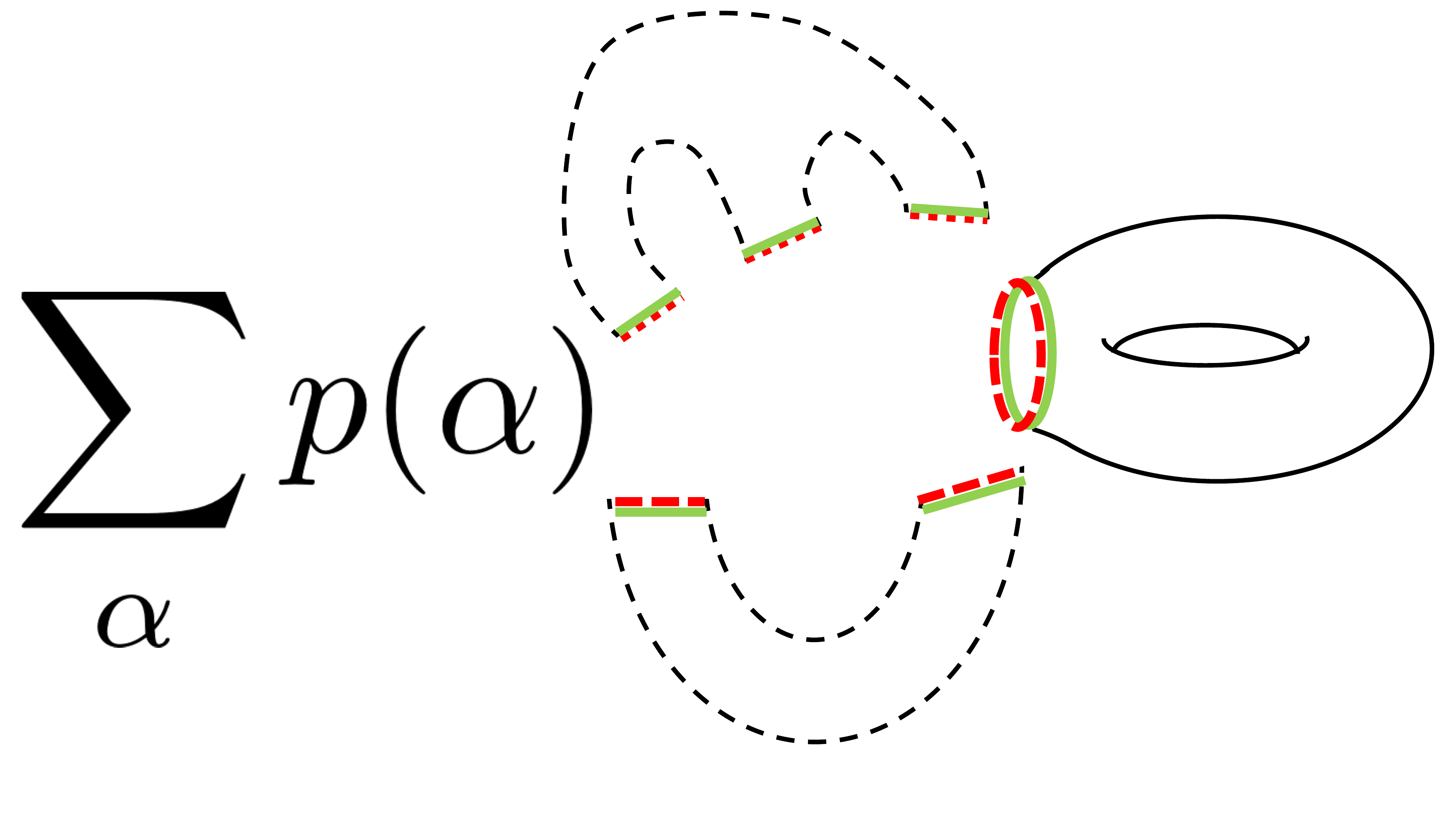}}}
    \]
    \caption{The gravity path integral where a defect line $\varepsilon$ (green solid lines) separates the gravitational and nongravitational regions, and the boundaries within the gravity region now have boundary conditions $a\otimes\varepsilon$. This gravity region is holographically dual to an ensemble of boundary conditions $\alpha$ (green lines with red dashed lines) that are enhanced with the $\mu_I$ degrees of freedom from the defect, so that $d_{\alpha I}=N_I$.}
    \label{fig:GPI_holography_defect}
\end{figure}

One solution fitting these requirements involves adding a defect line separating the gravitational and nongravitational regions in our gravity path integral.
We let this defect line, call it $\varepsilon$, have $\sum_I\mu_I$ degrees of freedom, and we take it to be coupled to the 2d TQFT so that there are $\mu_I$ degrees of freedom in each $I$ sector.
To explain what we mean by this, consider that with the addition of such a defect line our TQFT has additional open sector Hilbert spaces, namely those corresponding to intervals that cross the defect some number of times.
If the interval crosses the defect $\varepsilon$ once, the corresponding Hilbert space, call it $\CH_{a\varepsilon b}$ will be $\CH_{a\varepsilon b}\cong\bigoplus_I\left(\C^{d_{aI}}\otimes\C^{\mu_I}\otimes\C^{d_{bI}}\right)$.
In fact, each $I$ sector of the Hilbert space will have an additional tensor factor of $\C^{\mu_I}$ for each time the interval crosses a defect line $\varepsilon$.
Likewise, there are additional closed sector Hilbert spaces, for circles that intersect the defect line some number of times.
The simplest way to define a gravity path integral with the addition of such a defect line is to consider surfaces where the defect line is placed surrounding the gravity region, so that it runs along brane boundaries within the gravity region and along the interface between the gravity and nongravity regions.
To be precise, with the addition of the defect, a boundary $a$ within the gravity region gets replaced by the combined boundary and defect $a\otimes\varepsilon$, and at a triple junction of $a$, $\varepsilon$, and $a\otimes\varepsilon$
we contract the degrees of freedom in the natural way.
(See figure \ref{fig:open_sector_defect}.)
Calculating the resulting correlators of $\h{Z}_I$ and $\h{S}_{abIij}$ is straightforward.
For every boundary component, whether made by the $\h{Z}$, or a circle with boundary conditions $a$, or made from a combination of boundaries $a$ and interval operators $\h{S}$, the TQFT action now has an additional factor of $\mu_I$ for each $I$ sector.
The result is that \eqref{openclosedgenfunc} is modified to
\begin{align}\label{modifiedopenclosedgenfunc}
    F(u,t)
    =
    \prod_I
    e^{
        \lambda_I e^{K_I}
        \exp(
            u_I
            +
            \sum_{j=1}^{\infty}
            \frac{1}{j}\tr(T_{I}^j)
        )
    }
    =
    \prod_I
    \exp(
        \lambda_Ie^{K_I}
        e^{u_I}
        \det(\1-T_I)^{-1}
    )
    \;,
\end{align}
This is not hard to see: each $\h{Z}_I$ boundary component comes with an additional factor $\mu_I$, effectively giving $u_I\rightarrow\mu_Iu_I$; circles with boundary conditions $a$ gets an additional factor $\mu_I$ for each $I$ sector, giving $\lambda_Ie^{K_I/\mu_I}\rightarrow\lambda_Ie^{K_I}$; and finally, each boundary component made of $j$ connected $\h{S}_I$ intervals gets a factor of $\mu_I$, giving $\tr(T^j)\rightarrow\mu_I\tr(T^j)$.
This modified generating function for the correlators implies an ensemble of theories where $Z_I=d_{\alpha I}=N_I$ are random integers $N_I$, independently chosen from Poisson distributions with respective means $\lambda_Ie^{K_I}$.
Within the space of theories with given $Z_I=N_I$, the remaining probability distribution over the values of $S_{abIij}$ has the generating function $\det(\1-T_I)^{-N_I}$ for its moments.
This is the generating function for the moments of the Wishart distribution \cite{graczyk2003}.
Recognizing this, we can write it suggestively as
\begin{equation}
    \det(\1-T_I)^{-N_I}
    =
    \pi^{-K_IN_I}\int\!
    \prod_{a}
    \left(
        d\Psi_{\alpha aI}d\b{\Psi}_{\alpha aI}
        e^{-\tr(\Psi_{\alpha aI}^\dag\Psi_{\alpha aI})}
    \right)\,
    e^{
        \sum_{a,b}
        \tr(
            T_{abI}\Psi_{\alpha bI}^\dag\Psi_{\alpha aI}
        )
    }\;,
\end{equation}
where by $d\Psi_{\alpha aI}d\b{\Psi}_{\alpha aI}$ we mean integration over the $2d_{aI}N_I$-dimensional space of complex $N_I$ by $d_{aI}$ matrices $\Psi_{\alpha aI}$, and where $T_{abI}$ is the $d_{aI}$ by $d_{bI}$ matrix with entries $t_{abIij}$.
We can see immediately that the operator $\h{S}_{abI}$ takes the matrix value $S_{abI}=\Psi_{\alpha bI}^\dag\Psi_{\alpha aI}$ in a given alpha-state.
This is precisely consistent with the holographic picture of gravity as a boundary condition.
In that case locality implies that the matrix $S_{ab}$ should indeed factorize into the matrix multiplication of two boundary condition changing states, namely $\ket{\psi_{b\alpha}}=\bigoplus_I\Psi_{\alpha bI}^\dag$ and $\ket{\psi_{\alpha a}}=\bigoplus_I\Psi_{\alpha aI}$.
Thus we can replace the gravity region's fluctuating topology with an ensemble of boundary conditions $\alpha$, as illustrated in the example of figure \ref{fig:GPI_holography_defect}.
The $\alpha$ boundary conditions are, as expected completely characterized by the dimensions $d_{\alpha I}$ and the boundary condition changing states $\ket{\psi_{\alpha a}}$, where $d_{\alpha I}$ are drawn from Poisson distributions and the entries in the matrix representation of $\ket{\psi_{\alpha a}}$ are independent complex Gaussian random variables. 

There are many other ways we could choose to configure defect lines, and there may certainly be others that lead to a well-defined ensemble of boundary conditions.
For example, instead of the simple junctions pictured in figure \ref{fig:open_sector_defect}, we could allow the degrees of freedom to mix or to end at the junction.
Such different setups would lead to the inclusion of additional operators in the traces $\tr(T_I^j)$ in the double exponent of \eqref{openclosedgenfunc}.
The setup considered above is simply the most straightforward option, and it does, in fact, cure the problem of negative ensemble probabilities.
We speculate that in a TQFT that descends from a realistic gravity theory, a defect line separating gravitational and nongravitational regions may descend from the data that specify how the geometries of the gravity and nongravity regions are consistently glued together.
For example, in the construction of \cite{Almheiri:2019psf} and subsequent papers, wherein JT gravity with CFT matter has its boundary glued to a flat region without gravity,
the ``boundary graviton" mode of JT gravity appears as a reparametrization of the boundary, determining how the JT gravity region is glued to the non-gravitational flat region.

The alpha-states $\ket{\alpha}$, defined similarly to \eqref{alpha}, satisfy $\braket{\alpha'}{\alpha}\sim\delta_{\alpha',\alpha}p(\alpha)$.
The failure to obtain a well-defined ensemble with nonnegative probabilities is thus equivalent to the presence of negative norm states in the baby universe Hilbert space.
One possible solution, then, to the problem discussed in this section would be to simply project out the negative norm states.
The baby universe Hilbert space can be constructed by acting on $\ket{\text{HH}}$ with the single-boundary operators $\h{Z}$ and $\h{S}$.
A projection on the space of single-boundary operators would thus induce a projection on the baby universe Hilbert space.
It is possible for the states projected out by such an operation to include the offending negative-norm states.
In fact, the gravity model with defect separating gravity and nongravity regions is an example of precisely this.
The diagram pictured in figure \ref{fig:open_sector_defect}, when viewed from right to left, constitutes a projection.\footnotemark
\footnotetext{The analogous map on the closed sector Hilbert space $\CH_{S^1}$ does not project out any states. It is simply a rescaling, though, as explained above, a crucial one for the interpretation of $\h{Z}$ as a partition function in a 1d topological theory or as boundary conditions in a non-gravitational TQFT.}
Namely the large Hilbert space $\CH_{a\otimes\varepsilon,b\otimes\varepsilon}$ is mapped to the smaller Hilbert space $\CH_{ab}$.
The observables that non-gravitational observers have access to only include those built from the $\dim\CH_{ab}$ operators $\h{S}_{abIij}$, rather than the much larger number of operators that correspond to states in the $\CH_{a\otimes\varepsilon,b\otimes\varepsilon}$ Hilbert space.
Specifically, it is the defect degrees of freedom that are inaccessible.

\section{Future directions} \label{discuss}
In our work we extended the 2d topological gravity model of \cite{Marolf:2020xie} to  a broader class of topological actions.
The holographic duals of these gravity models are ensembles of 1d topological theories with random dimension.
This is, in retrospect, not terribly surprising, as all 2d TQFTs are in a sense direct sums of the simplest TQFT, whose Hilbert space is one-dimensional and whose action is proportional to the Euler characteristic, like in \cite{Marolf:2020xie}.

Perhaps the most obvious limitation of the present work, then, is our restriction to TQFTs as defined by Atiyah's axioms.
In particular, TQFTs satisfying Atiyah's axioms are always finite dimensional, so this restriction rules out many TQFTs of physical interest.
These include, such TQFTs as the A- and B-models of topological string theory (both examples of the broader class of ``topological conformal field theories").
Likewise, JT gravity has a description as a modified BF theory with gauge group $\operatorname{SL}(2,\R)$ \cite{Blommaert:2018iqz}, an infinite dimensional topological field theory.

Also in the spirit of working towards more realistic theories would be the extension to higher dimensional spacetimes.
Several recent works attempt to make connections between a 3d bulk gravity and an ensemble of 2d CFTs on the boundary \cite{Afkhami-Jeddi:2020ezh,maloney2020averaging,cotler2020ads3,Cotler:2020hgz}.
Making sense of gravity path integrals in three dimensions, however, runs up against the difficulty that the equivalent of the genus expansion, for 3d manifolds, is not so well-behaved.
Just as any 2d closed, connected, oriented manifold is the connected sum of some number of tori, any 3d closed, connected, oriented manifold can be uniquely written as the connected sum of so-called ``prime" manifolds.
In contrast to the simplicity of the genus expansion in 2d, the prime manifolds are infinite in number, and not completely and uniquely classified.

Besides the above extensions, there is, of course, always the possibility of considering models of surfaces with more complicated structures like defects, foliations, or (as in \cite{Balasubramanian:2020jhl}) spin structures.

Finally, the coupling of gravitational regions to non-gravitational TQFT regions opens other avenues for further study.
We find especially interesting the question of whether a version of the black hole information paradox can be phrased in this framework, perhaps analogously to the construction in \cite{Almheiri_2020}, or in some different way.
Also of interest in this framework is the question of bulk reconstruction.
For example, one could consider TQFT operators in the bulk gravity region that are appropriately ``gravitationally dressed" so as to be well-defined for spacetimes without fixed topology.
(A simple example of ``gravitational dressing" would be to specify a fixed boundary component that the observable must remain path connected to when we allow topology to fluctuate.)
Then the question arises of whether and how such bulk operators can be represented once we switch to the dual picture where gravity is an ensemble of boundary conditions.

\section*{Acknowledgements}
We wish to thank Per Kraus and Thomas Dumitrescu for helpful discussions. We also thank the Mani L. Bhaumik Institute for Theoretical Physics for its support. SM is grateful to the Alexander S. Onassis Foundation for its support.

\appendix
\section{State sum formulation of Dijkgraaf-Witten theory}\label{DWappendix}
Dijkgraaf-Witten theory can be equivalently formulated as a lattice gauge theory on a triangulated manifold.
The outputs of the theory formulated this way are so-called state-sum expressions for the partition function.
In \cite{doughterty20152dimensional}, this approach was generalized to DW theory with defects.
Their methods can be straightforwardly generalized to provide an independent method of obtaining our open/closed DW rules \eqref{r1}-\eqref{r5}, and at the same time provide directions for future work.

\subsection{Review of state sum for DW with defects}
Given a surface-curve pair $\Sigma \supset  C $, we can define a refine notion of triangulation as follows. 

\begin{figure}[t]
	\centering
	\includegraphics[page=2,scale=1]{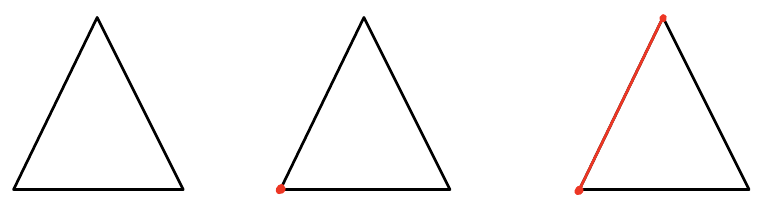}
	\caption{Basic building blocks of a flag-like traingulation.}
	\label{flag}
\end{figure}

\begin{definition}
	A triangulation $T$ of a surface with curve $\Sigma \supset  C $ is flag-like if C is a subcomplex, and if, for every 2-simplex $\sigma \in T$, the intersection of $\sigma$ and C is either a face (i.e.\ one vertex or one edge) or is empty.
\end{definition}
In other words a flag-like trangulation has to have the curve lie along faces, and only use triangles like the ones in figure \ref{flag}.

\begin{figure}[t]
	\centering
	\includegraphics[page=1,scale=0.85]{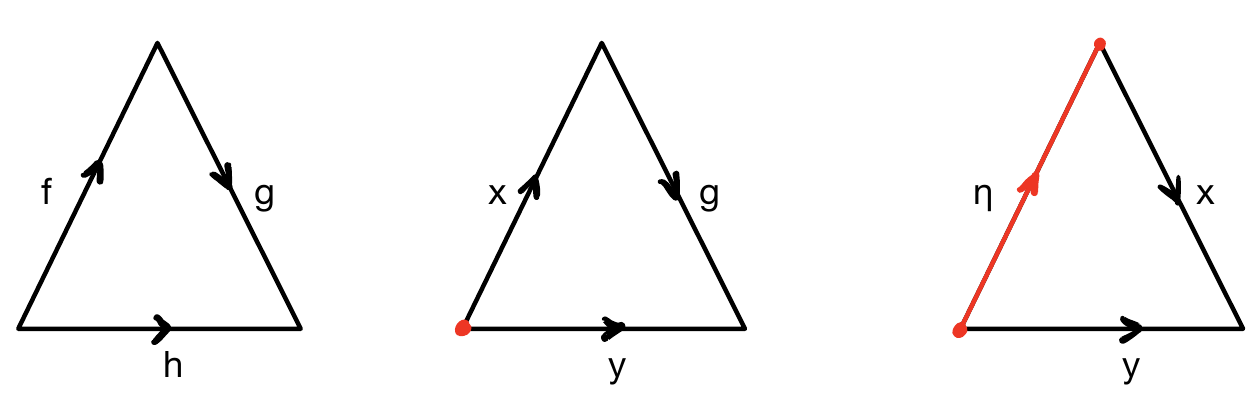}
	\caption{Admissible colorings of a flag-like triangulation.
		Here $f,g,h \in G$, $x,y \in X$ and $\eta \in H$.}
	\label{color}
\end{figure}

In DW theory without defects, the construction proceeds by choosing appropriate an appropriate group $G$ whose elements will label the edges of the triangles in the triangulation.
In the presence of defects, we need to pick two finite groups $G$, $H$ and a space $X$ with a right $G$-action and a left $H$-action.
Physically $G$ specifies the degrees of freedom in the bulk, $H$ the degres of freedom on the defect and $X$ is a way of coupling (or not) the two together.
Then we assign elements of $G$, $H$ and $X$ on edges of the triangles from figure \ref{flag}, in a consistent way as shown in fig \ref{color}.

Now we can define the partition function of DW theory.
\begin{definition}
	Let $G$, $H$ be finite groups and $X$ a set equipped with commuting group action of $G$ on the right and $H$ on the left.
	Then, for any flag -like triangulation $T$ of a surface-curve pair $\Sigma \supset  C $, the partition function of untwisted DW theory is defined to be
	\begin{align}
	Z_{(H,X,G)}(\Sigma \supset  C ) := |G|^{-\sigma} |H|^{-c} \kappa_{(H,X,G)}(T)
	\end{align}
	where $\kappa_{(H,X,G)}(T) $ is the number of admissible colorings of the triangulation, $\sigma$ is the number of bulk vertices, and $c$ is the number of vertices lying on the defect.
\end{definition}

The method of \cite{doughterty20152dimensional} for defining the partition function of 2d DW with 1d defects on a closed manifold $\Sigma$ begins by choosing two groups $G$, $H$ and a space $X$ on which $G$ has a right action and $H$ has a left action.
The physical interpretation of these choices are as the dofs of the DW, the dofs on the defects, and a coupling between these dofs, respectively.
Then they introduce a triangulation of $\Sigma$, that is made out of three basic triangles, see figure \ref{flag}.
A triangulation obtained from these triangles they call flag-like, and then the partition function of untwisted DW is essentially the counting of possible group element assignments on each edge of the triangulation, which is the lattice equivalent of counting isomorphism classes of bundles,
\begin{align}
Z_{(H,X,G)}(\Sigma \supset C) := |G|^{-|T^0_0|} |H|^{-|T^1_0|} \kappa_{(H,X,G)}(T) \label{Yetter}
\end{align}
where $T^k_n$ denote the set of n-simplices of $T$ with $k$ vertices on the curve, and $\kappa_{(H,X,G)}(T)$ is the number of admissible ${(H,X,G)}$-colorings of T.
Then they show that this is independent of the flag-like triangulation chosen, and hence a topological invariant of the surface-curve pair $\Sigma \supset C$.

\subsection{Generalization and importance for our methods}
Although, in \cite{doughterty20152dimensional} they restrict attention to closed manifolds $\Sigma$, their methods can be straightforwardly generalized to manifolds with fixed holonomies and/or parallel transports on their boundaries.
So using \eqref{Yetter}, we can reproduce all of the DW TQFT rules \eqref{r1}-\eqref{r5}.
The simplest choice for $X$ is the direct product $X=G \times H$.
With this choice, we label the brane boundaries by parallel transports $(h,1) \in X$, and since the brane boundaries are dynamical we will sum over all those $h$.
Each non-dynamical, open boundary has to be made up by at least two different edges, since there is no 2-simplex with two defect vertices and no defect edge.
Let $g \in G$ be the parallel transport specified along a non-dynamical, open boundary, then we will model this boundary by allowing a non zero $H$ parallel transport along it, so that the total transport is $(h,g) \in X$ for some $h \in H$.
Other than reproducing \eqref{r1}-\eqref{r5}, the lattice perspective helps visualize how a (small) gauge $G$ transformation wouldn't act on the boundary, since the brane boundaries are fixed to have parallel transport $(h,1)$.

\bibliographystyle{JHEP}
\bibliography{2dTQFTs_and_BUs}

\providecommand{\href}[2]{#2}\begingroup\raggedright\begin{thebibliography}{10}

\bibitem{Marolf:2020xie}
D.~Marolf and H.~Maxfield, \emph{{Transcending the ensemble: baby universes,
  spacetime wormholes, and the order and disorder of black hole information}},
  \href{https://doi.org/10.1007/JHEP08(2020)044}{\emph{JHEP} {\bfseries 08}
  (2020) 044} [\href{https://arxiv.org/abs/2002.08950}{{\ttfamily
  2002.08950}}].

\bibitem{maloney2020averaging}
A.~Maloney and E.~Witten, \emph{Averaging over narain moduli space},  2020.

\bibitem{cotler2020ads3}
J.~Cotler and K.~Jensen, \emph{Ads$_3$ gravity and random cft},  2020.

\bibitem{maxfield2020path}
H.~Maxfield and G.J.~Turiaci, \emph{The path integral of 3d gravity near
  extremality; or, jt gravity with defects as a matrix integral},  2020.

\bibitem{Coleman:1988cy}
S.R.~Coleman, \emph{{Black Holes as Red Herrings: Topological Fluctuations and
  the Loss of Quantum Coherence}},
  \href{https://doi.org/10.1016/0550-3213(88)90110-1}{\emph{Nucl. Phys. B}
  {\bfseries 307} (1988) 867}.

\bibitem{Giddings:1988wv}
S.B.~Giddings and A.~Strominger, \emph{{Baby Universes, Third Quantization and
  the Cosmological Constant}},
  \href{https://doi.org/10.1016/0550-3213(89)90353-2}{\emph{Nucl. Phys. B}
  {\bfseries 321} (1989) 481}.

\bibitem{Polchinski_1994}
J.~Polchinski and A.~Strominger, \emph{Possible resolution of the black hole
  information puzzle},
  \href{https://doi.org/10.1103/physrevd.50.7403}{\emph{Physical Review D}
  {\bfseries 50} (1994) 7403–7409}.

\bibitem{Maldacena:2004rf}
J.M.~Maldacena and L.~Maoz, \emph{{Wormholes in AdS}},
  \href{https://doi.org/10.1088/1126-6708/2004/02/053}{\emph{JHEP} {\bfseries
  02} (2004) 053} [\href{https://arxiv.org/abs/hep-th/0401024}{{\ttfamily
  hep-th/0401024}}].

\bibitem{McNamara:2020uza}
J.~McNamara and C.~Vafa, \emph{{Baby Universes, Holography, and the
  Swampland}},  \href{https://arxiv.org/abs/2004.06738}{{\ttfamily
  2004.06738}}.

\bibitem{maldacena2016conformal}
J.~Maldacena, D.~Stanford and Z.~Yang, \emph{Conformal symmetry and its
  breaking in two dimensional nearly anti-de-sitter space},  2016.

\bibitem{Kitaev_2018}
A.~Kitaev and S.J.~Suh, \emph{The soft mode in the sachdev-ye-kitaev model and
  its gravity dual},
  \href{https://doi.org/10.1007/jhep05(2018)183}{\emph{Journal of High Energy
  Physics} {\bfseries 2018} (2018) }.

\bibitem{saad2019jt}
P.~Saad, S.H.~Shenker and D.~Stanford, \emph{Jt gravity as a matrix integral},
  2019.

\bibitem{witten2020matrix}
E.~Witten, \emph{Matrix models and deformations of jt gravity},  2020.

\bibitem{Almheiri:2019qdq}
A.~Almheiri, T.~Hartman, J.~Maldacena, E.~Shaghoulian and A.~Tajdini,
  \emph{{Replica Wormholes and the Entropy of Hawking Radiation}},
  \href{https://doi.org/10.1007/JHEP05(2020)013}{\emph{JHEP} {\bfseries 05}
  (2020) 013} [\href{https://arxiv.org/abs/1911.12333}{{\ttfamily
  1911.12333}}].

\bibitem{penington2020replica}
G.~Penington, S.H.~Shenker, D.~Stanford and Z.~Yang, \emph{Replica wormholes
  and the black hole interior},  2020.

\bibitem{Afkhami-Jeddi:2020ezh}
N.~Afkhami-Jeddi, H.~Cohn, T.~Hartman and A.~Tajdini, \emph{{Free partition
  functions and an averaged holographic duality}},
  \href{https://arxiv.org/abs/2006.04839}{{\ttfamily 2006.04839}}.

\bibitem{Cotler:2020hgz}
J.~Cotler and K.~Jensen, \emph{{AdS$_3$ wormholes from a modular bootstrap}},
  \href{https://arxiv.org/abs/2007.15653}{{\ttfamily 2007.15653}}.

\bibitem{Balasubramanian:2020jhl}
V.~Balasubramanian, A.~Kar, S.F.~Ross and T.~Ugajin, \emph{{Spin structures and
  baby universes}},  \href{https://arxiv.org/abs/2007.04333}{{\ttfamily
  2007.04333}}.

\bibitem{Kapec:2019ecr}
D.~Kapec, R.~Mahajan and D.~Stanford, \emph{{Matrix ensembles with global
  symmetries and \textquoteright{}t Hooft anomalies from 2d gauge theory}},
  \href{https://doi.org/10.1007/JHEP04(2020)186}{\emph{JHEP} {\bfseries 04}
  (2020) 186} [\href{https://arxiv.org/abs/1912.12285}{{\ttfamily
  1912.12285}}].

\bibitem{Atiyah}
M.F.~Atiyah, \emph{Topological quantum field theory}, {\emph{Publications
  Math\'ematiques de l'IH\'ES} {\bfseries 68} (1988) 175}.

\bibitem{graczyk2003}
P.~Graczyk, G.~Letac and H.~Massam, \emph{The complex wishart distribution and
  the symmetric group},
  \href{https://doi.org/10.1214/aos/1046294466}{\emph{Ann. Statist.} {\bfseries
  31} (2003) 287}.

\bibitem{Dijkgraaf:1989pz}
R.~Dijkgraaf and E.~Witten, \emph{{Topological Gauge Theories and Group
  Cohomology}}, \href{https://doi.org/10.1007/BF02096988}{\emph{Commun. Math.
  Phys.} {\bfseries 129} (1990) 393}.

\bibitem{Freed:1991bn}
D.S.~Freed and F.~Quinn, \emph{{Chern-Simons theory with finite gauge group}},
  \href{https://doi.org/10.1007/BF02096860}{\emph{Commun. Math. Phys.}
  {\bfseries 156} (1993) 435}
  [\href{https://arxiv.org/abs/hep-th/9111004}{{\ttfamily hep-th/9111004}}].

\bibitem{jones_1998}
G.A.~Jones, \emph{Characters and surfaces: a survey},  in \emph{The Atlas of
  Finite Groups - Ten Years On}, R.T.~Curtis and R.A.~Wilson, eds., London
  Mathematical Society Lecture Note Series, p.~90–118, Cambridge University
  Press (1998), \href{https://doi.org/10.1017/CBO9780511565830.013}{DOI}.

\bibitem{mednykh_nonequivalent_1984}
A.D.~Mednykh, \emph{Nonequivalent coverings of {Riemann} surfaces with a
  prescribed ramification type},
  \href{https://doi.org/10.1007/BF00968900}{\emph{Siberian Mathematical
  Journal} {\bfseries 25} (1984) 606}.

\bibitem{Turaev_2007}
V.~Turaev, \emph{Dijkgraaf–witten invariants of surfaces and projective
  representations of groups},
  \href{https://doi.org/10.1016/j.geomphys.2007.08.009}{\emph{Journal of
  Geometry and Physics} {\bfseries 57} (2007) 2419–2430}.

\bibitem{Carqueville:2017fmn}
N.~Carqueville and I.~Runkel, \emph{{Introductory lectures on topological
  quantum field theory}}, \href{https://doi.org/10.4064/bc114-1}{\emph{Banach
  Center Publ.} {\bfseries 114} (2018) 9}
  [\href{https://arxiv.org/abs/1705.05734}{{\ttfamily 1705.05734}}].

\bibitem{lurie}
J.~Lurie, \emph{On the classification of topological field theories},  in
  \emph{Current Developments in Mathematics, 2008}, vol.~Volume 2008 of
  \emph{Current Developments in Mathematics}, (Boston, MA), pp.~129--280,
  International Press of Boston (2009),
  \href{https://projecteuclid.org/euclid.cdm/1254748657}{https://projecteuclid.org/euclid.cdm/1254748657}.

\bibitem{Durhuus_1994}
B.~Durhuus and T.~Jonsson, \emph{Classification and construction of unitary
  topological field theories in two dimensions},
  \href{https://doi.org/10.1063/1.530752}{\emph{Journal of Mathematical
  Physics} {\bfseries 35} (1994) 5306–5313}.

\bibitem{Carqueville_2018}
N.~Carqueville, \emph{Lecture notes on two-dimensional defect tqft},
  \href{https://doi.org/10.4064/bc114-2}{\emph{Banach Center Publications}
  {\bfseries 114} (2018) 49–84}.

\bibitem{Cardy:2004hm}
J.L.~Cardy, \emph{{Boundary conformal field theory}},
  \href{https://arxiv.org/abs/hep-th/0411189}{{\ttfamily hep-th/0411189}}.

\bibitem{Almheiri:2019psf}
A.~Almheiri, N.~Engelhardt, D.~Marolf and H.~Maxfield, \emph{{The entropy of
  bulk quantum fields and the entanglement wedge of an evaporating black
  hole}}, \href{https://doi.org/10.1007/JHEP12(2019)063}{\emph{JHEP} {\bfseries
  12} (2019) 063} [\href{https://arxiv.org/abs/1905.08762}{{\ttfamily
  1905.08762}}].

\bibitem{Blommaert:2018iqz}
A.~Blommaert, T.G.~Mertens and H.~Verschelde, \emph{{Fine Structure of
  Jackiw-Teitelboim Quantum Gravity}},
  \href{https://doi.org/10.1007/JHEP09(2019)066}{\emph{JHEP} {\bfseries 09}
  (2019) 066} [\href{https://arxiv.org/abs/1812.00918}{{\ttfamily
  1812.00918}}].

\bibitem{Almheiri_2020}
A.~Almheiri, R.~Mahajan, J.~Maldacena and Y.~Zhao, \emph{The page curve of
  hawking radiation from semiclassical geometry},
  \href{https://doi.org/10.1007/jhep03(2020)149}{\emph{Journal of High Energy
  Physics} {\bfseries 2020} (2020) }.

\bibitem{doughterty20152dimensional}
A.L.~Doughterty, H.~Park and D.N.~Yetter, \emph{On 2-dimensional
  dijkgraaf-witten theory with defects},  2015.

\end{thebibliography}\endgroup

\end{document}